\documentclass[showpacs,twocolumn,preprintnumbers,amsmath,amssymb,aps,nofootinbib]{revtex4}

\newcommand{\step}{\vspace{.5em}}
\newcommand{\smallstep}{\vspace{.0em}}

\def\di{\displaystyle}

\def\bg{\begin{eqnarray}\begin{array}{rcl}\displaystyle}
\def\eg{\end{array} &\di    &\di   \end{eqnarray}}
\def\bm#1{\begin{eqnarray}\begin{array}{#1}\di}
\def\bmo#1{\begin{eqnarray*}\begin{array}{#1}\di}
\def\bml#1#2{\begin{eqnarray}\begin{array}{#1}\label{#2}\di}
\def\bgo{\begin{eqnarray*}\begin{array}{rcl}\displaystyle}
\def\ego{\end{array} &\di    &\di \nonumber  \end{eqnarray*}}

\def\btensor#1#2{\renew\left#1\begin{array}{#2}\di}
\def\brtensor#1#2#3{\ren#3\left#1\begin{array}{#2}}
\def\botensor#1#2{\renew\left#1\begin{array}{#2}}
\def\etensor#1{\end{array}\right#1}

\def\eq#1{(\ref{#1})}
\def\Eq#1{Eq.~(\ref{#1})}

\def\d{{d}}
\def\tr{{\rm tr}}
\def\Tr{{\rm Tr}}

\def\id{1\!\mbox{l}}
\def\nin{\in\llap{\hspace{-.15cm}/\hspace{.15cm}}}

\def\s0#1#2{\mbox{\small{$ \frac{#1}{#2} $}}}
\def\0#1#2{\frac{#1}{#2}}

\def\p{\partial\llap{/}}

\def\dr{{D\!\llap{/}}\,}

\def\A{A\!\llap{/}}

\def\R{{{\rm l}\!{\rm R}}}
\def\N{{{\rm l}\!{\rm N}}}

\def\CA{{\mathcal A}}

\def\CF{{\mathcal F}}
\def\CG{{\mathcal G}}

\def\CI{{\mathcal I}}

\def\CO{{\mathcal O}}
\def\CP{{\mathcal P}}

\def\CS{{\mathcal S}}
\def\CT{{\mathcal T}}

\def\CV{{\mathcal V}}
\def\CW{{\mathcal W}}

\newcommand{\Gg}{{\mathfrak g}}

\newcommand{\Gn}{{\mathfrak n}}

\newcommand{\Gs}{{\mathfrak s}}

\renewcommand{\theequation}{\arabic{section}.\arabic{equation}}

\def\ren#1{\renewcommand{\arraystretch}{#1}}

\def\renew{\renewcommand{\arraystretch}{1}}

\begin{document}

\title{Aspects of the Functional Renormalisation Group}

\pacs{05.10.Cc,11.10.Gh,11.10.Hi,11.15.-q\vspace{-.2cm}}

\author{
Jan M. Pawlowski
}

\affiliation{
Institut f\"ur Theoretische Physik, Universit\"at Heidelberg, \\  
Philosophenweg 16, D-69120 Heidelberg, Germany.\\ 
\small\sf j.pawlowski@thphys.uni-heidelberg.de\\\vspace{-.39cm}}
\preprint{HD-THEP-05-28}
\preprint{hep-th/0512261}

\thispagestyle{empty}

\abstract{We discuss structural aspects of the functional
  renormalisation group. Flows for a general class of correlation
  functions are derived, and it is shown how symmetry relations of the
  underlying theory are lifted to the regularised theory. A simple
  equation for the flow of these relations is provided. The setting
  includes general flows in the presence of composite operators and
  their relation to standard flows, an important example being $N$PI
  quantities. We discuss optimisation and derive a functional optimisation
  criterion.
  
  Applications deal with the interrelation between functional flows
  and the quantum equations of motion, general Dyson-Schwinger
  equations. We discuss the combined use of these functional equations
  as well as outlining the construction of practical renormalisation
  schemes, also valid in the presence of composite operators.
  Furthermore, the formalism is used to derive various representations
  of modified symmetry relations in gauge theories, as well as
  to discuss gauge-invariant flows. We close with the construction and
  analysis of truncation schemes in view of practical optimisation.
\vspace{-.4cm}}}

\maketitle


\newpage
\pagestyle{plain}
\setcounter{page}{1}

\setcounter{equation}{0}
\setcounter{section}{1}

\pagenumbering{roman}

\tableofcontents

\setcounter{section}{0}
\ 
\newpage
\

\newpage 

\pagestyle{plain}

\setcounter{page}{0}

\pagenumbering{arabic}

\section{Introduction}
The Functional Renormalisation Group (FRG) in its continuum
formulation
\cite{Wilson:1971bg,Wilson:1971dh,Wilson:1973jj,Wegner:1972ih,Nicoll:1977hi,Polchinski:1983gv,Warr:1986we,Hurd:1989up,Keller:1990ej,Wetterich:1992yh,Bonini:1992vh,Ellwanger:1993mw,Morris:1993qb,Becchi:1996an}
has proven itself as a powerful tool for studying both perturbative
and non-perturbative effects in quantum field theory and statistical
physics, for reviews see
\cite{Fisher:1998kv,Litim:1998nf,Morris:1998da,Aoki:2000wm,Bagnuls:2000ae,Berges:2000ew,
  Polonyi:2001se,Salmhofer:2001tr,Delamotte:2003dw}. In this approach
a regularisation of a quantum theory is achieved by suppressing part
of the propagating degrees of freedom related to a cut-off scale $k$.
This results in regularised generating functionals such as the
effective action $\Gamma_{k}$ where part of the modes have been
integrated out.  The flow equation describes the response of the
generating functional to an infinitesimal variation of $k$, and can be
used to successively integrate-out modes. Hence, a generating
functional at some initial scale $\Lambda$ together with its flow
serve as a definition of the quantum theory. For example, the flow
equation allows us to calculate the full effective action $\Gamma$
from an initial effective action $\Gamma_{\Lambda}$ if the latter is
well under control. For an infrared momentum cut-off and sufficiently
large $\Lambda$ we have a good grip on $\Gamma_{\Lambda}$ as it can be
computed perturbatively.  \smallstep

The main advantages of such a formulation are its flexibility when it
comes to truncations of the full theory, as well as its numerical
accessibility.  Both properties originate in the same structural
aspects of such flows.  Quite generally functional flows are
differential equations that relate an infinitesimal $k$-variation of a
generating functional $Z$ with some functional of $Z$, its derivatives
and the regulator. The quantum theory, and hence the physics, is
solely specified by the boundary condition of such a flow.  Due to
this structure truncations are introduced on the level of the
generating functional itself which leads to self-consistent truncated
flows.  Moreover, a change of degrees of freedom also is done on the
level of the generating functional, and the structure of the flow
stays the same.  Last but not least, numerical stability of the flow
for a given problem and truncation is governed by the choice of
the specific regularisation procedure. \smallstep

In other words, the advantages are carried by the structural aspects
of the functional RG, whose understanding and further development is
the main purpose of the present work. It is not meant as a review and
for a more complete list of references we refer the reader to the
reviews already cited above,
\cite{Fisher:1998kv,Litim:1998nf,Morris:1998da,Aoki:2000wm,Bagnuls:2000ae,Berges:2000ew,
  Polonyi:2001se,Salmhofer:2001tr,Delamotte:2003dw}. We
close the introduction with an overview over the work. \smallstep

In section~\ref{sec:general} we evaluate functional equations of
quantum field theories, such as Dyson-Schwinger equations, symmetry
identities, such as Slavnov-Taylor identities (STIs), and introduce
some notation.  \smallstep

In section~\ref{sec:flows} flows are derived for general correlation
functions including those for the effective action and the Schwinger
functional.  We present a derivation of the flow equation which
emphasises the subtleties of renormalisation. Moreover, no use of the
path integral representation is made, the derivation solely relies on
the existence of a finite effective action or Schwinger functional for
the full theory. First we introduce the setting and notion of
regularisation.  This is used to derive the general flows
\eq{eq:flowIk} and \eq{eq:flowtIk} which comprise the main results of
this part. The flows discussed here include those for $N$-particle
irreducible ($N$PI) quantities as well as relations between the
different formulations. For general flows one has to carefully study
the boundary conditions. A comparison of results obtained for
different regularisations, in particular in view of optimisation,
requires the study of variations of the regulator.

In section~\ref{sec:RG-flows} we discuss the fate of RG equations of
the full theory displaying reparameterisation invariance in the
presence of a general regularisation. This is important when matching
the scale dependence of quantities in the presence of the
regularisation to that in the full theory without cut-off. The key RG
flows are \eq{eq:RGflowIk},\eq{eq:RGflowtIk} and are basically
generalisations of \eq{eq:flowIk} and \eq{eq:flowtIk}. \smallstep

The important aspect of optimisation is investigated in
section~\ref{sec:optimal}. In most situations one has to rely on
truncations to the full theory. Optimised flows should lead to results
as close as possible to the full theory within each order of a given
systematic truncation scheme. We develop a functional approach to
optimisation of general flows which allows us to systematically access
and develop optimisation criteria. We discuss the relation between
different optimisation ideas used in the literature. The definition of
an effective cut-off scale is introduced and a constructive
optimisation criterion is put forward in section~\ref{sec:functopt}.
Roughly speaking, optimal regulators are those, that lead to
correlation functions as close as possible to that in the full theory
for a given effective cut-off scale. \smallstep

The rest of the present paper deals with structural applications of
these findings. In section~\ref{sec:applications} we relate flows to
other functional methods such as Dyson-Schwinger equations or the use
of $N$PI effective actions. To that end we consider flows in the
presence of composite operators. In particular we construct practical
renormalisation schemes, the latter being of importance for the
renormalisation of Dyson-Schwinger equations and $N$PI effective
actions. \smallstep

A main motivation for the development of the present approach resides
in its application to gauge theories. In section~\ref{sec:gauge}
various structural aspects of gauge theories are investigated. We
discuss the formulation of gauge theories using appropriate degrees of
freedom. The modification of symmetry identities in the presence of
the regularisation and their different representations are evaluated.
The latter allow for a purely algebraic representation of the symmetry
identities. We also outline the construction of gauge-invariant flows
and discuss the fate of gauge symmetry constraints in these
formulations. We close with a brief evaluation of anomalous symmetries
in the presence of a regulator. \smallstep

In section~\ref{sec:optimalatwork} we discuss consequences of the 
functional optimisation criterion and the RG equations for the 
construction of truncation schemes and optimal regulators. It is shown
that a specific class of regulators preserves the RG scalings of the
underlying theory. We discuss the use of integrated flows that
constitute finite renormalised Dyson-Schwinger equations. These
integrated flows can be used in asymptotic regimes or a fixed point
analysis within the functional RG setting. The constructive
optimisation criterion developed in section~\ref{sec:optimal} is put
to work within a simple example. Further applications are outlined.
\step

\section{Preliminaries}\label{sec:general}
We consider the {\it finite renormalised} Euclidean Schwin\-ger
functional $W[J]$ of the theory under investigation, where we do not only
allow for source terms for the fundamental fields $\hat\varphi$ of the
theory, but also for sources for general
tensorial composite operators $\hat\phi(\hat\varphi)$ with
\begin{eqnarray}\nonumber 
  &&\hspace{-1cm}e^{W[J]}=\int d\mu[\hat\varphi]\,\exp\Bigl\{-S[\hat\varphi]+
  \sum_{n=1}^{n_{\rm max}} \int_{x_1,...,x_n} \\
  &&\hspace{-.6cm} 
  J^{\alpha_1\cdots \alpha_n}(x_1,...,x_n)
  \,\hat\phi_{\alpha_1\cdots \alpha_n}[\hat\varphi](x_1,...,x_n)\Bigr\}\,.  
\label{eq:prepathI}\end{eqnarray} 
Here $\alpha_i$ comprises possible Lorentz and gauge group indices and
species of fields. The measure $d\mu[\hat\varphi]$ ensures the
finiteness of the Schwinger functional and hence depends on some
renormalisation scale $\mu$, as well as on $S[\hat\varphi]$. For the
sake of simplicity, and for emphasising the structure of the results,
we use a condensed notation with indices $a,b$ that stand for an
integration over space-time and a summation over internal indices:
\begin{eqnarray}\label{eq:condense}
  J^a \hat\phi_a=\int d^d x\, J^\alpha(x)\hat\phi_\alpha(x)\,, 
\end{eqnarray} 
In \eq{eq:condense} we have implicitly defined the ultra-local metric
$\gamma^a{}_{a'}=\delta(x-x')\, \gamma^{\alpha}_{\alpha'}$, leaving the 
internal part $\gamma^{\alpha}_{\alpha'}$ undetermined. In case
$\hat\phi_a$ involves fermionic variables we have $J^a\hat\phi_a\neq
\hat\phi_a J^a$. The notation as well as some properties of the metric
$\gamma^{ab}$ are detailed in appendix~\ref{app:notation}. In the
general case \eq{eq:prepathI} we consider the coupling of $N$
tensorial fields with rank $n_i\leq n_{i+1}$ to the theory. We
substitute indices $a$ by multi-indices ${\bf a}=a_{11}\cdots
a_{1n_1}, \cdots, a_{N 1}\cdots a_{N n_N}$ with $n_N=n_{\rm max}$. In
the general case, different $a_{ij}$ can carry different internal
indices, e.g.\ different representations of a gauge group relating to
different species of fields. This is implicitly understood and we
identify $a_{ij}=a_j$ from now on in a slight abuse of notation.
Contractions read
\begin{eqnarray}\label{eq:genmetriccon}
  T_1{}^{\bf a}\, T_2\,{}_{\bf a}
  =\sum_{i=1}^N T_1{}^{a_1\cdots a_{n_i}}\, 
  T_2\,{}_{a_{1}\cdots a_{n_i}} \,,
\end{eqnarray}
and the generalised metric $\gamma^{\bf ab}$ is defined as
\begin{eqnarray}\label{eq:genmetricdef} 
  (\gamma^{\bf ab}) =\bigoplus_{i=1}^N 
  (\otimes \gamma)^{n_i}\,.
\end{eqnarray}
The definitions in \eq{eq:genmetriccon},\eq{eq:genmetricdef} are
nothing but the extension of the field space to include composite
operators $\hat\phi_{a_1\cdots a_{n}}$. The interest in such a general
setting is twofold: firstly, it allows us to formulate, at all scales,
the theory in terms of physically relevant degrees of freedom.
Secondly, it naturally includes the coupling to composite operators
and related flows.  The source term in the Schwinger functional
\eq{eq:prepathI} reads
\begin{eqnarray}\label{eq:gensource1} 
  J^{{\bf a}}\hat\phi_{{\bf a}}=\sum_{i=1}^N 
  J^{a_1\cdots a_{n_i}}\, 
  \hat\phi_{a_1\cdots a_{n_i}}\,.
\end{eqnarray}  
For $n_i=1$ for all $i$ the general source term \eq{eq:gensource1}
boils down to the standard source \eq{eq:condense}. A simple tensorial
example is given by ${\bf a}=a,a_1 a_2$ and $\hat\phi_{\bf
  a}=({\hat\phi}_a,\hat\phi_{a_1 a_2})=
({\hat\varphi}_a,\hat\varphi_{a_1}\hat \varphi_{a_2})$ with
$a=a_1=a_2=x$, a scalar field and its two-point function. This leads
to a source term
\begin{eqnarray}\nonumber 
  && \hspace{-1.3cm} J^{\bf a} \hat\phi_{\bf a}=\int d^d x\, 
  J(x)\hat\varphi(x)\\
  && \hspace{.8cm} +\int d^d x\,d^d y\, J(x,y)\hat\varphi(x) 
  \hat\varphi(y)\,. 
\label{eq:condense2PI}\end{eqnarray} 
The above example also emphasises that the sources $J^{\bf a}$ should
be restricted to those sharing the (index-) symmetries of the fields
$\hat\phi_{\bf a}$. We illustrate this within the above example of a
scalar field. The source term for $\hat\phi_{a_1
  a_2}=\hat\varphi_{a_1} \hat\varphi_{a_2}$ satisfies $J^{a_1
  a_2}\phi_{a_1}\phi_{a_2}=J^{(a_1 a_2)}\phi_{a_1}\phi_{a_2}$, where
$J^{(a_1 a_2)}=\s012 (J^{a_1 a_2}+J^{a_2 a_1})$ is the symmetric part
of $J$. The anti-symmetric part $J^{[a_1 a_2]}=\s012 (J^{a_1
  a_2}-J^{a_2 a_1})$ does not couple to the field, $J^{[a_1 a_2]} 
\phi_{a_1 a_2}=0$. Consequently we restrict the sources
to the symmetric ones. The symmetry properties of a function $J^{\bf
  a}$ or $\phi^{\bf a}$ are also carried by its derivatives.  Again we
illustrate this by the example introduced above: derivatives
w.r.t.\ the function $J^{{(\bf a)}}$ carry its symmetry properties.
This entails that
\begin{eqnarray}\label{eq:symmprops} 
  \0{\delta F[J]}{\delta J^{{(\bf a)}}}=F_{,{(\bf a)}}=\Bigl(F_{,a}\,,\,
  \s012\left( 
    F_{,{a_1 a_2}}+F_{,{a_2 a_1}}\right)\Bigr)\,,   
\end{eqnarray} 
where $J^{(a)}=J^a$.  The basic example is the derivative of $J$ w.r.t.\
$J$. It reads
\begin{eqnarray}\label{eq:basicsymmprops}
  \0{\delta J^{({\bf b})}}{\delta J^{({\bf a})}}=
  \delta^{({\bf b})}_{({\bf a})}=\Bigl(\delta^b_a\,,\,
  \s012\left( 
    \delta^{b_1}_{a_1}\delta^{b_2}_{a_2}+
    \delta^{b_1}_{a_2}\delta^{b_2}_{a_1}\right)\Bigr)\,, 
\end{eqnarray}
the second entry on the rhs is the identity kernel in the symmetric
subspace. We also have $J^{(\bf a)}_{,[\bf b]}=0$ with
$J^{[a]}=0$, and get $J^{[\bf a]}_{,[\bf b]}=(0\,,\,\s012\left(
  \delta^{b_1}_{a_1}\delta^{b_2}_{a_2}-
  \delta^{b_1}_{a_2}\delta^{b_2}_{a_1}\right))$. From now on we
suppress this detail. Derivatives are always taken within the
appropriate spaces defined by the corresponding projections, and carry
the related symmetry properties. \smallstep

Within the above conventions the Schwinger functional \eq{eq:prepathI}
reads
\begin{eqnarray}\label{eq:pathI}
  e^{W[J]}=\int d\mu[\hat\varphi]\,\exp\{-S[\hat\varphi]+J^{\bf a}
  \hat\phi_{\bf a}
  (\hat\varphi)\}\,.
\end{eqnarray} 
Many of the structural results presented here can be already
understood within a scalar theory with a single field. There we have
${\bf a}=a=x$ with the ultra-local metric
$\gamma^a{}_{a'}=\delta(x-x')$. In these cases one can simply ignore
the additional notational subtleties in the presence of fermions and
tensorial fields.  \smallstep

The definition \eq{eq:pathI} is rather formal. For most interacting
theories it is impossible to strictly prove the non-perturbative
existence of $d\mu[\hat\varphi]\exp\{ -S[\hat\varphi]\}$, not to
mention determining it in a closed form. Here we follow a bootstrap
approach in simply assuming that a finite $W[J]$ exists.  This
assumption is less bold than it seems at first sight. It is merely the
statement that the classical action $S[\hat\varphi]$ admits a
well-defined quantum field theory in terms of appropriately chosen
fields $\hat\phi(\hat\varphi)$. Then quite general normalised
expectation values $I[J]=\langle \hat I[J,\hat\phi]\rangle$ are
defined by
\begin{eqnarray}\label{eq:I0}
  I[J]=e^{-W[J]} \hat I[J,\s0{\delta}{\delta J}] e^{W[J]}\,. 
\end{eqnarray} 
The $I$ include correlation functions that relate to 
one particle irreducible (1PI) as well as connected and disconnected
Green functions in $\phi$. Subject to the definition of $\hat\phi$ 
 this may include $N$PI Green
functions in the fundamental fields $\hat\varphi$. As an important
sub-class included in \eq{eq:I0} we present normalised $N$-point
functions
\begin{subequations}\label{eq:moments}
\begin{eqnarray}\label{eq:momentsI}
  I^{(N)}_{{\bf a}_1\cdots {\bf a}_N}=
  \langle \prod_{i=1}^N\hat\phi_{{\bf a}_i} \rangle\,,
\end{eqnarray}
with
\begin{eqnarray}\label{eq:momentshatI}
  \hat I^{(N)}_{{\bf a}_1\cdots {\bf a}_N}=
  \prod_{i=1}^N\0{\delta}{\delta J^{{\bf a}_i}}\,.
\end{eqnarray} 
\end{subequations}
The correlation functions \eq{eq:moments} include all moments of the
Schwinger functional and their knowledge allows the construction of
the latter.  A simple example for \eq{eq:moments} is $\langle
\hat\phi\rangle$, the expectation value of the
operator $\hat\phi$ coupled to the current with $\hat
I^{(1)}=\s0{\delta}{\delta J}$. We briefly illustrate the construction
of connected or 1PI Green functions by an important example, the full
propagator. With the 1-point function $\langle\hat\phi\rangle$, the
propagator $W_{,{\bf a}_1 {\bf a}_2}[J]= \langle \hat\phi_{{\bf
    a}_1}\hat\phi_{{\bf a}_2}\rangle_{\rm 1PI}$ follows as $I_{{\bf
    a}_1 {\bf a}_2}^{(2)}- I_{{\bf a}_1}^{(1)}I_{{\bf a}_2}^{(1)}$.
\smallstep

Further important examples are correlation functions $I$ where $\hat
I[J,\s0{\delta}{\delta J}]$ generates a symmetry of the theory at
hand.  Let us first consider general Dyson-Schwinger (DS) equations,
formally given by
\begin{eqnarray}\label{eq:genDS} 
  \int \CG[\hat\varphi] \left(
    d\mu[\hat\varphi]\,\Psi[\hat\phi]\,e^{-S[\hat\varphi]+J^{\bf a}
      \hat\phi_{\bf a}[\hat\varphi]}
  \right)=0\,, 
\end{eqnarray} 
with bosonic operator $\CG$.  For \eq{eq:genDS} to hold the operator
$\CG[\hat\varphi]$ has to generate a symmetry of the path integral.
For infinitesimal transformations $\CG$, \eq{eq:genDS} translates into
\begin{subequations}\label{eq:DSgen}
\begin{eqnarray}\label{eq:DSI0}
  I[J]= 0\,,
\end{eqnarray}
with 
\begin{eqnarray}\label{eq:DShatI0}
  \hspace{-.7cm}\hat I=\left[ (\CG\Psi)-\Psi\,(\CG S) 
    +\Psi\,J^{\bf a}\, (\CG\hat\phi)_{\bf a}\right](\hat\phi=
  \s0{\delta}{\delta J})
  \,.
\end{eqnarray}
\end{subequations}
In \eq{eq:DSgen} we have assumed $(\CG\d\mu)=0$. We emphasise that
this can be easily achieved by reducing $\d\mu$ to the flat measure
with $\d\mu = \d\hat\varphi \,\Psi_1$. With $\d\mu'= \d\hat\varphi$
and $\Psi'=\Psi_1 \Psi$ we substitute $\d\mu \Psi=\d\mu'\Psi'$. The
simplest relevant example for \eq{eq:DSgen} is provided by the
standard DS equations.  They encode translation invariance of the flat
measure $d\hat\varphi$.  Accordingly, the standard DS equations are
obtained with $\hat\phi=\hat\varphi$,
$\CG[\hat\phi]=\s0{\delta}{\delta \hat\phi}$, $d\mu=d\hat\phi$ and
$\Psi=1$. Within this choice we arrive at
\begin{subequations}\label{eq:DS}
\begin{eqnarray}\label{eq:DSI}
  I_{\rm DSE}[J]=J-\langle 
  \0{\delta S }{\delta  \hat\phi}\rangle=0\,, 
\end{eqnarray}
with
\begin{eqnarray}\label{eq:DShatI}
  \hat I_{\rm DSE}=  
  J-\0{\delta S}{\delta \hat\phi}(\hat\phi=
  \s0{\delta}{\delta J})\,.  
\end{eqnarray}
\end{subequations}
\Eq{eq:DS} is the well-known functional Dyson-Schwinger equation. It
assumes a multiplicative renormalisation procedure preserving all
symmetries ($d\mu=d\hat\varphi$, $\Psi=1$).  When additive
renormalisation is required, or when we study a renormalisation
procedure breaking the symmetries of the classical action, this can be
captured in a non-trivial $\Psi$ \footnote{More precisely it is
  captured in a non-trivial $d\mu$ absorbed in $\Psi'$.}.

In case $\CG$ generates a symmetry of the action, $\CG S=0$, the above
relation simplifies. Restricting ourselves also to invariant
functionals $\Psi$ with $\CG \Psi=0$ we are led to
\begin{subequations}\label{eq:symgen}
\begin{eqnarray}\label{eq:symI0}
  I^\alpha[J]=0\,, 
\end{eqnarray}
with
\begin{eqnarray}\label{eq:symhatI0}
  \qquad & \hat I^\alpha=  \Psi\,J^{\bf a}\, \CG^\alpha\hat\phi_{\bf a}\,,  
\end{eqnarray} 
\end{subequations}
where $\alpha$ carries the group structure of the symmetry.  In
\eq{eq:symgen} we have used the bosonic nature of $\CG$ as well as
assuming that the symmetry is maintained within the quantisation:
$(\CG^\alpha d\mu)=0$. It is often possible and helpful to rewrite
symmetries in terms of derivative operators $\CG$ with
$\CG^2\hat\phi=0$.  This might necessitate the introduction of
auxiliary fields.  For example, in a gauge theory we deal with the
BRST symmetry with $\CG=\Gs$, the BRST derivative.  We add source
terms for $\CG\hat\phi$ with $J^{\bf a} \hat\phi_{\bf a}\to J^{\bf a}
\hat\phi_{\bf a}+Q^{\bf a} (\CG\hat\phi)_{\bf a}$.  The Schwinger functional
$W=W[J,Q]$ is a functional of both, $J$ and $Q$, and we are led to
\begin{eqnarray}\label{eq:brst} 
  &&\hspace{-.8cm}
  \hat I_\Gs= J^{\bf a} \0{\delta}{\delta Q^{\bf a}} \quad {\rm and} 
  \quad  I_\Gs=
  J^{\bf a} \0{\delta W[J,Q]}{\delta Q^{\bf a}}=0\,.
\end{eqnarray} 
We conclude that the set of $I$ defined in \eq{eq:I0} provides the
full information about the quantum theory as it spans the set of all
correlations functions $\{\CO\}$.  In this context we emphasise again
that not all correlation functions of interest are directly given by
the correlation functions $I$, a simple example being the propagator
$W_{,\bf ab}=I_{{\bf a}_1 {\bf a}_2}^{(2)}- I_{{\bf a}_1}^{(1)}I_{{\bf
    a}_2}^{(1)}$. \smallstep

The key object in the present approach is the Schwin\-ger functional of
the theory, or some related generating functional. Often one
concentrates on the Wilsonian effective action $S_{\rm eff}[\Phi]$,
the generating functional for amputated connected Green functions. It
is defined by 
\begin{eqnarray}\label{eq:S_eff}
  S_{\rm eff}[\Phi]:=-W[S^{(2)}[0]\,\Phi]\,, 
\end{eqnarray} 
where $S^{(2)}[0]=\delta^2 S/(\delta \Phi)^2[\Phi=0]$. The advantage
of working with the Schwinger functional $W$ or $S_{\rm eff}$ is that
it allows for the most straightforward derivation of functional
identities. However, a more tractable object is the effective action
$\Gamma$, the generating function of 1PI Green functions of
$\phi=\langle \hat\phi\rangle$. It is obtained as the Legendre
transform of $W$,
\begin{eqnarray}\label{eq:Gamma}
  \Gamma[\phi]=\sup_{\mbox{\tiny $ J$}} \left(J^{\bf a} \phi_{\bf a} 
    -W[J]\right)\,.  
\end{eqnarray}
\Eq{eq:Gamma} includes $N$PI effective actions \cite{Luttinger:1960ua,
  Baym:1962sx,Cornwall:1974vz} for an appropriate choice of $\phi_{\bf
  a}[\varphi]$. The definition \eq{eq:Gamma} leads to  
\begin{subequations}
  \begin{eqnarray} \label{eq:Jphi1}
  \Gamma^{,{\bf a}}[\phi]&=&\gamma^{\bf a}{}_{\bf b} J^{\bf b}(\phi)\,,\\
  W_{,{\bf a}}[J]& =&\phi_{\bf a}(J)\,,
  \label{eq:Jphi2} 
\end{eqnarray}
\label{eq:Jphi}\end{subequations}
implying that the field $\phi$ is the mean field,
$\phi=\langle\hat\phi\rangle$. In \eq{eq:Jphi} we have used that
$J^{\bf a}\phi_{\bf a} =\phi^{\bf a} J_{\bf a}= \phi_{\bf
  a}\gamma^{\bf a}{}_{\bf b }J^{\bf b}$. The derivatives in
\eq{eq:Jphi} are taken with respect to the variables of $\Gamma$ and
$W$ respectively, that is $\Gamma^{,{\bf a}}=\0{\delta\Gamma}{\delta
  \phi_{\bf a}}$ and $W_{,{\bf a}}[J]=\0{\delta W}{\delta J^{\bf a}}$.
Furthermore it follows that
\begin{eqnarray}\label{eq:props}
  \qquad W_{,{\bf ac}}
  \,\Gamma^{,{\bf cb}}=\gamma^{\bf b}{}_{\bf a}\,, 
\end{eqnarray}
The definition \eq{eq:I0} and the relation \eq{eq:DSgen} translate
into the corresponding equations in terms of 1PI quantities by using
\eq{eq:Jphi},\eq{eq:props} as well as
\begin{eqnarray} 
  W[J(\phi)]=\phi_{\bf a}
  \Gamma^{,{\bf a}}-\Gamma[\phi]\,,
  \label{eq:relation1}
\end{eqnarray}
and
\begin{eqnarray}
  \0{\delta}{\delta J^{\bf a}}
  =W_{,\bf ab} \0{\delta}{\delta \phi_{\bf b}}\,, 
  \label{eq:relation2} 
\end{eqnarray} 
For composite fields one usually splits up the irreducible part of
$\langle\hat\phi_{\bf a}\rangle$. As an example we study 2PI scalar
fields $\hat\phi_{\bf a}=
(\hat\varphi_a\,,\,\hat\varphi_{a_1}\hat\varphi_{a_2})$. There we have
$\phi_{a_1 a_2} =\langle \hat\varphi_{a_1}\hat\varphi_{a_2}\rangle
=\phi^{\rm ir}_{a_1 a_2}+\phi_{a_1}\phi_{a_2}$ with $\phi_a=\langle
\hat\varphi_a\rangle$. Here $\phi^{\rm PI}_{a_1 a_2}$ is the 1PI part
of $\phi_{a_1 a_2}$. This extends to general composite operators and
we parameterise $\Gamma^{\rm PI}[\phi^{\rm PI}]:=\Gamma[\phi(\phi^{\rm
  PI})]$. The $\phi^{\rm PI}$-derivative of $\Gamma^{\rm PI}$ reads
\begin{eqnarray}\label{eq:red2irred} 
  \Gamma^{{\rm PI},{\bf a}}[\phi^{\rm PI}]&=
  & \phi_{\bf c}{}^{,\bf a}(\phi^{\rm PI})\,\gamma^{\bf c}{}_{\bf b} 
  \, J^{\bf b}(\phi^{\rm PI})\,, 
\end{eqnarray}
where $\phi_{\bf c}{}^{,\bf a}(\phi^{\rm PI})$ stands for the
derivative of $\phi$ w.r.t.\ $\phi^{\rm PI}$. Within the above 2PI
example \eq{eq:red2irred} boils down to $(\Gamma^{{\rm PI},{\bf
    a}}[\phi^{\rm PI}])= (J^{a_1 a_2}\,,\, J^{a}+2 J^{a
  b}\phi_b)$, where we have sued that $J^{ab}=J^{(ab)}$. We close with
the remark that it does not make a difference in the relations of this
section whether we have tensorial multi-indices $\bf a$ or a vector
index $a$. \step

\setcounter{equation}{0}
\section{Flows}\label{sec:flows}
In interacting quantum theories it is hardly possible to compute
generating functionals, such as the Schwinger functional $W$, in a
closed form. In most situations one resorts to systematic expansion
schemes like perturbation theory or the $1/N$-expansion that come with
a small expansion parameter. In strongly interacting systems
truncations are not supported by a small expansion parameter and have
to be used with care. In general either case requires renormalisation 
\cite{Stueckelberg:1953dz,Gell-Mann:1954fq}. Renormalisation group
invariance encodes the independence of physics under general
reparameterisations of the theory, or, put differently, the physical
equivalence of (UV) cut-off procedures. RG invariance can be used to
resolve the momentum dependence of the theory by trading RG scaling
for momentum scaling.  RG transformations always imply the scaling of
all parameters of the theory, e.g.\ couplings and masses. In turn, the
change of a physical parameter is related to an RG rescaling. For
example, changing the mass-parameter of the theory leads to the
Callan-Symanzik equation \cite{Symanzik:1970rt,Callan:1970yg}.
Presented as a differential equation for a generating functional,
e.g.\ the Schwinger functional $W$ or the effective action $\Gamma$,
it constitutes a functional RG equation \cite{Symanzik:1970rt}. The
momentum dependence is more directly resolved by block-spinning on the
lattice \cite{Kadanoff:1966wm}. In the continuum theory this is
implemented with a momentum cut-off
\cite{Wilson:1971bg,Wilson:1971dh,Wilson:1973jj,Wegner:1972ih,Nicoll:1977hi,Polchinski:1983gv,Warr:1986we,Hurd:1989up,Keller:1990ej,Wetterich:1992yh,Bonini:1992vh,Ellwanger:1993mw,Morris:1993qb,Becchi:1996an}
leading to the Wilsonian RG. \smallstep

The strong interrelations between the different RG concepts as well as
their physical differences become apparent if presented as Functional
Renormalisation Group equations for generating functionals. FRG
formulations are also suitable for both discussing formal aspects as
well as practical applications. The FRG has been introduced with a
smooth momentum cut-off for simplifying proofs of perturbative
renormalisability and the construction of effective Lagrangians in
\cite{Polchinski:1983gv}, see also
\cite{Keller:1990ej,Keller:1991bz,Keller:1992by,Keller:1995qn}.  More
recently, there has been an increasing interest in FRG methods as a
computational tool for accessing both perturbative as well as
non-perturbative physics, initiated by
\cite{Wetterich:1992yh,Bonini:1992vh,Ellwanger:1993mw,Morris:1993qb,
  Becchi:1996an}. The recent success of FRG methods was also triggered
by formal advances that led to a deeper understanding of the FRG, and
here we aim at further progress in this direction. We close with a
brief overview on the literature in view of structural aspects:
general formal advances have been made in
\cite{Liao:1994cm,Ellwanger:1997tp,Pernici:1998tp,Bonini:1996bk,Comellas:1997ep,Bonini:2000wr,Latorre:2000qc,Polonyi:2000fv,Pawlowski:2001df,Pawlowski:2002eb,Bervillier:2004mf,Litim:2001hk,Litim:2001ky,Litim:2002xm,Litim:2002hj,Litim:2006nn,Stevenson:1981vj,Ball:1994ji,Liao:1999sh,Canet:2002gs,Canet:2004xe,Litim:2000ci,Litim:2001up,Litim:2001fd,Litim:2001dt,Litim:2002cf,Litim:2005us,JDL,Litinprep,Comellas:1996py,Comellas:1997tf,D'Attanasio:1997ej,Morris:2005ck,Oleszczuk:1994st,Liao:1996fp,Liao:1995nm,Floreanini:1995aj,Zappala:2002nx,Alexandre:1999nz,Alexandre:2000eg,Alexandre:2001wj,Jacquot:2004ae,Polonyi:2004pp,Gies:2001nw,Gies:2002hq,Harada:2005tw,Polonyi:2001uc,Schwenk:2004hm,Jaeckel:2002rm,Wetterich:2002ky,Schutz:2004rn,Salmhofer1,Dupuis:2005ij,Alexandre:1997gj,Alexandre:1998ts,Nandori:1999vi,Aoki:1996fn,Generowicz:1997wn,Aoki:1998um,Morris:1999ba,Papp:2000he,Schaefer:1999em,Blaizot:2004qa,Blaizot:2005xy}.
Progress in the construction of FRG flows in gauge theories has been
achieved in 
\cite{Reuter:1992uk,Reuter:1993kw,Reuter:1997gx,Bonini:1993kt,Bonini:1993sj,Ellwanger:1994iz,D'Attanasio:1996jd,Reuter:1996be,Pawlowski:1996ch,Bonini:1997yv,Bonini:1998ec,Falkenberg:1998bg,Litim:1998qi,Litim:1998wk,Litim:2002ce,Freire:2000bq,Igarashi:1999rm,Igarashi:2001mf,D'Attanasio:1996zt,D'Attanasio:1996fy,Ellwanger:1997wv,Ellwanger:1998th,Ellwanger:1995qf,Ellwanger:1996wy,Bergerhoff:1997cv,Pawlowski:2003hq,Pawlowski:2004ip,pawlowski,Fischer:2004uk,Gies:2002af,Gies:2003ic,Braun:2005uj,Branchina:2003ek,Pawlowski:2003sk,Morris:1999px,Morris:2000fs,Arnone:2005fb,Morris:2005tv,Rosten:2005ep,Simionato:1998te,Simionato:1998iz,Simionato:2000ut,Panza:2000tg,Ellwanger:2002sj}.
FRG flows in gravity are investigated
\cite{Reuter:1996cp,Lauscher:2001ya,Litim:2003vp,Bonanno:2004sy,Lauscher:2005xz,Fischer:2006fz}.
All these formal advances have been successfully used within
applications, see reviews
\cite{Fisher:1998kv,Litim:1998nf,Morris:1998da,Aoki:2000wm,Bagnuls:2000ae,Berges:2000ew,Polonyi:2001se,Salmhofer:2001tr,Delamotte:2003dw}.
\smallstep

\subsection{Setting}\label{sec:setting} 
The starting point of our analysis is the finite renormalised
Schwinger functional $W$ in \eq{eq:pathI}. So far we only assumed its
existence without offering a method of how to compute it.  We shall
turn the problem of computing the path integral \eq{eq:pathI} into the
task of successively integrating out modes, each step being
well-defined and finite. To that end we modify the Schwinger
functional as follows:
\begin{eqnarray}\label{eq:WR}
  e^{W[J,R]}=e^{-\Delta S[\s0{\delta}{\delta J},R]} 
  e^{W[J]} \,,
\end{eqnarray}
where 
\begin{eqnarray}\label{eq:reg} 
  \Delta S[\s0{\delta}{\delta J},R]=
  \sum_{n}
  R^{{\bf a}_{1}\cdots {\bf a}_{n}} \0{\delta}{\delta J^{{
        \bf a}_{1}}}\cdots  \0{\delta}{\delta J^{{
        \bf a}_{n}}}\,.  
\end{eqnarray} 
If used as a regulator, the operator $\exp-\Delta S$ in \eq{eq:reg} should
be positive (on $\exp W$),  
and $\Delta S[\s0{\delta}{\delta J},0]=0$.
For example, the standard setting is given by ${\bf a}=a$,
$\hat\phi^a= \hat\varphi^a$ and
\begin{eqnarray}\label{eq:standreg}
  \Delta S[\s0{\delta}{\delta J},R]
  =R^{\bf ab} \0{\delta}{\delta J^{\bf a}}\0{\delta}{\delta 
    J^{\bf b}}\,. 
\end{eqnarray} 
A factor $1/2$ on the rhs common in the literature is absorbed into
$R$. With the restrictions ${\bf a}=a$, $\hat\phi^a=\hat\varphi^a$,
and up to RG subtleties, \eq{eq:standreg} leads to a modification of
the kinetic term $S[\hat\varphi]$ in \eq{eq:pathI}: $S[\hat\varphi]\to
S[\hat\varphi]+R^{ab}\hat\varphi_a \hat\varphi_b$.  More generally,
\eq{eq:standreg} results in a modification of the propagation of the
field $\phi$ which is possibly composite. Such a modification can be
used to suppress the propagation of $\phi$-modes in the path integral.
In particular, it allows for a simple implementation of a smooth
momentum cut-off
\cite{Polchinski:1983gv,Wetterich:1992yh,Bonini:1992vh,Ellwanger:1993mw,Morris:1993qb,Becchi:1996an}.
An amplitude regularisation has been put forward in
\cite{Alexandre:1999nz,Alexandre:2000eg,Alexandre:2001wj,Jacquot:2004ae,Polonyi:2004pp}
and relates to $\Delta S\simeq S$ or parts of $S$, which ensures
positivity. A specifically simple flow of this type is the functional
Callan-Symanzik flow \cite{Symanzik:1970rt,Callan:1970yg}. In specific
theories, e.g.\ those with non-linear gauge symmetries, more general
regulator terms can prove advantageous. $\Delta S$ can also be used to
construct boundary RG flows, in particular thermal flows
\cite{D'Attanasio:1996zt,D'Attanasio:1996fy,Litim:1998nf}. \smallstep

General regulator terms $\Delta S$ according to \eq{eq:reg} involve
higher order derivatives and derivatives w.r.t.\ currents coupled to
composite operators. In this general setting a different point of view
is more fruitful: the operator $\exp-\Delta S$ adds source terms for
composite operators to the Schwinger functional. For example, in the
standard case with ${\bf a}=a$ and \eq{eq:standreg} a source term for
$\hat\varphi_a\hat\varphi_b$ with current $R^{ab}$ is introduced.  For
the class of positive regulator terms $\Delta S[\hat\phi,R]$ the
exponential $\exp-\Delta S$ is a positive operator with spectrum 
$[0,1]$ on $\exp W$ and the correlation functions \eq{eq:moments}.
Then, under mild assumptions the existence of $W[J,R]\leq W[J,0]$
follows from that of $W[J,0]=W[J]$.  Consequently $\exp-\Delta S$ can
be used for suppressing degrees of freedom, more precisely $J$-modes,
in the Schwinger functional $W[J]$. \smallstep

We add that $W[J,R]$ is not well-defined for general $R$. A simple
example is a mass-like $R$ with $R^{ab}= m^2 \delta^{ab}$ for a scalar
theory. Such an insertion leads to an un-renormalised Callan-Symanzik
flow \cite{Symanzik:1970rt,Callan:1970yg}.  The required
renormalisation can be added explicitly via a redefinition of $R^{ab}$
that generates appropriate subtractions. This amounts to an explicit
construction of a BPHZ-type renormalisation which is one way to render
the Callan-Symanzik flow finite. From now on such a redefinition of
$R$ is assumed whenever it is necessary; in most cases, however, the
regulators $R$ generate finite $W[J,R]$ from the outset. A necessary
condition for the latter is a sufficiently fast decay of $R$ in the
ultraviolet. \smallstep

Within this general setting the regulators $R^{{\bf a}_1\cdots {\bf
    a}_n}$ in \eq{eq:reg} can be (partially) fermionic, even though
$\Delta S$ should be kept bosonic (even number of fermions involved).
A simple example is provided by $R^a$ coupling to a fermion
$\hat\phi_a$. It is in general not possible to commute $J$-derivatives
and regulators $R^{{\bf a}_1\cdots {\bf a}_n}$. Due to the (anti-)
commutation relations of the currents $J^{\bf a}$ only specific tensor
structures have to be considered for the $R$:
\begin{eqnarray}\label{eq:Rsyms}
  R^{{\bf a}_1\cdots {\bf a}_i {\bf a}_{i+1}\cdots {\bf a}_n}=
  (-1)^{{\bf a}_i {\bf a}_{i+1}} 
  R^{{\bf a}_1\cdots {\bf a}_{i+1} {\bf a}_i\cdots {\bf a}_n}\,, 
\end{eqnarray} 
where $(-1)^{{\bf a}_i {\bf a}_{i+1}}$ is defined in
appendix~\ref{app:notation}. \Eq{eq:Rsyms} expresses the fact that
fermionic currents anti-commute, $J^{{\bf a}_i} J^{{\bf
    a}_{i+1}}=-J^{{\bf a}_{i+1}} J^{{\bf a}_i}$, whereas bosonic
currents commute with both, bosonic and fermionic currents, leading to
$J^{{\bf a}_i} J^{{\bf a}_{i+1}}=(-1)^{{\bf a}_i {\bf a}_{i+1}}
J^{{\bf a}_{i+1}} J^{{\bf a}_i}$. This symmetry structure carries over
to derivatives of $J^{\bf a}$. Hence, in \eq{eq:reg} only that part of
$R$ carrying the tensor structure expressed in \eq{eq:Rsyms}
contributes. \smallstep

For illustration, we again study this setting for the standard
regulator \eq{eq:standreg} providing a modification of the propagator.
There it follows from \eq{eq:Rsyms} that for bosonic variables only
the symmetric part of the tensor $R^{\bf ab}$ contributes. For the
fermionic part only the anti-symmetric part is relevant. Here we do
not allow for mixed (fermionic-bosonic) parts and \eq{eq:Rsyms}
reduces to
\begin{subequations}\label{eq:standRsyms}
  \begin{eqnarray}\label{eq:standRsymbos}
    R_{\rm bosonic}^{\bf ab}=R_{\rm bosonic}^{\bf ba}\,, 
  \end{eqnarray} 
and 
\begin{eqnarray}\label{eq:Rsymferm}
  R_{\rm fermionic}^{\bf ab}=-R_{\rm fermionic}^{\bf ba}\,.  
\end{eqnarray} 
\end{subequations}
The corresponding $\Delta S$ are bosonic. \smallstep 

So far we have discussed a modification of the Schwin\-ger functional.
The Schwinger functional $W[J,R]$ is only one, if important,
correlation function. We seek an extension of \eq{eq:I0} consistent with
\eq{eq:WR}: it should define general normalised expectation values in
the regularised theory as well as allowing for a straightforward
extension of the symmetry relations $I[J]=0$ as given in \eq{eq:DSI0}.
A natural extension is
\begin{eqnarray}\label{eq:preIR}
  I[J,R]=e^{-W[J,R]}\, e^{-\Delta S[\s0{\delta}{\delta J},R]} 
  \,\hat I[J,\s0{\delta}{\delta J}]\, e^{W[J]}\,.
\end{eqnarray}
\Eq{eq:preIR} entails that $I[J,0]=I[J]$ and guarantees well-defined
initial conditions $I[J,\infty]$. Moreover, applying the extension
\eq{eq:preIR} to a relation $I[J]= 0$ we are led to
\begin{eqnarray}\label{eq:symrelR}
  I[J]= 0 &\quad \rightarrow \quad & I[J,R]= 0, \quad \forall R\,.  
\end{eqnarray} 
Hence a symmetry relation $I[J]= 0$ is lifted to a symmetry relation
$I[J,R]= 0$ in the presence of the regulator. \Eq{eq:preIR} can be
rewritten solely in terms of $W[J,R]$ as
\begin{subequations}\label{eq:IR}
\begin{eqnarray}\label{eq:defIR}
  I[J,R]=e^{-W[J,R]}\,
  \hat I[J,\s0{\delta}{\delta J},R]\, e^{W[J,R]}\,,
\end{eqnarray}
with 
\begin{eqnarray}\label{eq:hatIR}
  \hat I[J,\s0{\delta}{\delta J},R] =
  e^{-\Delta S[\s0{\delta}{\delta J},R]}\,
  \hat I[J,\s0{\delta}{\delta J}]\,
  e^{\Delta S[\s0{\delta}{\delta J},R]}\,,  
\end{eqnarray}
\end{subequations}
see also \cite{Polonyi:2001se}. In case $\hat I[J,\s0{\delta}{\delta
  J}]$ only contains a polynomial in $J$ we can easily determine $\hat
I[J,\s0{\delta}{\delta J},R]$ in a closed form. As for $R=0$, the set
of all correlation functions $\{\CO[J,R]\}$ can be constructed from the set
$\{I[J,R]\}$.  A general flow describes the response of the theory to
a variation of the source $R$ and, upon integration, resolves the
theory. Such flows are provided by derivatives w.r.t.\ $R$ of
correlation functions $\CO[J,R]$ in the presence of the regulator
\begin{eqnarray}\label{eq:gendiff} 
  \delta R^{{\bf a}_1\cdots {\bf a}_n}\0{\delta\CO[J,R]}{\delta R^{{
        \bf a}_1\cdots {\bf a}_n}}\,.
\end{eqnarray} 
Here $\delta R^{{\bf a}_1\cdots {\bf a}_n}$ is a small variation.
Basic examples for correlation functions $\CO$ are the Schwinger
functional $W[J,R]$ and the expectation values $I[J,R]$ defined in
\eq{eq:IR}.  \smallstep

In case we define one-parameter flows $R(k)$ that are trajectories in
the space of regulators $R$ and hence in theory space, the general
derivatives \eq{eq:gendiff} provide valuable information about the the
stability of the chosen one-parameter flows, in particular if these
flows are subject to truncations. Stable one-parameter flows can be
deduced from the condition
\begin{eqnarray} 
  \left.\delta R_\bot^{{\bf a}_1\cdots {\bf
      a}_n} \0{\delta \CO[J,R]}{\delta R_{{\bf a}_1\cdots {\bf
        a}_n}}\right|_{R_{\rm stab}}=0 \,,
  \label{eq:stable} 
\end{eqnarray} 
where $\{R_\bot\}$ is the set of operators that provide a
regularisation of the theory at some physical cut-off scale
$k_{\rm eff}$, and $R_{\rm stab}\in \{R_\bot\}$.  \Eq{eq:stable}
ensures that the flow goes in the direction of steepest descent in
case \eq{eq:stable} describes a minimum. If flows are studied within
given approximations schemes, the stability condition \eq{eq:stable}
can be used to optimise the flow.  Note that \eq{eq:stable}, in
particular in finite approximations, does not necessarily lead to a
single $R_{\rm stab}$. Then \eq{eq:stable} defines a hypersurface of
stable regulators. We also emphasise that \eq{eq:stable} cannot vanish
in all directions $\delta R$ except at a stable fixed point in theory
space.  Consequently one has to ensure within an optimisation
procedure that the variations $\delta R_\bot$ considered are
orthogonal to the direction of the flow. If this is not achieved, no
condition is obtained at all.  We shall come back to the problem of
optimisation in section~\ref{sec:optimal}.\smallstep

\subsection{One-parameter flows}\label{sec:one-parameterflows}

\subsubsection{Derivation}

In most cases we are primarily interested in the underlying theory at
$R=0$, that is $\CO[J]=\CO[J,0]$, e.g.\ in $W[J]=W[J,0]$, the Schwinger
functional of the full theory and its moments. Total functional
derivatives \eq{eq:gendiff} with arbitrary $\delta R^{ab}$ scan the
space of theories given by $W[J,R]$. For computing $W[J]$ it is
sufficient to study one-parameter flows with regulators $R$ depending
on a parameter $k\in [\Lambda,0]$ with $R(k=0)\equiv 0$ and
$W[J,R_{\rm in}]$, $ \CO[J,R(\Lambda)]$ well under control. These
one-parameter flows derive from \eq{eq:gendiff} as partial derivatives
due to variations
\begin{eqnarray}\label{eq:oneparameter} 
  \delta R = 
  dt\, \partial_t R\,, 
\end{eqnarray} 
where $t=\ln (k/k_0)$ is the logarithmic cut-off scale. The
normalisation $k_0$ is at our disposal, and a standard choice is
$k_0=\Lambda$ leading to $t_{\rm in} =0$. In the following we shall
drop the normalisation. The flows with \eq{eq:oneparameter} lead to
correlation functions $\CO_k$ that connect a well-known initial
condition at $\Lambda$ with correlations functions $\CO=\CO_0$
in the full theory. In most cases a well-defined initial condition is
obtained for large regulator $R\to \infty$. This is discussed in
section~\ref{sec:multi-loop}.  \smallstep

The most-studied one-parameter flow relates to a successive
integration of momentum modes of the fields $\varphi$, that is $k$ is
a momentum scale. More specifically, we discuss regulators leading to
an infrared regularisation with IR scale $k$ of the theory under
investigation, the scale $k$ providing the parameter $k\in [k_{\rm
  in},0]$. To that end we choose regulator terms $\Delta
S[\varphi]=R^{ab} \varphi_a \varphi_b$ for a scalar theory with
\begin{eqnarray}\label{eq:exampleR}
  R=R(p^2)\delta (p-p')\,, 
\end{eqnarray} 
with the properties

\begin{itemize}

\item[(i)] it has a non-vanishing infrared limit, $p^2/k^2 \to
  0$, typically $R\to k^2$ for bosonic fields.
 
\item[(ii)] it vanishes for momenta $p^2$ larger than the cut-off
  scale, for $p^2/k^2 \to \infty$ at least with $(p^2)^{(d-1)/2} R\to
  0$ for bosonic fields.

\item[(ii)'] (ii) implies that it vanishes in the limit $k\to 0$. In
  this limit, any dependence on $R$ drops out and all correlation
  functions $\CO_k$ reduce to the correlation functions in the full
  theory $\CO=\CO_0$, in particular the Schwinger functional $W_k$ and
  the Legendre effective action $\Gamma_k$.
  
\item[(iii)] for $k\to \infty$ (or $k\to \Lambda$ with $\Lambda$ being
  some UV scale much larger than the relevant physical scales), $R$
  diverges. Thus, the saddle point approximation to the path integral
  becomes exact and correlation functions $\CO_k$ tend towards their
  classical values, e.g.\ $\Gamma_{k\to\Lambda}$ reduces to the
  classical action $S$.

\end{itemize} 
Property (i) guarantees an infrared regularisation of the theory at
hand: for small momenta the regulator generates a mass. Property (ii)
guarantees the (ultraviolet) definiteness of $W[J,R]$. The insertion
$\Delta S$ vanishes in the ultraviolet: no further ultraviolet
renormalisation is required, though it might be convenient.  It is
property (ii) that facilitates perturbative proofs or
renormalisability. Properties (ii)' and (iii) guarantee well defined
initial conditions, and ensure that the full theory as the end-point
of the flow.  In most cases the regulator $R=p^2 r(p^2/k^2)$ is a
function of $x=p^2/k^2$, up to the prefactor carrying the dimension.
For such regulators the condition (iii) follows already from (i). For
regulators \eq{eq:exampleR} with the properties (i)-(iii) we can study
flows from a well-known initial condition, the classical theory or
perturbation theory, to the full theory.  Integrating the flow
resolves the quantum theory. The properties (i),(ii) guarantee that
the flow is local in momentum space leading to well-controlled limits
$x\to0,\infty$. In turn, mass-like regulators violate condition (ii):
additional renormalisation is required. Moreover, the flow spreads
over all momenta which requires some care if taking the limits
$k^2\to0,\infty$, see e.g.\ \cite{Litim:1998nf}. \smallstep

General one-parameter flows are deduced from \eq{eq:WR}, \eq{eq:IR} by
inserting regulators $R(k)$ where $k\in [\Lambda,0]$. The condition
$R(0)\equiv 0$ guarantees that the endpoint of such a flow is the full
theory. For one-parameter flows, \eq{eq:WR} reads
\begin{eqnarray}\label{eq:Wk} 
  e^{W_k[J]}=e^{-\Delta S_k[\s0{\delta}{\delta J}]} 
  e^{W[J]} 
\end{eqnarray}
with 
\begin{eqnarray*}
  \Delta S_k[\s0{\delta}{\delta J}]=\Delta S[
  \s0{\delta}{\delta J},R(k)]\,, 
\end{eqnarray*}
and $\Delta S$ is defined in \eq{eq:reg}. Similarly we rewrite
\eq{eq:IR} as
\begin{subequations}\label{eq:Ik}
\begin{eqnarray}\label{eq:defIk}
  I_k[J]=e^{-W_k[J]}\,
  \hat I_k[J,\s0{\delta}{\delta J}]\, e^{W_k[J]}
\end{eqnarray}
with 
\begin{eqnarray}\label{eq:hatIk}
  \hat I_k[J,\s0{\delta}{\delta J}] =
  e^{-\Delta S_k[\s0{\delta}{\delta J}]}\,
  \hat I[J,\s0{\delta}{\delta J}]\,
  e^{\Delta S_k[\s0{\delta}{\delta J}]}\,.  
\end{eqnarray}
\end{subequations}
We also recall that \eq{eq:Ik} entails that 
$I_0[J]=I[J]$ and 
\begin{eqnarray}\label{eq:symrel}
  I[J]= 0 &\quad \rightarrow 
  \quad & I_k[J]= 0\quad \forall k, 
\end{eqnarray}
that is a symmetry relation $I[J]= 0$ is lifted to a relation $I_k[J]=
0$ in the presence of the cut-off. The flow of $k$-dependent
quantities $I_k$, $\partial_t I_k$ with $t=\ln k$ at fixed current $J$
allows us to compute $I[J]$, if the initial condition $I_{\Lambda}$ is
under control. For momentum flows, this input is the high momentum
part of $I$ at some large initial scale $\Lambda$. Perturbation theory
is applicable for large scales, and hence $I_{ \Lambda}[J]$ is well
under control. The flow equation $\partial_t I_k$ can be evaluated
with \eq{eq:preIR} for $R(k)$.  However, for later purpose it is more
convenient to approach this question as follows. Let us study the
operators
\begin{eqnarray}\label{eq:hatF0}
  \hat F[J,\s0{\delta}{\delta J}]=\partial_t \hat
  I[J,\s0{\delta}{\delta J}]\,, 
\end{eqnarray}
and
\begin{eqnarray}\label{eq:hatDeltaI0}
  \Delta\hat I
  =[\partial_t\,,\, \hat I]\,.
\end{eqnarray} 
Here the $t$-derivative acts on everything to the right, i.e.\
$\partial_t \hat I G[J]=(\partial_t \hat I) G[J]+ \hat I
\partial_tG[J]$, and is taken at fixed $J$. The notation for partial
derivatives is explained in appendix~\ref{app:derivatives}.  The
functionals $I$, $F$ and $\Delta I$ fall into the class of functionals
\eq{eq:I0} and can be lifted to their $R$-dependent analogues
\eq{eq:IR}, and in particular to $F_k,I_k,\Delta I_k$ as defined in
\eq{eq:Ik}. The full Schwinger functional $W[J]=W_0[J]$ is independent 
of $t$, $\partial_t W=0$, and we derive from
\eq{eq:preIR} that $F=\Delta I$ and consequently
\begin{eqnarray}\label{eq:F=DeltaI}
  F_k=\Delta I_k\,.
\end{eqnarray}
Moreover, the most interesting $I$ are expectation values in the full
theory and do not depend on $t$. For this class we have $\Delta \hat I
=0$ leading to $F_k=0$. Still, the consideration of more general $F_k$
will also prove useful so we do not restrict ourselves to $F_k=0$. The
general $\hat F_k$ is derived from \eq{eq:hatIk} with help of
\begin{eqnarray}\label{eq:tcom}
  \hspace{-.4cm}[\partial_t\,,\,  R^{{\bf a}_1\cdots {\bf a}_n} 
  \s0{\delta}{\delta J^{{\bf a}_1}}
  \cdots \s0{\delta}{\delta J^{{\bf a}_n}}]
  =\dot  R^{{\bf a}_1\cdots {\bf a}_n} \s0{\delta}{\delta J^{{\bf a}_1}}
  \cdots \s0{\delta}{\delta J^{{\bf a}_n}}
  . 
\end{eqnarray} 
In \eq{eq:tcom} we have used that $[\partial_t\,,\,\s0{\delta}{\delta
  J} ]=0$ as $\partial_t=\partial_t|_J$. The rhs of \eq{eq:tcom}
commutes with $\Delta S_k[\s0{\delta}{\delta J}]$ and we conclude that
$(\partial_t + \Delta S[\s0{\delta}{\delta J}\,,\, \dot R]) \exp
-\Delta S_k=(\exp -\Delta S_k )\partial_t $. Inserting $\hat F$ into
\eq{eq:hatIk} and using \eq{eq:tcom} we are led to $\hat F_k$ with
\begin{eqnarray}\label{eq:hatFk} 
  \hat F_k=\left(\partial_t + \Delta S[\s0{\delta}{\delta J}\,,\, 
    \dot R]
  \right) \hat I_k\,. 
\end{eqnarray} 
The expression in the parenthesis in \eq{eq:hatFk} is an operator
acting on everything to the right. Inserting \eq{eq:hatFk} into
\eq{eq:defIk} we arrive at
\begin{eqnarray}\label{eq:Fk} 
  e^{-W_k} \left(\partial_t + \Delta S[\s0{\delta}{\delta J}\,,\, 
    \dot R] \right) e^{W_k}\, I_k=\Delta I_k\,,  
\end{eqnarray} 
valid for general $I_k$ given by \eq{eq:Ik}. $\Delta I_k$ on the right
hand side carries the explicit $t$-scaling of the
operator $\hat I$ and vanishes for $t$-independent $\hat I$. In order
to get rid of the exponentials in \eq{eq:Fk} we use that
$\s0{\delta}{\delta J} e^{W_k}= e^{W_k}(\s0{\delta}{\delta J}+
\s0{\delta W_k}{\delta J})$. With this relation \eq{eq:Fk} turns into
\begin{eqnarray}\label{eq:preflowIk} 
  \left(\partial_t + \dot W_k+\Delta S[\s0{\delta}{\delta J}+\phi\,,\, 
    \dot R] \right)\, I_k=\Delta I_k,  
\end{eqnarray} 
where we have introduced the expectation value
$\phi=\langle\hat\phi\rangle_J$ of the operator coupled to the current
\begin{eqnarray}\label{eq:defofphi} 
  \phi_{\bf a}[J]:=W_{k,\bf a}[J]\,.
\end{eqnarray} 
\Eq{eq:preflowIk} involves the flow of the Schwinger functional, $\dot
W_k$, reflecting the normalisation of $I_k$. Independent flows of
$I_k$ are achieved by dividing out the flow of the Schwinger
functional. The flow $\dot W_k$ is extracted from \eq{eq:preflowIk} for 
the choice $I_k=1$ with $\Delta I_k=0$, following from $\hat I=1$ and
$\Delta\hat I=[\partial_t\, ,\, \hat I]=0$.  Then, \eq{eq:preflowIk}
boils down to $ \dot W_k+(\Delta S[\s0{\delta}{\delta J}+\phi\,,\,
\dot R])=0$, where both expressions are functionals and not operators.
More explicitly it reads
\begin{eqnarray} \label{eq:preflowWk} &&\
  \hspace{-1cm}\Bigl(\partial_t+\sum_n \dot R^{{\bf a}_1\cdots
    {\bf a}_n} \\
  &&\hspace{-.0cm} \times \bigl(\s0{\delta}{\delta J}+\phi\bigr)_{{\bf
      a}_1} \cdots \bigl(\s0{\delta}{\delta J}+ \phi \bigr)_{{\bf
      a}_{n-1}}\,\s0{\delta}{\delta J^{{\bf a}_{n}}} \Bigr)
  W_{k}[J]=0\,.  \nonumber
\end{eqnarray} 
\Eq{eq:preflowWk} is the flow equation for the Schwinger functional.
It links the flow of the Schwinger functional, $\dot W_k$, to a
combination of connected Green functions $W_{k,{\bf a}_1\cdots {\bf
    a}_n}$. For quadratic regulators \eq{eq:standreg} we obtain the
standard flow equation for the Schwinger functional,
\begin{eqnarray}\label{eq:standpreflowWk}
  \left(\partial_t+\dot R^{\bf ab} \s0{\delta}{\delta J^{\bf a}} 
    \s0{\delta}{\delta J^{\bf b}}+ \dot R^{\bf ab}\phi_{\bf a} 
    \s0{\delta}{\delta J^{\bf b}} 
  \right) W_k[J]=0\,.
\end{eqnarray} 
We remark for comparison that the standard notation involves a factor
$\s012$ in the $\dot R$-terms.  It has been shown in
\cite{Litim:2002xm} that \eq{eq:standpreflowWk} is the most general
form of a one loop equation.  \Eq{eq:preflowWk} makes this explicit in
a more general setting as the one considered in \cite{Litim:2002xm}.
Only flows depending on $W_{k,{\bf a}_1\cdots {\bf a}_n}$ with $n\leq
2$ contain one loop diagrams in the full propagator.  Note in this
context that $J$ couples to a general operator $\phi$, not necessarily
to the field.  \smallstep

\Eq{eq:preflowWk} is the statement that the flow operator $\Delta S_1[J,\dot
R]=\dot W_k+ \Delta S[\s0{\delta}{\delta J}+\phi\,,\, \dot R]$ with
\begin{eqnarray}\label{eq:DelS1}
  \hspace{-.3cm}\Delta S_1[J,\dot R]=\Delta S
  [\s0{\delta}{\delta J}+\phi\,,\,\dot R]-(\Delta S
  [\s0{\delta}{\delta J}+\phi\,,\,\dot R])\,, 
\end{eqnarray}
is given by all terms in $\Delta S[\s0{\delta}{\delta J}+\phi\,,\,
\dot R]$ with at least one derivative $\s0{\delta}{\delta J}$ acting
to the right. For later use we also define $\Delta S_n[J,\dot R]$ as
the part of $\Delta S$ with at least $n$ $J$-derivatives. Their
definitions and properties are detailed in appendix~\ref{sec:Delta
  S_n}. The operator of interest here, $\Delta S_1$, can be written
with an explicit $J$-derivative as 
\begin{eqnarray}\label{eq:rdDeltaS}
 \Delta S_1[J,\dot R]= \Delta S^{\bf
  a}[J,\dot R]\s0{\delta}{\delta J^{\bf a}}\,.
\end{eqnarray}
The operator $\Delta S^{\bf a}[J,\dot R]$ is defined in 
\eq{eq:DelSi}. Using \eq{eq:preflowWk} and
the definition \eq{eq:rdDeltaS} in \eq{eq:preflowIk} we arrive at
\begin{eqnarray}\label{eq:flowIk}
  \left(\partial_t +\Delta S^{\bf a}[J\,,\, 
    \dot R]\0{\delta}{\delta J^{\bf a}}
  \right)\, I_k=\Delta I_k\,,  
\end{eqnarray}
valid for general $I_k,\Delta I_k$ given by \eq{eq:Ik}. $\Delta
I_k$ carries the explicit $t$-scaling of $\hat I$ and is derived from
\eq{eq:hatDeltaI0}. The partial $t$-derivative is taken at fixed
current $J$. The flow of a general functional $I_k$ requires the
knowledge of $\phi_{\bf a}[J]=W_{k,\bf a}[J]$ and $\Delta I_k$. Only
for those $I_k$ that entail this information in a closed form,
$\phi=\phi[I_k]$ and $\Delta I_k=\Delta I_k[I_k]$, the flow equation
\eq{eq:flowIk} can be used without further input except that of
$I_\Lambda$. \smallstep

\subsubsection{Flow of the Schwinger functional} 
We proceed by describing the flow \eq{eq:flowIk} for correlation
functions \eq{eq:Ik} within basic examples. To begin with, we study
the flow of the Schwinger functional $W_k$. First we note that its
flow \eq{eq:preflowWk} was derived from \eq{eq:preflowIk} with $I=1$.
The final representation \eq{eq:flowIk} was indeed achieved by
dividing out \eq{eq:preflowWk}. Nonetheless, the latter should follow
from the general flow equation \eq{eq:flowIk}.  Naively one would
assume that $I_k=W_k$ can be obtained from a $t$-independent operator
$\hat I$, that is $\Delta \hat I=0$. However, inserting the assumption
$I_k=W_k$ into the flow \eq{eq:flowIk} and using \eq{eq:preflowWk} we
are led to
\begin{eqnarray}\label{eq:contradict} 
  \Delta I_k=\Delta S^{\bf a}[J\,,\, 
  \dot R] \phi_{\bf a}-(\Delta S[\s0{\delta}{\delta J}+\phi,\dot R])\,.
\end{eqnarray}
which does not vanish for all $J$, e.g.\ for quadratic regulators it
reads $\Delta I_k=\dot R^{\bf ab}\phi_{\bf a}\phi_{\bf b}$. Hence
\eq{eq:contradict} proves that $I_k=W_k$ implies $\Delta \hat I\neq
0$. Indeed in general \eq{eq:contradict} cannot be deduced from a
$\Delta \hat I$ that is polynomial in the current and its derivatives.
The above argument highlights the necessity of the restriction of
\eq{eq:flowIk} to functionals $I_k$ constructed from \eq{eq:Ik}. Still
the flow equation for $W_k$ can be extracted as follows. Let us study
the flow of $(I_k)_{\bf a}=W_{k,\bf a}=\phi_{\bf a}$ which also is of
interest as $\phi$ is an input in the flow \eq{eq:flowIk}.  $I_k=\phi$
falls into the allowed class of $I_k$ as
\begin{eqnarray}\label{eq:Wk,a}
  \hat I_{\bf a}=(\hat I_k)_{\bf a}=\s0{\delta}{\delta J^{\bf a}}
  \quad \rightarrow \quad (I_k)_{\bf a}=W_{k,\bf a}=\phi_{\bf a}\,.
\end{eqnarray}
Moreover, $\Delta I_k=0$.  Consequently, the flow of the functional
$I_k$ introduced in \eq{eq:Wk,a} reads
\begin{eqnarray}\nonumber  
\dot W_{k,\bf a}+\Delta S[\s0{\delta}{\delta J}+\phi\,,\,
    \dot R]\phi_{\bf a}
 -(\Delta S[\s0{\delta}{\delta J}+\phi\,,\,
  \dot R])\phi_{\bf a}=0\,.\\ 
\label{eq:step1dWk}  \end{eqnarray}
With $\s0{\delta}{\delta J} 1=0$ the second term on the left hand side can be 
rewritten as follows 
\begin{eqnarray}\nonumber 
  \hspace{-1cm}\Delta S[\s0{\delta}{\delta J}+\phi\,,\,
  \dot R] \phi_{\bf a}&=&
  \Delta S[\s0{\delta}{\delta J}+\phi\,,\,
  \dot R](\s0{\delta}{\delta J}+\phi)_{\bf a}\\ 
  &=&
  (\s0{\delta}{\delta J}+\phi)_{\bf a}
  \Delta S[\s0{\delta}{\delta J}+\phi\,,\,\dot R] \,.
\label{eq:observe1}\end{eqnarray} 
We emphasise that the first line in \eq{eq:observe1} is not an
operator identity. For the second line in \eq{eq:observe1} we have
used the bosonic nature of the regulator term and the representation
$\s0{\delta}{\delta J}+\phi= e^{-W}\s0{\delta}{\delta J} e^{W}$. This
also entails that $\phi_{\bf a} (\Delta S [\s0{\delta}{\delta
  J}+\phi\,,\,\dot R])=(\Delta S [\s0{\delta}{\delta J}+\phi\,,\,\dot
R])\phi_{\bf a} $. We have already mentioned that $\partial_t
\phi_{\bf a}[J]\neq 0$ as the $t$-derivative is taken at fixed $J$.
For the same reason we can commute $t$-derivatives with
$J$-derivatives: $\partial_t W_{k,\bf a}[J]= (\partial_t
W_{k}[J])_{,\bf a}$.  We conclude that the flow of $W_{k,\bf a}$ can
be written as a total derivative
\begin{eqnarray}\label{eq:totalWk} 
  \left[\partial_t W_k +(\Delta S [
    \s0{\delta}{\delta J}+\phi\,,\,\dot R])\right]_{,\bf a}=0\,,
\end{eqnarray} 
which upon integration yields 
\begin{eqnarray}\label{eq:flowWk} 
  \partial_t W_k +(\Delta S [
  \s0{\delta}{\delta J}+\phi\,,\,\dot R])=0\,.  
\end{eqnarray} 
\Eq{eq:flowWk} agrees with \eq{eq:preflowWk} \footnote{We have fixed
  the integration constant to precisely match \eq{eq:preflowWk}.}.
\smallstep

\subsubsection
{Standard flow}
For its importance within applications we also discuss the standard
quadratic flow. In this case the flow \eq{eq:flowIk} reduces to
\begin{eqnarray}\label{eq:standflowIk}
  \left(\partial_t+ \dot R^{\bf ab} \s0{\delta}{\delta J^{\bf a}}
    \s0{\delta}{\delta J^{\bf b}}+ 2  \dot R^{\bf ab}\phi_{\bf a}
    \s0{\delta}{\delta J^{\bf b}} 
  \right) I_k[J]=0\,,  
\end{eqnarray}
and \eq{eq:contradict} turns into $\Delta I_k= \dot R^{\bf
  ab}\phi_{\bf b}\phi_{\bf a}$ which does not vanish for $\phi\neq 0$.
That proves that there is no $\hat I$ leading to $I_k=W_k$. The flow
of $(I_k)_{\bf a}=W_{k,\bf a}$ follows as
\begin{eqnarray}\nonumber 
  \hspace{-1cm}(\partial_t W_{k}[J])_{,\bf a}
  &=&-\left( \dot R^{\bf bc} \s0{\delta}{\delta J^{\bf b}}
    \s0{\delta}{\delta J^{\bf c}}+2  \dot R^{\bf bc}\phi_{\bf b}
    \s0{\delta}{\delta J^{\bf c}} 
  \right) W_{k,\bf a}\\
  &=&-\left[\dot R^{\bf bc}\left( W_{k,\bf bc}
      +\phi_{\bf b}\phi_{\bf c}
    \right)\right]_{,\bf a}.   
\label{eq:flowWk,a}\end{eqnarray} 
Both sides in \eq{eq:flowWk,a} are total derivatives w.r.t.\ $J^{\bf
  a}$.  Integration leads to
\begin{eqnarray}\label{eq:standardflowWk}   
  \dot W_k[J]=-\dot R^{\bf ab}  \left( W_{k,\bf ab}
    +\phi_{\bf a}\phi_{\bf b}
  \right) \,,
\end{eqnarray}
where we have put the integration constant to zero. For the reordering
in \eq{eq:standardflowWk} we have used that the regulator $R^{\bf ab}$
is bosonic.  \Eq{eq:standardflowWk} agrees with
\eq{eq:standpreflowWk}. It also follows straightforwardly from
\eq{eq:flowWk} for quadratic regulators. \smallstep

\subsubsection
{Flow of amputated correlation functions} 
The results of the previous sections translate directly into similar
ones for amputated correlation functions $I_k[J(\phi)]$ with the
following $k$-dependent choice of the current
\begin{equation}
J^{\bf a}(\Phi)=\left[S+\Delta S
\right]^{,\bf ba}_{\Phi=0}\, \Phi_{\bf b}\,, \qquad   
\Phi_{\bf a}= (P_k)_{\bf ba} 
J^{\bf b}\,, 
\label{eq:Jk}\end{equation}
introducing the classical propagator $P_k$. With \eq{eq:Jk} the flow for
general correlation functions $O_k[J(\Phi)]$ is computed as
\begin{equation}\label{eq:ampflow}
  \partial_t O_k[J(\Phi)]=\left[\partial_t O_k[J]+
    \Phi_{\bf a} (\partial_t \Delta S)^{,\bf ab}_{\Phi=0} O_{k,\bf b} 
  \right]_{J=J(\Phi)}\,, 
\end{equation} 
in particular valid for $O_k=I_k$. The $t$-derivative on the lhs of
\eq{eq:ampflow} is taken at fixed $\Phi$: the first term on the rhs of
\eq{eq:ampflow} is the flow \eq{eq:flowIk} at fixed $J$, and the
second term stems from the $k$-dependence of $J(\Phi)$.  For
example, in the presence of a regulator the effective Lagrangian
$S_{\rm eff}[\Phi]$ \eq{eq:S_eff} turns into
\begin{eqnarray}\label{eq:S_k}
  S_{{\rm eff}_k}[\Phi]:=-W_k[J(\Phi)]\,, 
\end{eqnarray}
and hence has the flow \eq{eq:ampflow} with \eq{eq:preflowWk}. This
flow further simplifies for quadratic regulators $R^{\bf
  ab}\hat\phi_{\bf a}\hat\phi_{\bf b}$. For this choice we arrive at
\begin{equation}\label{eq:dS_eff}
  \partial_t S_{{\rm eff}_k}[\Phi]= 
  \0{1}{2}(\dot P_{k})_{\bf ab} 
  \left(S_{{\rm eff}_k}^{,\bf ab}-S_{{\rm eff}_k}^{,\bf a}\,
    S_{{\rm eff}_k}^{,\bf b}-2 
    J^{\bf a} S_{{\rm eff}_k}^{,\bf b}\right) \,.
\end{equation}
Often \eq{eq:dS_eff} is rewritten in terms of the interaction part
of the effective Lagrangian defined as $S_{{\rm int}_k}=S_{{\rm eff}_k}+\s012 
\left[S+\Delta S \right]^{,\bf ba}_{\Phi=0}\Phi_{\bf a}\Phi_{\bf b}$.
The flow of $S_{{\rm int}_k}$ follows as  
\begin{equation}\label{eq:polch}
\partial_t S_{{\rm int}_k}[\Phi]= 
\0{1}{2}(\dot P_{k})_{\bf ab} 
\left(S_{{\rm int}_k}^{,\bf ab}-S_{{\rm int}_k}^{,\bf a}\,
  S_{{\rm int}_k}^{,\bf b}\right)\,, 
\end{equation}
where we dropped the $\Phi$-independent term $ -(\partial_t \ln
P_k)^{\bf a}{}_{\bf a}$. Flows for $S_{{\rm eff}_k}$ and its $N$-point
insertions can be found e.g.\ in
\cite{Polchinski:1983gv,Keller:1990ej,Ellwanger:1993mw,Morris:1993qb,Pernici:1998tp}.
They are closely related to Callan-Symanzik equations for $N$-point
insertions for $R\propto k^2$ with a possible mass renormalisation,
see also \cite{Zinn-Justin:1993wc}. The flows \eq{eq:dS_eff},
\eq{eq:polch} can be extended to $\Phi$-dependent $P_k$ by using the
general DS equations \eq{eq:genDS} in the presence of a regulator, see
e.g.\ \cite{Latorre:2000qc,Morris:1999px}. Then it also
nicely encodes reparameterisation invariance.\smallstep

We close this section with a remark on the structure of the flows
\eq{eq:flowIk},\eq{eq:ampflow}. They equate the scale derivative of a
correlation function to powers of field derivatives of the same
correlation function. The latter are unbounded, and the boundedness of
the flow must come from a cancellation between the different terms.
Hence, within truncations the question of numerical stability of these
flows arises, see \cite{Litim:2005us}.

\subsection{Flows in terms of mean fields}\label{sec:1PIflows} 
\subsubsection{Derivation}
In most situations it is advantageous to work with the flow of 1PI
quantities like the effective action, formulated as functionals of the
mean field $\phi_{\bf a}=W_{,\bf a}$. In other words, we would like to
trade the dependence on the current $J$ and its derivative
$\s0{\delta}{\delta J}$ in \eq{eq:flowIk} for one on the expectation
value $\phi$ and its derivative $\s0{\delta}{\delta \phi}$. Similarly
to \eq{eq:Gamma} we define the effective action
$\Gamma=\Gamma[\phi,R]$ as
\begin{eqnarray}\label{eq:GammaR} 
  \Gamma[\phi,R]=\sup_{\mbox{\tiny $ J$}} \left( J^{\bf a}\phi_{\bf a}-
    W[J,R]\right)
  -\Delta S'[\phi,R]\,. 
\end{eqnarray} 
where 
\begin{eqnarray}\label{eq:DelS'}
  \Delta S'[\phi,R] =
  \sum_{n\geq 2} R^{{\bf a}_1\cdots {\bf a}_n}\phi_{{\bf a}_1}\cdots 
  \phi_{{\bf a}_n}\,. 
\end{eqnarray}
The exclusion of the linear regulator terms in $\Delta S'_k$ is
necessary as they simply would remove the dependence on the linear
regulator.  $\Gamma[\phi,R]$ is the Legendre transform of $W[J,R]$,
where the cut-off term has been subtracted for convenience. For $R\to
0$ \eq{eq:GammaR} reduces to \eq{eq:Gamma}. The definition
\eq{eq:GammaR} constrains the possible choices of the operators
coupled to $J$ to those which at least locally admit a Legendre
transform of $W[J,R]$. \Eq{eq:GammaR} implies
\begin{eqnarray}\label{eq:Jphik} 
  \hspace{-1cm}\gamma^{\bf a}{}_{\bf b} J^{\bf b}=
  (\Gamma +\Delta S')^{,\bf a}\,,  & 
  \qquad & \phi_{\bf a}=W_{,\bf a}
  \,, \end{eqnarray} 
as well as 
\begin{eqnarray}\label{eq:propk}
  G_{\bf ac} (\Gamma+\Delta S)^{,\bf cb}=
  \gamma^{\bf b}{}_{\bf a}\,,
\end{eqnarray} 
with 
\begin{eqnarray}\label{eq:defofpropk}
  G_{\bf ac}=
  W_{,\bf ac}\,.
\end{eqnarray}
Here $\gamma^{\bf b}{}_{ \bf a}$ leads to the minus sign in fermionic
loops, see appendix~\ref{app:notation}. For quadratic regulators
\eq{eq:standreg} the above relations read
\begin{eqnarray}\label{eq:standpropk1} 
  \gamma^{\bf a}{}_{\bf b} 
  J^{\bf b}=\Gamma^{,\bf a} +2 R^{\bf ab}\phi_{\bf b}\,, 
\end{eqnarray}
and 
\begin{eqnarray}\label{eq:standpropk2}
  G_{\bf ac} (\Gamma^{,\bf cb}+2
  R^{\bf bc})= \gamma^{\bf b}{}_{\bf a}.
\end{eqnarray}
For \eq{eq:standpropk1},\eq{eq:standpropk2} we have used
\eq{eq:standRsyms} and the bosonic nature of $R^{\bf bc}$. The
operator $G[\phi]$ in \eq{eq:propk} is the full field dependent
propagator. With \eq{eq:propk} we are able to relate derivatives
w.r.t.\ $J$ to those w.r.t\ $\phi$ via
\begin{eqnarray}\label{eq:djphi}  
  \0{\delta}{\delta J^{\bf a}}=
  G_{\bf ab}\0{\delta}{\delta \phi_{\bf b}}\,,
\end{eqnarray}
where we have used that $\phi_{{\bf b},\bf a}=W_{,\bf ab}=G_{\bf ab}$. 
As in the case of the Schwinger functional we are not only interested
in the flow of $\Gamma$ but in that of general correlation functions
$\tilde I$ as functions of $\phi$. This is achieved by defining
$I[J,R]$ as a functional of $J[\phi]$:
\begin{eqnarray}\label{eq:tildeIk} 
  \tilde I[\phi,R]=I[J(\phi),R]. 
\end{eqnarray}
We emphasise that $\tilde I$ is not necessarily 1PI, it only is
formulated in terms of such quantities. Still, all 1PI quantities can
be constructed from the class of $\tilde I$. \smallstep

One-parameter flows for $\tilde I$ are derived by using trajectories
$R(k)$. We extend the notation introduced in the last section for
flows of $\tilde I$ with
\begin{eqnarray}\label{eq:ItotIk}
  \tilde I_k[\phi]&=& \tilde I[\phi,R(k)]\,,
\end{eqnarray}
and
\begin{eqnarray}\label{eq:Gammak}
  \Gamma_k[\phi]&=& \Gamma[\phi,R(k)]\,. 
\end{eqnarray} 
For reformulating \eq{eq:flowIk} in terms of $\tilde I_k$ we need the
relation between $\partial_t\tilde I_k=\partial_t|_\phi \tilde I_k$
and $\partial_t I_k=\partial_t|_J I_k$, see also
appendix~\ref{app:derivatives}. With \eq{eq:djphi} we rewrite
$I_{k,{\bf a}}=G_{\bf ab} \tilde I_k{}^{,{\bf b}}$, and it follows
from \eq{eq:tildeIk} that
 \begin{eqnarray}\label{eq:resolve}
   \partial_t \tilde I_k [\phi]=\partial_t I_k[J]+
   (\partial_t J^{\bf a}[\phi])\,G_{\bf ab} \tilde I_k{}^{,{\bf b}}\,,
\end{eqnarray}
with $\partial_t J^{\bf a}[\phi]=\partial_t|_\phi J^{\bf a}[\phi]$.
Now we insert the flow for $I_k$, \eq{eq:flowIk}, in \eq{eq:resolve}.
With \eq{eq:djphi} the operator $\Delta S_1[J,\dot R]= \Delta S^{\bf
  a}[J,\dot R]\s0{\delta}{\delta J^{\bf a}}$ is rewritten in terms of
$G_{\bf ab}\s0{\delta}{\delta \phi_{\bf b}}$. As it is more convenient
to use an expansion in plain derivatives $\s0{\delta}{\delta \phi_{\bf
    b}}$ we also employ the identity $\Delta S_1[J,\dot R]= 
\Delta S_1[\phi,\dot R]$, the terms
that contain at least one derivative w.r.t.\ $J$ are equivalent to
those containing at least one derivative w.r.t.\ $\phi$. Note that
this fails to be true for higher derivative terms, $\Delta S_n$ with
$n> 1$.  Together with \eq{eq:djphi}, \eq{eq:resolve} the above
considerations lead to the flow \eq{eq:flowIk} as an equation for
$\tilde I_k$
\begin{equation}
\left(
  \partial_t -\left(
    (\partial_t J^{\bf a}) G_{\bf ab} +
    \Delta S_{\bf b}[\phi,\dot R]\right)
  \s0{\delta}{\delta \phi_{\bf b}}\right) \tilde I_k[\phi]
=\Delta \tilde I_k[\phi]\,,\\
\label{eq:preflowtIk}\end{equation}
where $\Delta S_{\bf b}$ is defined in \eq{eq:DelSi2}. It can be
easily computed for general regulators. However, the higher the order
of derivatives is in the regulator term, the more loop terms are
contained in $\Delta S_1$. For further illustration we have detailed
the simplest case of the standard flow in
appendix~\ref{app:quadratic1PI}. We proceed by evaluating
\eq{eq:preflowtIk} for a specific simple $\tilde I_k$: we use $\tilde
I_k[\phi]=\phi$ already introduced via $\hat I_{\bf
  a}=\s0{\delta}{\delta J^{\bf a}}$ in \eq{eq:Wk,a}. For this choice
we have $\Delta \tilde I_k=0$ and $\partial_t \tilde I_k =0$, and the
flow \eq{eq:preflowtIk} reads
 \begin{eqnarray}\label{eq:flowofJ} 
   \partial_t J^{\bf a} G_{\bf ab}   
   -(\Delta S_{\bf b}[\phi\,,\,\dot R]) =0\,. 
\end{eqnarray}
Here $(\Delta S_{\bf b}[\phi,R])$ is the linear
expansion coefficient of $\Delta S_1$ in a power expansion in
derivatives w.r.t.\ $\phi$, see also \eq{eq:DelSi}.  Note that
\eq{eq:flowofJ} already comprises the flow equation for $\Gamma_k$: 
it follows from the definition of the current in \eq{eq:Jphik} that
$J^{\bf b}\gamma^{\bf a}{}_{\bf b} = (\Gamma+\Delta S')^{,\bf a}$.
Moreover $\partial_t (\Gamma_k{}^{,\bf a}) =(\partial_t\Gamma_k)^{,\bf
  a}$ as the partial $t$-derivative is taken at fixed $\phi$. Then
\eq{eq:flowofJ} contracted with $ (\Gamma_k+\Delta S')^{,\bf ba}$
comprises $\partial_t (\Gamma_{k,\bf b})$ and is a total derivative
w.r.t.\ $\phi$ which can be trivially integrated. This can be best
seen for quadratic regulators \eq{eq:standreg} for which
\eq{eq:flowofJ} boils down to
 \begin{eqnarray}\label{eq:standflowofJ} 
   \partial_t J^{\bf a}+  (G\,\dot R\, G)_{\bf bc}  
   \Gamma_{k}{}^{,\bf cbd}\,\gamma^{\bf a}{}_{\bf d}-\phi_{\bf b} \dot R^{\bf 
     ba}=0\,, 
\end{eqnarray}
see also \eq{eq:standpreflowtIk} in appendix~\ref{app:quadratic1PI}.
We also remark that an alternative derivation of the identity
\eq{eq:flowofJ} solely makes use of structural considerations which
prove useful for general flows: for 1PI $\tilde I_k$ the related term
in \eq{eq:preflowtIk} is not 1PI, whereas the other terms are.
Accordingly these terms have to vanish separately \footnote{Strictly
  speaking, one also has to use that the span of 1PI $\tilde I$ 
generates all 1PI
  quantities.}, which implies that the expression in the parenthesis
has to vanish leading to \eq{eq:flowofJ}. With $\partial_t J^{\bf
  a}\, G_{\bf ab}=-(\Delta S_{\bf b}[\phi,\dot R])$ the coefficient
of $\s0{\delta}{\delta \phi_{\bf b}}$ in \eq{eq:preflowtIk} takes the 
form 
\begin{eqnarray}
\Delta S_{\bf
    b}[\phi,\dot R]-(\Delta S_{\bf
    b}[\phi,\dot R])=\Delta S_{\bf ab}[\phi,\dot R]
  \s0{\delta}{\delta \phi_{\bf a}}\,,  
\label{eq:DelSab}\end{eqnarray} 
where $\Delta S_{\bf ab}\s0{\delta}{\delta \phi_{\bf a}}$ is the part
of the operator $\Delta S_{\bf b}$ containing at least one
$\phi$-derivative. $\Delta S_{\bf ab}$ follows from \eq{eq:DelSab}, 
see also \eq{eq:DelSi2}. With \eq{eq:DelSab} the operator in the flow 
\eq{eq:preflowtIk} is 
\begin{eqnarray}
  \Delta S_2[\phi,\dot R]= 
  \Delta S_{\bf ab}[\phi,\dot R]\0{\delta^2}{\delta
    \phi_{\bf a}\delta \phi_{\bf b}}\,,
\label{eq:DelS2der}\end{eqnarray} 
that part of $\Delta S[G\s0{\delta}{\delta \phi}+\phi,\dot R]$
containing at least two $\phi$-derivatives, and we arrive at
\begin{eqnarray}\label{eq:flowtIk}
  \left(\partial_t+ \Delta S_2[\phi,\dot R] \right)
  \tilde I_k[\phi]=\Delta\tilde I_k\,,
\end{eqnarray}
for general functionals $\tilde I_k$ as defined with \eq{eq:Ik} and
\eq{eq:tildeIk}. The functional $\Delta \tilde I_k$ originates in the
explicit $t$-scaling of $\hat I$.  The partial $t$-derivative on the
left hand side of \eq{eq:flowtIk} is taken at fixed $\phi$, and the
operator $\Delta S_2$, \eq{eq:DelS2der}, accounts for inserting the
regulator $\dot R$ into the Green functions contained in correlation
functions $\tilde I_k$. We also provide a representation of $\Delta
S_2[\phi,\dot R]$ that only makes direct use of $\Delta
S[G\s0{\delta}{\delta \phi}+\phi\,,\,\dot R]$,
\begin{eqnarray}\label{eq:DelS2} 
  &&\hspace{-.7cm} \Delta S_2[\phi,\dot R]=\Delta S
  [G\s0{\delta}{\delta \phi}+\phi\,,\,\dot R]\\ 
  &&\hspace{-.5cm} -([\Delta S[G\s0{\delta}{\delta \phi}+
  \phi,\dot R]\,,\,\phi_{\bf b}])
  \s0{\delta}{\delta \phi_{\bf b}}-(\Delta S
  [G\s0{\delta}{\delta \phi}+\phi\,,\,\dot R])\,, 
\nonumber \end{eqnarray}
where $(G\s0{\delta}{\delta \phi})^{\ }_{\bf b}=G_{\bf bc}
\s0{\delta}{\delta \phi_{\bf c}}$. The relatively simple insertion
operator $\Delta S_2$ in terms of derivatives w.r.t.\ $\phi$ is
related to the structural dependence of $\tilde I_k$ on $\phi$ and $R$
that is fixed by the definitions \eq{eq:Ik},\eq{eq:tildeIk}. In turn,
changing the definition of $I_k$, $\tilde I_k$ leads to different
flows. The construction of $I_k,\tilde I_k$ is a natural one as it
includes general Green functions $\langle \phi^n\rangle$ as building
blocks.  Still, it might be worth exploring the flows of different
correlation functions for specific problems, whose setting admit more
natural variables than the $\tilde I_k$.\smallstep

Let us now come back to the remark on numerical stability at the end
of section~\ref{sec:one-parameterflows}. In contradistinction to the
flows \eq{eq:flowIk},\eq{eq:ampflow} the flow \eq{eq:flowtIk} relates
the scale derivative of a correlation function to a polynomial of the
full propagator, field derivatives of the effective action and the
correlation function itself. In most cases both sides of the flow
\eq{eq:flowtIk} are bounded, ensuring numerical stability and hence
better convergence towards physics \cite{Litim:2005us}. A notable
exception is the case where the Legendre transform from $W_k$ to
$\Gamma_k+\Delta S_k'$ is singular. This either hints at a badly chosen
truncation, or it relates to physical singularities that show up in
the propagator $G$, see also \cite{Litim:2006nn}. In the scale-regime
where such a singularity occurs one might switch back to the flow of
$W_k$ or $S_{{\rm eff}_k}$ \cite{Alexandre:1998ts}. In the vicinity of
$S_{{\rm eff},\bf ab}\approx 0$ the flows \eq{eq:dS_eff},\eq{eq:polch}
are bounded. \smallstep

\subsubsection{Flow of the effective action} 
As in the case of the flow equation for $I_k$ we describe the content
and the restrictions of \eq{eq:flowtIk} within basic examples. From
its definition \eq{eq:GammaR} it follows that its flow is closely related 
to that of $W_k$, 
\begin{eqnarray}\label{eq:preflowGk}
\partial_t \Gamma_k[\phi] -\partial_t W_k[J]-\Delta S'[\phi,\dot R]=0\,,
\end{eqnarray}
where we have used \eq{eq:Jphik} for $\dot J^{\bf a}(\phi_{\bf a}
-W_{k,\bf a}[J])=0$. Inserting the flow \eq{eq:flowWk} for the 
Schwinger functional we are led to
\begin{eqnarray}\label{eq:flowGk} 
  \partial_t\Gamma_k[\phi]-(\Delta S[G\s0{\delta}{\delta \phi}
  +\phi\,,\,\dot R])+\Delta S'[\phi,\dot R]=0\,.
\end{eqnarray}
More explicitly it reads
\begin{eqnarray}\nonumber 
  \hspace{-.5cm}&&\partial_t\Gamma_k[\phi]-\dot 
  R^{\bf a}\phi_{\bf a}-\sum_{n\geq 2}
  \dot R^{ {\bf a}_1\cdots {\bf a}_n} \\\nonumber 
  &&\hspace{-.3cm}\times \left[
    (G\s0{\delta}{\delta \phi}+\phi)_{{\bf a}_1}\cdots 
    (G\s0{\delta}{\delta \phi}+\phi)_{{\bf a}_{n-1}}-
    \phi_{{\bf a}_1} \cdots \phi_{{\bf a}_{n-1}}\right]
  \phi_{{\bf a}_{n}}\\
  &&=0\,.
\label{eq:flowGkexplicit} \end{eqnarray} 
The explicit form of the flow \eq{eq:flowGkexplicit} allows us to read off 
the one particle irreducibility of $\Gamma_k[\phi]$ as a consequence 
of that of the classical action $S[\phi]$: the flow preserves 
irreducibility and hence it follows recursively from that of 
$S[\phi]$. \smallstep 

As for the Schwinger functional there is no $\hat I$ with $\Delta
\tilde I=0$ leading to $\tilde I_k=\Gamma_k$. The related consistency
equation reads
\begin{eqnarray}\nonumber 
&&\hspace{-1.4cm}
\Delta \tilde I_k[\phi]=\Delta S_2[\phi,\dot R]
  \Gamma_k
\\  && 
+(\Delta S[G\s0{\delta}{\delta \phi}+\phi\,,\,\dot R])
  -\Delta S'[\phi,\dot R]\,.
  \label{eq:contradict2} \end{eqnarray} 
The right hand side of \eq{eq:contradict2} does not vanish for all
$\phi$ implying $\Delta \tilde I_k\neq 0$. Moreover, in general
\eq{eq:contradict2} cannot be deduced from a $\Delta \hat I$
polynomial in the current $J$ and its derivatives. Again this
highlights the necessity of restricting $\tilde I_k$ to those
constructed from \eq{eq:Ik} and \eq{eq:tildeIk}.\smallstep 

Similarly to the derivation of the flow of $W_k$ we can derive the
flow \eq{eq:flowGk} from that of its derivative, $\Gamma_k^{,\bf a}$.
We use $\hat I^{\bf a}= \gamma^{\bf a}{}_{\bf b} J^{\bf b}$.  The
corresponding $\hat I_k$ derived from \eq{eq:hatIk} as $\hat I_k^{\bf
  a}= \gamma^{\bf a}{}_{\bf b} J^{\bf b}- \Delta S^{,\bf a}[J\,,\,R]$.
The second operator $\Delta S^{,\bf a}$ originates from the commutator
term $\gamma^{\bf a}{}_{\bf b} [\Delta S,J^{\bf b}]$. The commutator
gives the right $\phi$-derivative of $\Delta S[\s0{\delta }{\delta
  J}+\phi,R]$ at fixed $J$, see appendix~\ref{sec:Delta S_n}.
Contracted with $\gamma^{\bf a}{}_{\bf b}$ we arrive at the left
derivative, where we have also used the bosonic nature of $\Delta S$. 
The corresponding $\tilde I_k$
reads with \eq{eq:Jphik}
\begin{eqnarray}\label{eq:choicetIk}
\hspace{-.3cm}  \tilde
  I_k=\Gamma_k{}^{,\bf a}+\Delta S'^{,\bf a}[\phi\,,\,R]-(\Delta
  S^{,\bf a}[G\s0{\delta }{\delta
  \phi}+\phi\,,\,R])\,,
\end{eqnarray} 
Moreover, $ \Delta \tilde I_k=0$. The choice \eq{eq:choicetIk}
boils down to $\tilde I_k=\Gamma_k{}^{,\bf a}$ in the standard case.
For general flows the last term on the right hand side of
\eq{eq:choicetIk} is non-trivial by itself. Indeed, its flow can 
be separately studied and follows from $\hat I=\Delta S^{,\bf
  a}[\s0{\delta}{\delta J}\,,\,R]$ and $\hat F=\partial_t \Delta
S^{,\bf a}[\s0{\delta}{\delta J}\,,\,R]$. This leads to $\Delta\hat I=
\Delta S^{,\bf a}[\s0{\delta}{\delta J}\,,\,\dot R]$ and $\Delta
\tilde I_k= (\Delta S^{,\bf a}[G\s0{\delta}{\delta \phi}+\phi\,,\,\dot
R])$.  Inserting this into the flow \eq{eq:flowtIk} we are led to 
\begin{eqnarray}\nonumber 
  &&  \hspace{-1.7cm}
  \left(\partial_t+\Delta S_2[\phi,\dot R]\right)
  (\Delta S^{,\bf a}[G\s0{\delta}{\delta \phi}+\phi\,,\,R])\\ 
  && \hspace{1.7cm} 
  = (\Delta S^{,\bf a}[G\s0{\delta}{\delta \phi}+\phi\,,\,\dot R])\,.
\label{eq:flowDelta}\end{eqnarray}
The above equation describes the flow of the functional $(\Delta
S^{,\bf a}[G\s0{\delta}{\delta \phi}+\phi\,,\,R])$ at fixed second
argument $R$.  Using \eq{eq:flowDelta} within the flow of $\tilde I_k$
of \eq{eq:choicetIk} it reads
\begin{eqnarray} \nonumber \hspace{-1cm}\partial_t\Gamma_{k}{}^{,\bf
    a}&=& - \Delta S_2[\phi,\dot R]\left(\Gamma_k+\Delta S'
  \right)^{,\bf a}\\
  &&-\Delta S'^{,\bf a}[\phi,\dot R]+ (\Delta S^{,\bf
    a}[G\s0{\delta}{\delta \phi}+\phi\,,\,\dot R]) \,.
  \label{eq:flowdGk0} \end{eqnarray} 
\Eq{eq:flowdGk0} looks rather complicated. However, 
note that $\Delta S_2$ acts on the current as 
$(\Gamma_{k} +\Delta S_k')^{,\bf a}=\gamma^{\bf a}{}_{\bf b} J^{\bf b}$, 
see \eq{eq:Jphik}. Hence the evaluation of \eq{eq:flowdGk0} is simplified 
if representing $\Delta S_2[\phi,\dot R]$ in terms of $J$-derivatives as  
all higher $J$-derivatives vanish. To that end we use that the sum of all 
derivative terms in either $\phi$ or $J$ coincide as in both cases it is 
given by the operator $\Delta S- (\Delta S)$. The latter can be written as 
the sum of all terms with two and more derivatives, $\Delta S_2$ and 
the linear derivative terms, $\Delta S_{\bf a}\s0{\delta}{\delta 
  \phi_{\bf a}}$ and $\Delta S^{\bf a}[J,\dot R]
\s0{\delta}{\delta J^{\bf a}}$ respectively. This leads us to 
\begin{eqnarray}\nonumber 
  &&  \hspace{-1.5cm}\Delta S_2[\phi,\dot R] =
  \Delta S_2[J,\dot R]\\ 
  &&-
  (\Delta S_{\bf a}[\phi\,,\,\dot R])\s0{\delta}{\delta \phi_{\bf a}}+
  (\Delta S^{\bf a}[J\,,\,\dot R])\s0{\delta}{\delta J^{\bf a}}\,.
\label{eq:dj-dphi}\end{eqnarray}
The validity of \eq{eq:dj-dphi} follows from the above considerations,
but also can be directly proven by inserting \eq{eq:djphi} in the
first term on the right hand side. Using the representation
\eq{eq:dj-dphi} of $\Delta S_2[\phi,\dot R]$ in \eq{eq:flowdGk0}, only
the terms in the second line of \eq{eq:dj-dphi} survive as $ (\Delta
S_2[J,\dot R]\, J)=0$. Furthermore $\Delta S^{\bf c}[J,\dot
R]\s0{\delta}{\delta J^{\bf c}} \gamma^{\bf a}{}_{\bf b} J^{\bf b}
=(\Delta S^{,\bf a}[G\s0{\delta}{\delta \phi}+\phi\,,\,\dot R])$, and
\eq{eq:flowdGk0} reduces to
\begin{eqnarray}\nonumber 
\partial_t\Gamma_{k}{}^{,\bf a}= 
  (\Delta S_{\bf b}[\phi\,,\,\dot R])\left(\Gamma_k+\Delta S'
  \right)^{,\bf ba}-\Delta S'^{,\bf a}[\phi,\dot R]\,.\\    
\label{eq:flowdGk1}  \end{eqnarray} 
Both terms on the right hand side of \eq{eq:flowdGk1} are total
derivatives w.r.t.\ $\phi_{\bf a}$. For the first term this follows
with \eq{eq:propk} and it reduces to $(\Delta S[G\s0{\delta}{\delta
  \phi}+\phi\,,\,\dot R])^{,\bf a}$. With this observation we arrive
at
\begin{eqnarray}\label{eq:flowofdGk}
  \partial_t\Gamma_{k}{}^{,\bf a}= \left[ 
    (\Delta S[G\s0{\delta}{\delta \phi}+\phi\,,\,\dot R])
    -\Delta S'[\phi,\dot R]\right]^{,\bf a}\,,
\end{eqnarray}
which upon integration yields \eq{eq:flowGk}. \smallstep

\subsubsection{Standard flow}
The standard flow relates to regularisations $\Delta S_k[\phi]$
quadratic in the fields $\phi_{\bf a}$. We also restrict ourselves to
bosonic $R$'s, that is no mixing of fermionic and bosonic fields in
the regulator. Then, the flow of $\tilde I_k$ can be directly read off
from \eq{eq:flowtIk}
\begin{eqnarray}\label{eq:standflowtIk}
  \partial_t\tilde I_k[\phi]+ (G\,\dot R\, G)_{\bf bc}  
  \tilde I_{k}{}^{,\bf c b}[\phi]=0\,. 
\end{eqnarray} 
The flow equation for $\Gamma_k$ is extracted from $\tilde
I_{k}{}^{\bf a}=\Gamma_{k}{}^{,\bf a}$.  This $\tilde I_k$ can be
constructed from $\hat I^{\bf a}=\gamma^{\bf ba} J_{\bf b}$: we get
$\hat I_k{}^{\bf a} = \gamma^{\bf ba} J_{\bf b}- R^{\bf ab}
\s0{\delta}{\delta J^{\bf b}}$.  Inserting this operator into
\eq{eq:Ik} we arrive at $\tilde I_{k}{}^{\bf a}=\gamma^{\bf ba} J_{\bf
  b} -R^{\bf ab}\phi_{\bf b} = \Gamma_{k}{}^{,\bf a}$. Its flow is
read off from \eq{eq:flowtIk} as
\begin{eqnarray}\label{eq:standflowdGk}
 \partial_t\Gamma_{k}{}^{,\bf a}= - (G\,\dot R\, G)_{\bf bc} 
\,\Gamma_{k}{}^{,\bf cba}= 
\bigl[ G_{\bf bc}\dot R^{\bf bc}\bigr]^{,\bf a}\,,   
\end{eqnarray} 
where we again have used \eq{eq:propk}. The flow \eq{eq:standflowdGk}
matches \eq{eq:flowofJ} and can be trivially integrated in $\phi$, 
\begin{eqnarray}\label{eq:standflowGk}
  \dot \Gamma_k= 
  G_{\bf bc}\dot R^{\bf bc}\,, 
\end{eqnarray} 
where we have put the integration constant to zero.
\Eq{eq:standflowGk} is the standard flow equation of $\Gamma_k$ as
derived in \cite{Wetterich:1992yh} (up to the normalisation $\s012$
absorbed in $R$). It matches the flow of $W_k$, \eq{eq:flowWk}, when
using \eq{eq:propk} and the definition of $\Gamma_k$ in \eq{eq:Gammak}
\begin{eqnarray}\label{eq:standflowWG} 
  \hspace{-.4cm}\partial_t \Gamma_k[\phi]= 
  -\partial_t W_k[J]-\dot 
  R^{\bf ab}\phi_{\bf a} \phi_{\bf b} =\dot R^{\bf bc}\,G_{\bf bc}\,. 
\end{eqnarray} 
In \eq{eq:standflowWG} we have used that $(\partial_t J^{\bf a})
(\phi_{\bf a}-W_{k,\bf a})=0$, see \eq{eq:defofphi}. Note that we
could have used \eq{eq:standflowWG} instead of evaluating $\tilde
I_k=\phi$ for deriving \eq{eq:standflowofJ} with help of
\begin{eqnarray}\nonumber 
  \hspace{-1cm}\partial_t \gamma^{\bf ba} J_{\bf b}[\phi]&=&
  \partial_t (\Gamma_k{}^{,\bf a})
  +\dot 
  R^{\bf ab}\phi_{\bf b}\\
  &=& 
  (\partial_t \Gamma_k{})^{,\bf a}+\dot R^{\bf ab}\phi_{\bf b}\,. 
\label{eq:Jid}\end{eqnarray}
The derivatives in \eq{eq:Jid} commute as the partial $t$-derivative
is taken at fixed $\phi$. Indeed, it is the flow of the Schwinger
functional $W_k$ which is at the root of both derivations. The flow of
$W_k$ equals that of the effective Lagrangian $S_{{\rm eff}_k}$
generating amputated connected Green functions. The relation between
the flows $\partial_t \Gamma_k$ and $\partial_t S_{{\rm eff}_k}$, in
particular the (in-)equivalence within truncations, has been explored
in
\cite{Ellwanger:1993mw,Morris:1993qb,D'Attanasio:1997ej,Comellas:1997ep,Litim:2005us,Morris:2005ck},
see also the reviews
\cite{Morris:1998da,Litim:1998nf,Berges:2000ew,Bagnuls:2000ae,Polonyi:2001se,Salmhofer:2001tr}.
The numerical stability of the flows has been compared in
\cite{Litim:2005us}.  \smallstep

Finally let us study the consistency condition \eq{eq:contradict2} in
the present case. It reads $\partial_t\Gamma_k+ (G\,\dot R\, G)_{\bf
  bc} \Gamma_k^{,\bf bc}=0$, which does not match \eq{eq:standflowGk}.
Hence there is no $\hat I_k$ leading to $\tilde I_k=\Gamma_k$ and
$\Delta \tilde I_k=0$.  Again this underlines the importance of
\eq{eq:Ik} for devising flows: first one constructs an $I_k$ or
$\tilde I_k$ from \eq{eq:Ik}. Their flow is given by \eq{eq:flowIk}
and \eq{eq:flowtIk} respectively.\smallstep 

\subsubsection
{Initial condition for general flows}\label{sec:multi-loop} 

For 1PI correlation functions $\tilde I_k$ the lhs of \eq{eq:flowtIk}
consists of 1PI graphs in the full propagator $G$.  Furthermore,
\eq{eq:flowtIk} is only one loop exact if $\partial_x^2 \Delta S_k[x]$
does not depend on $x$, that is for $n\leq 2$, see also
\cite{Litim:2002xm}. For $n=2$ the flow \eq{eq:flowtIk} boils down to
the flow \eq{eq:standflowtIk}, whereas $\partial_t \tilde I_k=0$ for
$n=1$.  For $n>2$ we have higher loop terms in \eq{eq:flowtIk}.
Appropriately chosen $R^{{\bf a}_1\cdots {\bf a}_n}$ render all loops
finite. In the class of $R$ that provide momentum cut-offs, these
loops can be localised about the cut-off scale. Then the flows 
\eq{eq:flowtIk} are finite and numerically tractable, sharing most of the
advantages with the standard flow \eq{eq:standflowtIk} with $n=2$.
Indeed, for specific physical problems, in particular theories with
non-linear symmetries, the general choice in \eq{eq:flowtIk} can pay-off.
However, we emphasise that for general flows the limit $R\to \infty$
has to be studied carefully. Here, it is understood that $R\to \infty$
entails a specific limit procedure characterised by some parameter,
i.e.\ the standard momentum regularisation $R_k\to\infty$ for
$k\to\Lambda$. For practical purposes an accessible limit of the
effective action $\Gamma_k$ is required as it usually serves as the
initial condition for the flow.  In particular regulator terms $\Delta
S_k[\phi]$ that, after appropriate field rescaling, tend towards
finite expressions which are more than quadratic in the fields require
some care. The general case can be classified as follows. For a
regularisation $\Delta S$ of a theory with classical action
$S[\hat\varphi]$ and a given limit procedure $R\to \infty$ we can find
field transformations $\hat\phi\to f(R)\hat\phi$ with $f(R\to \infty)=
0$ that render $S+\Delta S$ finite:
\begin{eqnarray}\label{eq:genlim2} 
  \lim_{R\to \infty} \left(S+\Delta S\right)[f(R)\hat\phi]
  =\hat S[\hat\phi].
\end{eqnarray} 
For $R$ that diverge for all $\hat\phi$-modes $\hat S$ only depends on
$\Delta S$. In the standard case with $\hat\phi=\hat\varphi$ and $\hat
S[\hat\phi]= \hat S^{\bf ab}\hat\phi_{\bf a} \hat\phi_{\bf b}$ with
field-independent $\hat S^{\bf ab}$, the effective action $\Gamma_k$
tends towards the classical action $S$ of the theory \footnote{More
  precisely all power-counting irrelevant couplings tend to zero.}.  In
general, the corresponding effective action $\Gamma_k$ tends towards
\begin{subequations}\label{eq:Ginfty}
\begin{eqnarray}\label{eq:Ginfty1}
  \Gamma[\phi,R]\to S[\varphi(\phi)]+\left|\det\0{\partial\phi}
{\partial\varphi}(\phi)\right|
  +\hat\Gamma[\phi]\,, 
\end{eqnarray} 
where $\hat\Gamma$ is given by
\begin{eqnarray}\label{eq:hatG1}
  \hspace{-.6cm}\hat\Gamma[\phi] =- \ln\,\left(\int 
    [d\hat\phi]\,e^{
      -\hat S[\hat\phi+\phi]+\hat S[\phi]+\hat\phi_{\bf a}
      \hat S^{,\bf a}[\phi]
    }\right)\,, 
\end{eqnarray} 
\end{subequations}
and $[d\hat\phi]$ is the flat $\hat\phi$-measure including
renormalisation effects. The term in the exponent comprises the Taylor
expansion of $\hat S[\phi+\hat\phi]$ about $\phi$ leaving out the
first two terms,
\begin{eqnarray}\label{eq:hatG2}
  -\sum_{n\geq 2}\0{1}{n!} \hat\phi_{{\bf a}_n}\cdots \hat\phi_{{\bf a}_1}
  \hat S^{{\bf a}_1\cdots {\bf a}_n}\,.
\end{eqnarray}
The representation \eq{eq:hatG1} relates $\hat\Gamma$ to a Wilsonian
effective action. We emphasise that $\langle\hat\phi\rangle\neq \phi$, 
the mean field computed from the path integral \eq{eq:hatG1} is not the 
original mean field. Indeed we compute 
\begin{eqnarray}\label{eq:hatexample} 
\hat\Gamma^{,\bf a}[\phi]=(\phi_{\bf a}-\langle \hat\phi_{\bf b}\rangle ) 
\hat S^{,\bf b\bf a}\,.
\end{eqnarray}
\Eq{eq:hatexample} also entails that $\hat\Gamma$ has no classical
part due to the classical action $\hat S[\phi]$ in the exponent in
\eq{eq:hatG1}. Only those limits $(\Delta S_k,\hat S)$ admitting the
computation of the effective action $\hat\Gamma$ in \eq{eq:hatG1}
provide suitable initial conditions for the flow \eq{eq:flowtIk}. They
lead to consistent flows as defined in \cite{Litim:2002xm}. The
standard limit $(\Delta S_k, S^{\bf a\bf b} \phi_{\bf a} \phi_{\bf
  b})$ leads to a $\phi$-independent $\hat\Gamma$: the explicit
integration of \eq{eq:hatG1} gives $\s012 \ln\det S^{\bf a\bf b}$ up
to renormalisation terms stemming from $[d\hat\phi]$. Such a flow was
coined complete flow in \cite{Litim:2002xm} as it connects the
classical action with the full effective action. The flow provides for
the complete integration of quantum effects, and the theory is
determined by the parameters in the classical action $S$. The
requirement of convexity of the effective action constrains the set of
parameters in $S$, which can be evaluated with help of the regulator
dependence, see \cite{Litim:2006nn}. The general case
\eq{eq:Ginfty} with non-trivial, but accessible $\hat\Gamma$ was
coined consistent flow.  \Eq{eq:Ginfty} also covers the interesting
class of proper-time flows
\cite{Oleszczuk:1994st,Liao:1996fp,Liao:1995nm,Floreanini:1995aj,Zappala:2002nx},
where $\hat\Gamma$ comprises a full non-trivial quantum theory
\cite{Litim:2001hk,Litim:2001ky,Litim:2002xm,Litim:2002hj}. A detailed
discussion of the general situation will be given elsewhere.
\smallstep

\subsection{General variations}\label{sec:genvar} In the previous
sections we have studied one-parameter flows \eq{eq:oneparameter}.
These flows can be used to compute observables in the full theory
starting from simple initial conditions like the classical or
perturbation theory. For the question of stability of the flow or its
dependence on background fields present in the regulator we are also
interested in general variations \eq{eq:gendiff} of the regulator.  In
particular functional optimisation as introduced in
section~\ref{sec:optimal} is based on studying general variations
w.r.t.\ $R$. These variations are also useful for the investigation of
physical instabilities \cite{Litim:2006nn}. They can be
straightforwardly derived with the generalisation of \eq{eq:hatF0}:
\begin{eqnarray}\label{eq:genhatF0}
  \hat F[J,\s0{\delta}{\delta J}]=\delta R^{{\bf a}_1\cdots {\bf a}_n}
  \0{\delta}{\delta R^{{
        \bf a}_1\cdots {\bf a}_n}} \hat I[J,\s0{\delta}{\delta J}]\,,
\end{eqnarray}
with
\begin{eqnarray}
  \Delta \hat I[J,\s0{\delta}{\delta J}]=
  \left[\delta R^{{\bf a}_1\cdots {\bf a}_n}
  \0{\delta}{\delta R^{{
        \bf a}_1\cdots {\bf a}_n}}\,,\, \hat I[J,\s0{\delta}{\delta J}] 
  \right]\,.
\end{eqnarray}
The corresponding $\hat F[J,\s0{\delta}{\delta J},R]$ follows with the
commutator
\begin{eqnarray}\nonumber 
  &&\hspace{-.8cm}\left[\delta R^{{\bf a}_1\cdots {\bf a}_n}
  \0{\delta}{\delta R^{{
        \bf a}_1\cdots {\bf a}_n}}\,,\,  R^{{\bf b}_1\cdots {\bf b}_n} 
  \s0{\delta}{\delta J^{{\bf b}_1}}
  \cdots \s0{\delta}{\delta J^{{\bf b}_n}}\right]\\
  &&\hspace{-.9cm}=\delta R^{{\bf a}_1\cdots {\bf a}_n} \s0{\delta}{
    \delta J^{{\bf a}_1}}\cdots \s0{\delta}{\delta J^{{\bf a}_n}}
  \, 
\label{eq:gencom}\end{eqnarray} 
as 
\begin{eqnarray}\label{eq:genhatFk}
  &&\hspace{-.5cm}\hat F[J,\s0{\delta}{\delta J},R]\\\nonumber 
  &&\hspace{-.5cm}
  =\left(\delta R^{{\bf a}_1\cdots {\bf a}_n}\0{\delta}{\delta R^{{
          \bf a}_1\cdots {\bf a}_n}} + \Delta S[\s0{\delta}{\delta J}\,,\, 
    \delta R]
  \right) \hat I[J,\s0{\delta}{\delta J},R]\,. 
\end{eqnarray} 
With \eq{eq:genhatF0} and \eq{eq:genhatFk} the derivation of
one-parameter flows in the previous sections directly carries over to
the present case. Therefore we read off the response of $I_k$ and
$\tilde I_k$ to general variations from \eq{eq:flowIk} and
\eq{eq:flowtIk} respectively:
\begin{eqnarray}\nonumber 
  &&\hspace{-1.6cm}\left(\delta R^{{\bf a}_1\cdots {\bf a}_n}
    \0{\delta}{\delta R^{{
          \bf a}_1\cdots {\bf a}_n}} +\Delta S_1[J, 
    \delta R]
  \right)\, I[J,R]\\
  &&\hspace{3.7cm}=\Delta I[J,\delta R]\,,  
\label{eq:genflowIk}\end{eqnarray}
and 
\begin{eqnarray}\nonumber 
  &&\hspace{-1.5cm}\left(\delta R^{{\bf a}_1\cdots {\bf a}_n}
    \0{\delta}{\delta R^{{
          \bf a}_1\cdots {\bf a}_n}}+\Delta S_2[\phi,
    \delta R]\right)\tilde I[\phi,R]\\
  &&\hspace{3.5cm}=\Delta \tilde I[\phi,\delta R]\,,
\label{eq:genflowtIk}\end{eqnarray}
with $\Delta \tilde I[\phi,\delta R]=\Delta I[J(\phi),\delta R]$. For
the choice $R=R(k)$ and $\delta R=d t \,\dot R$ the flows
\eq{eq:genflowIk}, \eq{eq:genflowtIk} reduce to the one-parameter
flows \eq{eq:flowIk}, \eq{eq:flowtIk}. \step

\setcounter{equation}{0}
\section{Renormalisation group flows}\label{sec:RG-flows}
\subsection{RG flows of general correlation
  functions}\label{sec:RG-forgens} 

The flows \eq{eq:flowIk} and \eq{eq:flowtIk} comprise the successive
integrating-out of degrees of freedom in a general quantum theory. The
standard example is the integration of momentum modes, but the
formalism allows for more general definition of modes. The current $J$
and the regulator $R$ couple to $\hat\phi(\hat\varphi)$, which is not
necessarily the fundamental field $\hat\phi=\hat\varphi$. In any case,
with $R$ we have introduced a further scale $k$, thus modifying the RG
properties of the theory. Moreover, at any infinitesimal flow step
$k\to k-\Delta k$ there is a natural $k$-dependent reparameterisation
of the degrees of freedom. Taking this reparameterisation into account
should improve numerical stability. Hence the appropriate choice of
$\tilde I_\Lambda$ at the initial scale $\Lambda$ is affected by the
proper book keeping of the anomalous scaling, which becomes crucial in
the presence of fine-tuning problems. It also is relevant for studying
fixed point solutions of the flow. Hence the representation of RG
rescalings in the presence of a regulator is a much-studied subject,
e.g.\
\cite{Liao:1994cm,Ellwanger:1997tp,Pernici:1998tp,Bonini:1996bk,Comellas:1997ep,Bonini:2000wr,Latorre:2000qc,Polonyi:2000fv,Pawlowski:2001df,Pawlowski:2002eb,Bervillier:2004mf,Igarashi:1999rm,Igarashi:2001mf}.
From the formal point of view canonical transformations on the
functional space govern both RG rescalings and general flows presented
here. This point of view shall be developed elsewhere. In most
practical applications an appropriate $k$-dependent RG rescaling is
simply incorporated by hand, see reviews
\cite{Fisher:1998kv,Litim:1998nf,Morris:1998da,Aoki:2000wm,Bagnuls:2000ae,Berges:2000ew,Polonyi:2001se,Salmhofer:2001tr,Delamotte:2003dw}.
We emphasise that contrary to claims in the literature the
incorporation of RG rescalings is not a matter of consistency but
rather one of numerical stability and optimisation. We will come back
to this issue later in chapter~\ref{sec:optimal}. \smallstep

The formalism introduced in the previous chapter allows us to derive
RG equations in the presence of the regulator. In general we deal with
theories that depend on a number of fundamental couplings $\vec g$,
which also includes mass parameters.  We are interested in the
response of the theory to an infinitesimal total scale change of some
scale $s$, e.g.\ $s=k$, the flow parameter, or $s=\mu$, where $\mu$ is
the renormalisation group scale of the full theory. The couplings and
the currents may depend on this scale, $\vec g=\vec g(s)$, $J^{\bf
  a}=J^{\bf a}(s)$. An infinitesimal variation is introduced by the
operator $s\s0{d}{d s}$.  Here we consider a general linear operator
$D_s$ with
\begin{eqnarray}\label{eq:Ds} 
  D_s=s\partial_s + \gamma_{g}{}^{i}{}_j g_i\partial_{g_j}+ 
  \gamma^{ }_{\mbox{\tiny $ J$}}{}^{\bf a}{}_{\bf b} J^{\bf b}
  \s0{\delta}{\delta J^{\bf a}}\,, 
\end{eqnarray}
with
\begin{eqnarray*}
D_{s} W=0\,, 
\end{eqnarray*} 
where the partial $s$-derivative is taken at fixed $J$ (and $\vec
g$), see appendix~\ref{app:derivatives}, and the anomalous dimensions
$\gamma_{\mbox{\tiny $ J$}}$ do not mix fermionic and bosonic
currents. With $J$-independent matrices $\gamma$ we only
consider linear dependences of the currents. More general
relations are easily introduced but should be studied separately in
the specific situation that requires such a setting. Still we remark
that non-linear relations can be reduced to linear ones by coupling
additional composite operators to the currents. A relevant non-trivial
example for \eq{eq:Ds} is $s\partial_s=\mu\partial_\mu $ with
renormalisation scale (or cut-off scale) $\mu$ of $W[J,0]$ and
$\gamma_g,\gamma_{\mbox{\tiny $ J$}}$ the corresponding anomalous
dimensions of couplings and fields respectively.  We also could use
$s\partial_s=\mu\partial_\mu+\partial_t$.  We emphasise that the
operator $D_s$ accounts for more than multiplicative renormalisation.
The matrices $\gamma_g,\gamma_{\mbox{\tiny $ J$}}$ are not necessarily
diagonal and the multi-index $\bf a$ possibly includes composite
operators.  Hence \eq{eq:Ds} naturally includes the renormalisation of
composite operators (e.g.\ in $N$PI flows) or effects due to additive
renormalisation. The operator $D_s$ does not commute with derivatives
w.r.t.\ $J$. Still $D_s W=0$ can be easily lifted to identities for
general $N$-point functions with
\begin{eqnarray}\nonumber 
  &&\hspace{-1.5cm}D_s W_{,{\bf a}_1\cdots {\bf a}_N}= (D_s W)_{,{\bf a}_1
    \cdots {\bf a}_N}\\
  &&-\sum_{i=1}^N 
  \gamma^{ }_{\mbox{\tiny $ J$}}{}^{\bf b}{}_{{\bf a}_i}  
  W_{,{\bf a}_1\cdots {\bf a}_{i-1}\, {\bf b}\, {\bf a}_{i+1} 
    \cdots {\bf a}_N}, 
  \label{eq:Dsn} \end{eqnarray} 
where we have used the commutator 
\begin{eqnarray}\label{eq:dsdj}
[D_s\,,\,\s0{\delta }{\delta J^{\bf a}}]=-
\gamma^{ }_{\mbox{\tiny $ J$}}{}^{\bf b}{}_{\bf a}\s0{\delta }{
  \delta J^{\bf b}}\,.
\end{eqnarray}
The derivation of $D_s$-flows for functionals $I_k$ is done along the
same lines as that of the $t$-flow in section~\ref{sec:flows}. First we
define an operator $\hat F$ similarly to \eq{eq:hatF0} with
\begin{eqnarray}\label{eq:RGhatF0} 
  \hat F=D_s \hat I &\qquad {\rm and }\qquad \Delta \hat I=[D_s\,, \hat I]. 
\end{eqnarray} 
With $D_sW_k=0$ it follows that $F_k=\Delta I_k$ which does not vanish
in general. We shall use that still $\Delta I_k=0$ for $\hat I=1$.
The only further input needed is the commutator of the regulator term
$\Delta S$ with the differential operator $D_s$ defined in \eq{eq:Ds}.
For its determination we compute
\begin{eqnarray}\nonumber  
  &&\hspace{-2cm}[ \gamma^{ }_{
    \mbox{\tiny $ J$}}{}^{\bf a}{}_{\bf b} J^{\bf b} 
  \s0{\delta}{\delta J^{\bf a}}
  \,,\,R^{{\bf a}_1\cdots {\bf a}_n} 
  \s0{\delta }{\delta J^{{\bf a}_1} }
  \cdots \s0{\delta }{\delta J^{{\bf a}_n} }]\\ 
  &&\hspace{.2cm}
  =- n\gamma^{ }_{\mbox{\tiny $ J$}} {}^{{\bf a}_1}{}_{\bf b}  R^{{\bf b}
    {\bf a}_2\cdots {\bf a}_n}  \s0{\delta }{\delta J^{{\bf a}_1}}\cdots  
  \s0{\delta }{\delta J^{{\bf a}_n}}\,. 
  \label{eq:comandim} 
\end{eqnarray} 
where we have used the symmetry properties \eq{eq:Rsyms} of $R$.
\Eq{eq:comandim} enables us to compute the commutator $[D_s,\Delta
S]$.  For the sake of brevity we introduce a short hand notation for
the symmetrised contraction of $\gamma$ with $R$,
\begin{eqnarray}\label{eq:defofGphi}
  (\gamma^{ }_{\mbox{\tiny $ J$}}\, T)^{{\bf a}_1\cdots {\bf a}_n}=
  \sum_{i=1}^n 
  \gamma^{ }_{\mbox{\tiny $ J$}}{}^{{\bf a}_i}{}_{\bf b} T^{{\bf a}_1
    \cdots 
    {\bf a}_{i-1}{\bf b}\,{\bf a}_{i+1}\cdots {\bf a}_n}\,, 
\end{eqnarray} 
for a given $n$. The commutator of $\Delta S$ with the differential
operator $D_s$ takes the simple form
\begin{eqnarray}\label{eq:RGcom} \Bigl[D_s\,,\,\Delta S[\s0{\delta }{\delta
    J},R]\Bigr]= \Delta S[\s0{\delta}{\delta J}\,,\,(D_s-\gamma^{
  }_{\mbox{\tiny $ J$}} ) R]\,.
\end{eqnarray} 
With the above preparations the derivation of the RG flow boils down
to simply replacing $\dot R$ in the commutator \eq{eq:tcom} with
$(D_s-\gamma^{ }_{\mbox{\tiny $ J$}}) R$ and allowing for a non-zero
$F_k=\Delta I_k$. We finally arrive at
\begin{eqnarray}\label{eq:RGflowIk}
  \left(D_s +\Delta S_1[\s0{\delta}{\delta J}\,,\, 
    (D_s-\gamma^{ }_{\mbox{\tiny $ J$}}) R]\right)\, I_k=\Delta I_k\,,  
\end{eqnarray}
where $\Delta \hat I=[D_s, \hat I]$. The term $\Delta I_k$ contains
the $s$-scaling inflicted by the operator $\hat I$, and $\Delta S_1
I_k$ contains the additional scaling inflicted by the operator $\Delta
S$. In summary \eq{eq:RGflowIk} comprises general scalings in the
presence of the regulator, and reduces to the flow \eq{eq:flowIk} for
$s=k$, up to an additional $k$-dependent RG rescaling. We also
emphasise that for the derivation of \eq{eq:RGflowIk} only the
linearity of the operator $D_s$ has been used. \smallstep

An explicit example for the content of \eq{eq:RGflowIk} is provided by
the RG equation of $N$-point functions $I^{(N)}_k=\langle \phi_{{\bf
    a}_1 } \cdots \phi_{{\bf a}_N} \rangle$ as defined in
\eq{eq:moments}. Then $D_s =D_\mu$, implementing RG rescalings in the
full theory.  Furthermore we assume that the operator $\Delta S$ does
not spoil the RG invariance of the theory, i.e.\ the commutator
\eq{eq:RGcom} vanishes. The requirements on the regulator $R$ leading
to a vanishing commutator are further evaluated in the next
section~\ref{sec:RG-flowsI}. The RG equation for $I_k$ is read off
from \eq{eq:RGflowIk} as $(D_\mu+N \gamma_{\mbox{\tiny $ J$}}{}^{{\bf
    b}}{}_{{\bf a}_1}) (I_k^{(N)})_{{\bf b} {\bf a}_2 \cdots {\bf
    a}_N}=0$, where $\Delta I_k$ produces the explicit scaling
$N\gamma$ of the $N$-point function. This is the usual RG equation for
$N$-point functions as expected. For connected $N$-point functions it
is put down in \eq{eq:Dsn}. \smallstep

The general equation \eq{eq:RGflowIk} simplifies in the case of 
quadratic regulators, 
\begin{eqnarray}\nonumber 
  &&\hspace{-1.3cm}\Bigl(D_s +\s012 \left[\left(D_s-
      \gamma^{ }_{\mbox{\tiny $ J$}}\right) R\right]^{\bf ab} 
  \s0{\delta }{\delta J^{\bf a}}\s0{\delta }{\delta J^{\bf b}}\\
  &&+\phi^{\bf a} \left[\left(D_s- \gamma^{ }_{\mbox{\tiny $ J$}}\right) 
    R\right]^{\bf ab}
  \s0{\delta }{\delta J^{\bf b}}
  \Bigr) I_k=\Delta I_k, 
\label{eq:standRGflowIk}\end{eqnarray}
where
\begin{eqnarray}\label{eq:DsR} 
[(D_s -\gamma_{\mbox{\tiny $ J$}}) R]^{\bf ab}=
D_s R^{\bf ab}-2 \gamma^{ }_{\mbox{\tiny $ J$}}{}^{\bf a}{}_{\bf c}\,  
R^{\bf cb}\,. 
\end{eqnarray} 
In the last equality in \eq{eq:DsR} we have used $R^{\bf ab}=(-1)^{\bf
  ab} R^{\bf ba}$ and the fact that $\gamma_{\mbox{\tiny $ J$}}$ does
not mix fermionic and bosonic currents. The general $s$-scaling of the
Schwinger functional for quadratic regulator is derived similarly to
the flow \eq{eq:standardflowWk}: we use $\hat I=\s0{\delta }{\delta
  J^{\bf a}}$ which leads to $\Delta I_{k}=-\gamma^{ }_{\mbox{\tiny $
    J$}}{}^{\bf b}{}_{\bf a} W_{k,\bf b}$.  Moreover we have $D_s
W_{k,\bf a}+\gamma^{ }_{\mbox{\tiny $ J$}}{}^{\bf b}{}_{\bf a}
W_{k,\bf b}=(D_s W_k)_{,\bf a}$.  Inserting this into \eq{eq:RGflowIk}
we arrive at
\begin{eqnarray}\label{eq:RGflowdW}
  \hspace{-.6cm}  \left[D_s W_k+\s012 (G_{\bf bc}
    +\phi_{\bf b}\phi_{\bf c})\, 
    [(D_s - \gamma^{ }_{\mbox{\tiny $ J$}}) 
    R]^{\bf bc}\right]_{,\bf a}=0. 
\end{eqnarray}
Upon integration we are led to
\begin{eqnarray}\label{eq:RGflowWk}
  D_s W_k=-\s012 (G_{\bf ab}+\phi_{\bf a}\phi_{\bf b} )\,[
  (D_s - \gamma^{ }_{\mbox{\tiny $ J$}}) R]^{\bf ab}. 
\end{eqnarray} 
\Eq{eq:RGflowWk} entails the response of the theory to a general
scaling including the flow \eq{eq:flowIk} as well as RG rescalings.
For $s=\mu$ \eq{eq:RGflowWk} expresses the modification of the
RG equation $D_\mu W[J,0]=0$ in the presence of the regulator.\smallstep 

\subsection{RG flows in terms of mean fields}
\label{sec:RG-flowsI}
We proceed by turning \eq{eq:RGflowIk} into an equation formulated in
terms of 1PI quantities and fields. This is done by repeating the
steps in the derivation of \eq{eq:flowtIk}, and hence we shorten the
details.  First we lift \eq{eq:resolve} to operators $D_s$. This
requires the definition of the action of $D_s$ on functionals
$F[\phi]$ as provided in appendix~\ref{app:derivatives}:
\begin{eqnarray}\label{eq:1PIDs}
  D_s =(s\partial_s + \gamma_{g}{}^{i}{}_j  g_i\partial_{ g_j}+ 
  \gamma_\phi{}{}^{\bf b}{}_{\bf a} \phi_{\bf b}
  \,\s0{\delta}{\delta \phi_{\bf a}} )\,.
\end{eqnarray}
With $\tilde I_k=I_k[J(\phi)]$ we rewrite $D_s\tilde
I_k$ in terms of $\tilde I_k$ as 
\begin{eqnarray}\nonumber 
  D_s \tilde I_k[\phi]= D_s I_k[J]
  +\bigl((
  D_s-\gamma^{ }_{\mbox{\tiny $ J$}}) 
  J[\phi]\bigr)^{\bf a} G_{\bf ab}\,
  \tilde I_{k}{}^{,\bf b}[\phi]\,,\\ 
\label{eq:resolveDs}\end{eqnarray}
where $(\gamma^{ }_{\mbox{\tiny $ J$}} J)^{\bf a}= \gamma^{
}_{\mbox{\tiny $ J$}}{}^{\bf a}{}_{\bf b} J^{\bf b}$.  In
\eq{eq:resolveDs} we have used that $D_s|_J I_k= D_s I_k -(\gamma^{
}_{\mbox{\tiny $ J$}} J[\phi])^{\bf a} I_{k,\bf a}$.  In case $D_s$
stands for a total derivative w.r.t.\ $s$, the second term on the
right hand side of \eq{eq:resolveDs} has to vanish, $(D_s-\gamma^{
}_{\mbox{\tiny $ J$}}) J=0$.  Then, keeping track of dependences on
$\phi$ or $J$ is irrelevant. With \eq{eq:resolveDs} we get
\begin{subequations}\label{eq:preRGflowtIk}
  \begin{eqnarray} \nonumber \left(D_s+\Delta S_2[\phi, 
(D_s-\gamma^{ }_{\mbox{\tiny $ J$}})
      R]\right)\tilde I_k[\phi]
    =\Delta \tilde I_k-\Delta_{\bf b}\, \tilde I_{k}{}^{,\bf b}\,, \\
\label{eq:preRG1}
\end{eqnarray}
where 
\begin{eqnarray}
  \hspace{-.7cm}\Delta_{\bf b} =
  (D_s J)^{\bf a} G_{\bf ab}
  +(\Delta S_{\bf b}[\phi\,,\,(D_s-\gamma^{ }_{\mbox{\tiny $ J$}}) R])\,.  
\label{eq:preRG2}
\end{eqnarray} 
\end{subequations}
We emphasise that $\gamma_\phi$ in $D_s\tilde I_k$ is at our
disposal.  Now, as in the case of the $t$-flow for $\tilde I_k$, we
simplify the above equation by solving it for $\tilde I_k=\phi$
following from $\hat I_k= \s0{\delta}{\delta J} $. Then, $D_s \tilde
I_k= \gamma_\phi \phi $ and $\Delta \tilde I_k=-\gamma^{
}_{\mbox{\tiny $ J$}} \phi$. This leads to
\begin{eqnarray}
\Delta_{\bf b} =-(\gamma_\phi+
\gamma^{ }_{\mbox{\tiny $ J$}})^{\bf a}{}_{\bf b}\phi_{\bf a}\,. 
\label{eq:RGzero}
\end{eqnarray} 
Inserting this into \eq{eq:preRGflowtIk} the $D_s$-flow equation for
$\tilde I_k$ reads
\begin{eqnarray}\nonumber
  (D_s+\Delta_{\bf a} \s0{\delta}{\delta \phi_{\bf a}})
  \tilde I_k+\Delta S_2[\s0{\delta}{\delta \phi}\,,\,
  (D_s-\gamma^{ }_{\mbox{\tiny $ J$}}) R]\tilde 
  I_{k}=\Delta \tilde I_k\,,\\  
\label{eq:preRG}\end{eqnarray} 
where 
\begin{eqnarray}\label{eq:defDs+Del}
  D_s+\Delta_{\bf a} \s0{\delta}{\delta \phi_{\bf a}}=
  s\partial_s +\gamma_{g}{}^{i}{}_j g_i \partial_{ g_j}- 
  \gamma^{ }_{\mbox{\tiny $ J$}}{}^{\bf b}{}_{\bf a} \, 
  \phi_{\bf b}  \s0{\delta}{\delta \phi_{\bf a}}\,.
\end{eqnarray}
The dependence on $\gamma_\phi$ has completely dropped out. Its r$\hat
{\rm o}$le has been taken over by $-\gamma^{ }_{\mbox{\tiny $ J$}}$.
In other words, however we choose the fields $\phi$ to scale under
$D_s$, the RG flow \eq{eq:preRGflowtIk} shows its natural RG scaling
induced by $D_s W=0$ and $D_s J=\gamma^{ }_{\mbox{\tiny $ J$}} J$. For
the $t$-flows studied in section~\ref{sec:flows} this translates into
$\partial_t\phi=0$, corresponding to the natural choice
$\gamma_\phi=0$. As $\gamma_\phi$ is at our disposal we take the
natural choice
\begin{eqnarray}\label{eq:natural} 
  \gamma_\phi=-\gamma^{ }_{\mbox{\tiny $ J$}}\,,
\end{eqnarray}
for which $\Delta_{\bf b}\equiv 0$. With the choice \eq{eq:natural} we
arrive at
\begin{eqnarray}\label{eq:RGflowtIk}
  \left(D_s+\Delta S_2[\phi,
    (D_s+\gamma_\phi) R]\right)\tilde I_k[\phi]=\Delta \tilde I_k[\phi]\,, 
\end{eqnarray}
where $\Delta \tilde I$ derived from \eq{eq:Ik} with 
$\Delta \hat I=[D_s, \hat I]$. \Eq{eq:RGflowtIk} is the
$\phi$-based representation of \eq{eq:RGflowIk}, and hence comprises
general explicit and implicit scalings in the presence of the
regulator. A special case are those $s$-scalings with
$(D_s+\gamma_\phi)R=0$ leading to $\Delta S_2 \tilde I_k=0$. For these
choices of the pairs $(R\,,\, D_s)$ the $s$-scaling of the regularised
theory remains unchanged in the presence of the regulator. If $D_s$
stands for a scale-symmetry of the full theory such as the
RG invariance with $s=\mu$, regulators with $(D_s+\gamma_\phi)R=0$
preserve the RG properties of the full theory, see
\cite{Pawlowski:2001df,Pawlowski:2002eb}. We shall discuss this
interesting point later in section~\ref{sec:secRG&opt}. \smallstep

The above equations \eq{eq:RGflowIk},\eq{eq:RGflowtIk} can be
straightforwardly lifted to include general variations
\eq{eq:genflowIk},\eq{eq:genflowtIk} by
\begin{eqnarray}\label{eq:D_R}
  D_s\to D_R=\left.\delta R^{{\bf a}_1\cdots {\bf a}_n}
    \0{\delta}{\delta R^{{
          \bf a}_1\cdots {\bf a}_n}}\right|_s +\delta s\,D_s\,
\end{eqnarray}
with variations $\delta R(k)$ about $R(k)$ and $\delta s(R,\delta
R),s(R)$.  The operator $D_R$ stands for the total derivative w.r.t.\
$R$, hence using $D_R$ in \eq{eq:genflowtIk} simply amounts to
rewriting a total derivative w.r.t\ $R$ in terms of partial
derivatives.  These general variations are important if it comes to
stability considerations of the flow as well as discussing fixed point
properties. \smallstep

We close this section by illustrating the content of the RG flow
\eq{eq:RGflowtIk} within some examples. First we note that by
following the lines of the derivation for the $t$-flow of $\Gamma_k$,
\eq{eq:flowGk}, we can derive the RG flow of the effective action. It
is given with the substitutions $\partial_t\to D_s$ and $\dot R\to
(D_s+\gamma_\phi)R$ in \eq{eq:flowGk}.  For quadratic regulators
\eq{eq:RGflowtIk} reduces to
\begin{eqnarray}\label{eq:standRGflowtIk}
  D_s \tilde I_k+\s012  (G\,\left[(D_s+ 
    \gamma_\phi) R\right]\, G)_{\bf ab} \tilde 
  I_{k}{}^{,\bf ab}=\Delta \tilde I_k\,, 
\end{eqnarray}
where
\begin{eqnarray}\label{eq:DsRphi} 
  [(D_s -\gamma_{\phi}) R]^{\bf ab}=
  D_s R^{\bf ab}+2 \gamma_{\phi}{}^{\bf a}{}_{\bf c}\,  
  R^{\bf cb}\,. 
\end{eqnarray} 
The $D_s$-flow of the effective action $\Gamma_k$ is derived with the
choice $\hat I^{\bf a} =\gamma^{\bf a}{}_{b} J^b$.  This leads to
$\tilde I_k^{\bf a}=\Gamma_{k}{}^{,\bf a}$ and $\Delta\tilde I_k^{\bf a}
=\gamma^{
}_{\mbox{\tiny $ J$}}{}^{\bf a}{}_{\bf b} \Gamma_{k}{}^{,\bf b}$.  By
also using the commutator $[D_s\,, \s0{\delta}{\delta \phi_{\bf
    a}}]= \gamma^{ }_{\mbox{\tiny $ J$}}{}^{\bf
  a}{}_b\s0{\delta}{\delta \phi_{\bf b}}$ we are led to
\begin{eqnarray}\label{eq:preRGflow1PI}
  \left[D_s \Gamma_k-\s012 G_{\bf bc} [(D_s + \gamma_\phi) R]^{\bf bc}
  \right]_{,\bf a}=0\,. 
\end{eqnarray}
This is trivially integrated and we arrive at 
\begin{eqnarray}\label{eq:RGflow1PI}
  D_s \Gamma_k=\s012 G_{\bf bc} [(D_s + \gamma_\phi) R]^{\bf bc}\, ,
\end{eqnarray}
where we have set the integration constant to zero. The lhs of
\eq{eq:RGflow1PI} can be projected onto the anomalous dimensions
$\gamma$ with appropriate derivatives w.r.t.\ fields and momenta. Then
the rhs is some linear combination of $\gamma$'s. These relations can
be solved for the $\gamma$'s, see e.g.\
\cite{Pernici:1998tp,Pawlowski:2001df,Pawlowski:2002eb}.
With the choice $s=\mu$ and $(D_\mu+\gamma_\phi)R=0$ we are led to the
equation $D_\mu \Gamma_k=0$, the regularised effective action
satisfies the RG equation of the full theory. This interesting case is
further discussed in section~\ref{sec:secRG&opt}.\step

\setcounter{equation}{0}
\section{Optimisation}\label{sec:optimal}
An important aspect concerns the optimisation of truncated flows.
Optimised flows should lead to results as close as possible to the
full theory within each order of a given systematic truncation scheme.
This is intimately linked to numerical stability and the convergence
of results towards physics as already mentioned in the context of RG
rescalings in the last section.  By now a large number of conceptual
advances have been accumulated
\cite{Ball:1994ji,Liao:1999sh,Litim:2000ci,Litim:2001up,Litim:2001fd,Litim:2001dt,Litim:2002cf,Litim:2005us,Canet:2002gs,Canet:2004xe,JDL,Litinprep},
and are detailed in sections~\ref{sec:PMS},~\ref{sec:stability}. In
particular \cite{Litim:2000ci} offers a structural approach towards
optimisation which allows for a construction of optimised regulators
within general truncation schemes. Still a fully satisfactory set-up
requires further work. In the present section we take a functional
approach, which allows us to introduce a general setting in which
optimisation can accessed.  This is used to derive a functional
optimisation criterion, which admits the construction of optimised
regulators as well as providing a basis for further advances.
\smallstep

\subsection{Setting}\label{sec:optintro}
The present derivation of flows is based on the existence of a {\it
  finite} Schwinger functional $W$ and finite correlation functions
$\CO[\phi]$ for the full theory. These quantities are modified by the
action of an $R$-dependent operator, $\CO[\phi]\to \CO[\phi,R]$ with
$\CO[\phi]=\CO[\phi,0]$, see section~\ref{sec:setting}. One-parameter
flows \eq{eq:genflowtIk} connect initial conditions, that are well
under control, with the full theory.  For most theories these flows
can only be solved within approximations. Typically truncated results
for correlation functions $\CO[\phi,0]$ show some dependence on the
chosen flow trajectory $R(k)$ not present for full flows by
definition.  Naturally the question arises whether we can single out
regulators $R(k)$ that minimise this non-physical regulator
dependence.  \smallstep

Consider a general systematic truncation scheme: at each order of this
systematic expansion we include additional independent operators to
our theory, thus successively increasing the number of independent
correlation functions. At each expansion step these correlation
functions take a range of regulator-dependent values.  This regulator
dependence should be rather small if the truncation scheme is well
adapted to the physics under investigation. In extremal cases the
truncation scheme may only work for a sub-set of well-adapted
regulators but fail for others. An optimisation of the truncation
scheme is achieved if at each successive expansion step and for the
set of correlation functions included in this step we arrive at values
that are as close as possible to the physical ones of the full theory.
In all cases such an optimisation of the truncation scheme is wished
for as it increases the reliability and accuracy of the results, in
the extremal case discussed above it even is mandatory. \smallstep

General correlation functions $\CO[\phi]$ are either given directly by
$\tilde I[\phi]$ or can be constructed from them as the $\tilde I$
include all moments of the Schwinger functional, $\tilde I^{(N)}$, see
\eq{eq:moments}. From now on we restrict ourselves to $\tilde
I[\phi]$.  Most relations directly generalise to correlation functions
$\CO[\phi]$, in particular to physical observables, except those whose
derivation exploits the flows of $\tilde I$. The constraint of
quickest convergence can be cast into the form of an equation on the
single iteration steps within a given truncation scheme. We expand a
correlation function $\tilde I[\phi,R]$ in orders of the truncation
\begin{eqnarray}\label{eq:truncexpansion}
  \tilde I_k^{(i)}[\phi,R]=\tilde I_k^{(i-1)}[\phi,R]+\Delta^{(i)} \tilde 
  I_k[\phi,R]
  \,,
\end{eqnarray} 
where $\Delta^{(i)} \tilde I$ adds the contribution of the $i$th
order.  With adding the subscript ${}_k$ and keeping the variable $R$
we wish to make explicit the two qualitatively different aspects of
the $R$-dependence of $\tilde I^{(i)}[\phi,R]$. Firstly, the $\tilde
I^{(i)}[\phi,R]$ depend on the functional form of $R(k)$ that singles
out a path in theory space. Secondly, $k$ is specifying that point on
the path belonging to the value $k$ of the cut-off scale ranging from
$k/\Lambda\in [0,1]$. If we could endow the space of theories with a
metric, optimisation could be discussed locally as a stationary
constraint at each $k$. The resulting flows are geodesic flows, and
$k$ turns into a geodesic parameter. For now we put aside the problem
of defining a natural metric or norm on the space of theories, but we
shall come back to this important point later. \smallstep

The full correlation function in the physical theory is given by $\tilde
I[\phi]=\tilde I^{(\infty)}_0[\phi,R]$ and shows no $R$-dependence
except for a possible $R$-dependent renormalisation group
reparameterisation, not present for RG invariant quantities.
Therefore, optimisation of a correlation function $\tilde I$ at a given order
$i$ of an expansion scheme is simply minimising the difference
\begin{eqnarray}\label{eq:mintI} 
  \min_{R(k)} \|\tilde I[\phi]-\tilde I_0^{(i)}[\phi,R]\|=\min_{R(k)}  \|
  \sum_{n=i+1}^\infty\Delta^{(n)}\tilde I_0 \|\,, 
\end{eqnarray} 
on the space of one-parameter flows $R(k)$. An optimal trajectory
$R_{\rm opt}(k)$ is one where the minimum \eq{eq:mintI} is achieved.
As already mentioned in the last paragraph, for the general
discussion we leave aside the subtlety of specifying the norm
$\|.\|$. The constraint \eq{eq:mintI} also fixes the freedom of
RG rescalings for a given $\tilde I$ with fixed RG scheme in the full
theory.  \smallstep

How can such an optimisation \eq{eq:mintI} be achieved? A priori we
cannot estimate how close to physics the results are, that were
obtained with a specific regulator and truncation step. If we could,
we knew the physical results in the first place and there would be no
need for any computation. Hence an optimisation of the $i$th order
within a general truncation scheme has to be based either on
structural aspects of the flow or on an evaluation of successive
truncation steps; both procedures allow to evaluate \eq{eq:mintI}
within the given $i$th order.  For correlation functions $\tilde I$ with
\begin{eqnarray}\label{eq:condpos}
  \|\sum_{n=i+1}^\infty \Delta^{(n)}\tilde I \|=
  \sum_{n=i+1}^\infty \|\Delta^{(n)}\tilde I \|\,,
\end{eqnarray}
we can reduce \eq{eq:mintI} to a constraint on $\tilde I^{(i)}$ at a
given order $i$. The minimum in \eq{eq:mintI} is approached for
regulators minimising each term $\|\Delta^{(n)}\tilde I \|$
separately. In this case optimised regulators $R_{\rm opt}(k)$ are
those with
\begin{eqnarray}\label{eq:stability}
  \|\Delta^{(i)} \tilde I_0[\phi,R_{\rm opt}(k)]\|= \min_{R(k)}
  \|\Delta^{(i)} \tilde I_0[\phi,R(k)]\|\,, 
\end{eqnarray} 
for almost all $i,\phi$.  \Eq{eq:stability} is the wished for relation
applicable at each order of the truncation. Note that
\eq{eq:stability} also eliminates the freedom of a $k$-dependent
RG scaling of general correlation functions. It picks out that
implicit RG scaling which minimises the norm of $\Delta^{(i)} \tilde
I_0$. One could argue that an optimisation with \eq{eq:stability}
possibly gives close to optimal convergence even if \eq{eq:condpos} is
not strictly valid: in the vicinity of optimal regulators sub-leading
orders $\tilde I-\tilde I^{(i+1)}$ are small in comparison to the
leading rest term $\Delta^{(i)} \tilde I$ and a partial cancellation
between them should not have a big impact on the optimisation. Still
it is dangerous to rely on such a scenario. For its importance we
discuss the general situation more explicitly: assume that we deal
with $m_{\rm max}$ observables $\lambda_m^{\rm phys}$, $m=1,...,m_{\rm max}$,
built off some set of $\tilde I[\phi,R]$'s.  Examples are critical
exponents, physical masses, particle widths etc.. Within the $i$th
order of a given truncation scheme and a flow trajectory $R(k)$ we get
$\lambda_{m}^{(i)}[R]$ taking values in an interval $ [{\lambda_m^{\rm
    min}}^{(i)},{\lambda_m^{\rm max}}^{(i)}]$. By construction the
extremisation picks out either ${\lambda_m^{\rm min}}^{(i)}$ or $
{\lambda_m^{\rm max}}^{(i)}$. This procedure entails an optimisation
if $\lambda_m^{\rm phys}\nin [{\lambda_m^{\rm
    min}}^{(i)},{\lambda_m^{\rm max}}^{(i)}]$ (subject to the correct
choice of the closest extremum).  In turn, if $\lambda_m^{\rm phys}\in
[{\lambda_m^{\rm min}}^{(i)},{\lambda_m^{\rm max}}^{(i)}]$ a procedure
picking out the boundary points decouples from optimisation, only by
chance it provides close to optimal results.  Indeed this scenario is
likely to be the standard situation at higher order of the truncation
scheme.  An indication for this case is the failure of finding
coinciding extrema for all observables, in particular if these extrema
are far apart. The resolution of this problem calls for an
observable-independent optimisation based on \eq{eq:mintI}. \smallstep
 
The evaluation of the optimisation \eq{eq:mintI} is more convenient in
a differential form. This equation can be directly derived from
\eq{eq:mintI}. However, there exists an alternative point of view
which might also be fruitful: truncated flows may be amended with
functional relations valid in the full theory. The hope is to carry
over some additional information from the full theory that is not
present in the truncation of the flow. This is the idea behind the use
of symmetry relations such as STIs together with flows.  In the
context of optimisation the key relation is the regulator independence
of the full theory. By construction the end-points of one-parameter
flows $\tilde I[\phi]=\tilde I_0[\phi,R]$ are correlation functions in
the full theory, being trivially independent of the path $R(k)$ in
regulator space: $k$ is a further variable of $R$ and any local
variation of such a path about a regulator $R^{{\bf a}_1\cdots {\bf
    a}_n k}$ does not change $\tilde I_0[\phi,R]$.  Moreover, in
section~\ref{sec:RG-flows} we have seen that there is the freedom of
$k$-dependent RG scalings of the full theory, and the apparent
independence of $\tilde I[\phi]=\tilde I_0[\phi,R]$ on the path $R(k)$
for full flows is expressed in the relation
\begin{eqnarray}
  &&\hspace{-1cm}\delta R^{{\bf a}_1\cdots {\bf a}_n k}
  \0{\delta \tilde I_0[\phi,R]}{\delta R^{{\bf a}_1\cdots {\bf a}_n k}}
  = \delta\left(\ln\mu\right)\,D_{\mu}\tilde I_0[\phi,R] \,,
\label{eq:Rindep}\end{eqnarray} 
for all $\tilde I[\phi]$.  The variation on the lhs of \eq{eq:Rindep}
stands for the total derivative w.r.t.\ $R^{{\bf a}_1\cdots {\bf a}_n
  k}$ also including possible $R$-dependent RG scalings as in $D_R$,
\eq{eq:D_R}.  The rhs of \eq{eq:Rindep} accounts for a possible
integrated $R$-dependence of the renormalisation scheme at $k=0$:
$\delta \mu(R,\delta R),\mu(R)$.  For RG invariant $\tilde I[\phi]$
the rhs of \eq{eq:Rindep} vanishes.  For RG variant $\tilde I[\phi]$
the rhs can always be absorbed in an appropriate redefinition of the
variation w.r.t.\ $R$, though technically this might be difficult. The
relation of \eq{eq:Rindep} to the optimisation \eq{eq:mintI} is
provided by enforcing \eq{eq:Rindep} already for the $i$th order of
the truncation scheme and absorbing the RG scaling on the rhs in an
appropriate redefinition of the $R$-variation. Also assuming
\eq{eq:condpos} we are led to
\begin{eqnarray}\label{eq:RindepdI}
  \delta R^{{\bf a}_1\cdots {\bf a}_n k}\,
  \0{\delta \|\tilde I_0[\phi,R]-\tilde I^{(i)}_0[\phi,R]\|}
  {\delta R^{{\bf a}_1\cdots {\bf a}_n k}} 
  = 0 \,,  
\end{eqnarray} 
which is the differential form of \eq{eq:mintI}. \Eq{eq:Rindep} is an
integrability condition for the flow. Its relation to
reparameterisations of the flow and the initial condition $\tilde
I[\phi,R(\Lambda)]$ become more evident by using
\begin{eqnarray}\label{eq:intflow} 
  \tilde I_0[\phi,R]=\tilde I_{\Lambda}[\phi,R]+\int_{\Lambda}
  ^0\frac{d k}{k} 
  \partial_t \tilde I_k[\phi,R]\,.
\end{eqnarray} 
Inserting \eq{eq:intflow} in \eq{eq:Rindep} leads to
\begin{eqnarray}\nonumber
  &&\hspace{-.7cm}\left.D_R\tilde I[\phi,R]\right|_{R({\Lambda})}+
  \int_{\Lambda}
  ^0\frac{d k}{k}\,\partial_t \left[D_R\tilde I[\phi,R]\right]_{R(k)}\\
  &&
  = 
  \delta\left(\ln\mu\right)\,D_{\mu}\tilde I_0[\phi,R]\,,  
\label{eq:Rindep1}\end{eqnarray}
with $D_R$ defined in \eq{eq:D_R}. The integrand in \eq{eq:Rindep1} is
a total derivative, and with using that $\delta R|_{R=0}=\delta \mu$
the lhs in \eq{eq:Rindep1} equals the rhs. A variation of the initial
regulator $R(\Lambda)$ in general entails that $\tilde
I[\phi,R(\Lambda)]$ cannot be kept fixed by adjusting an appropriate
RG scaling. For example, a different momentum dependence of
$R(\Lambda)$ leads to different composite operators coupled to the
theory via $\Delta S$, and hence physically different theories.  For
sufficiently large regulators these differences are usually
sub-leading. Neglecting this subtlety we conclude that in general a
change of regulator with a vanishing rhs and fixed initial conditions
$\tilde I[\phi,R(\Lambda)]$ entails a $k$-dependent RG scaling of the
flow.  \smallstep

\subsection{Principle of Minimum Sensitivity}\label{sec:PMS}

For the sake of simplicity we only discuss couplings $\lambda$'s and
not general functionals $\tilde I$ or $\CO$. \Eq{eq:RindepdI},
evaluated for one or several observables $\lambda_m$, $m=1,...,m_{\rm
  max}$, at some order $i$ of a given truncation scheme can be viewed
as a constraint for truncated flows. This implies the search for local
extrema of observables $\lambda_m$ in regulator space. However, not
knowing $\lambda_{\rm phys}$ we have to resort to \eq{eq:Rindep}, most
conveniently written as
\begin{eqnarray}\label{eq:Rindep||}
  \delta R^{{\bf a}_1\cdots {\bf a}_n k}
  \0{\delta \lambda_m}{\delta R^{{\bf a}_1\cdots {\bf a}_n k}} 
  = 0\,.
\end{eqnarray}
\Eq{eq:Rindep||} can be seen as a symmetry constraint as suggested in
the last section or as an optimisation with the assumption
\eq{eq:condpos}.  As a constraint, \eq{eq:Rindep||} can have several
solutions or none (the extremum could be a point on the boundary in
regulator space).  \Eq{eq:Rindep||} in its integral form, only
allowing global changes along the full flow trajectory, is related to
the principle of minimum sensitivity (PMS) \cite{Stevenson:1981vj},
which has been introduced to the functional RG in \cite{Ball:1994ji},
for further applications see
\cite{Liao:1999sh,Canet:2002gs,Canet:2004xe}. Its
limitations have been discussed in \cite{Litim:2001fd}. Practically
such a PMS extremum has been evaluated by computing observables
$\lambda_1,..., \lambda_{m_{\rm max}}$ for a class of regulators
$R(\alpha_1,...,\alpha_j)$ labelled with $\alpha_1,...,\alpha_j$.
Strictly speaking, $m_{\rm max}$ should increase with the order $i$ of
the truncation, as the number of observables increase with the order
$i$ of the truncation scheme. The functional derivatives w.r.t.\ $R$
turn into ordinary ones and we are left with the problem of finding a
coinciding extremum for these $\lambda$. As already mentioned before,
even if they exist at all, these extrema need not coincide.  There are
several options of how to proceed in such a situation. We can
constrain the set of regulators by fixing the value of some
$\lambda_1,...,\lambda_r$ to their physical value to all orders of the
truncation, thereby sacrificing a part of the predictive power. Such a
procedure resolves (if $r$ is big enough) the above mentioned problem
and the optimisation is done for the other observables
$\lambda_{r+1},...,\lambda_{m_{\rm max }}$ in this smaller set of
regulators, see \cite{Ball:1994ji}. One also could argue that
optimised values for each of these variables are obtained at their
extrema. A regulator that optimises the flow of $\lambda_1$ is not
necessarily optimising that for other $\lambda_m$.  This idea has been
used in \cite{Canet:2004xe} and in general requires the use of
supplementary constraints. Both procedures have to be used with care
as already discussed in general in the last
section~\ref{sec:optintro}.  Within the present explicit procedure
this analysis hints at several short-comings: firstly, fixing the
values of $r$ observables does not necessarily lead to small flow
operators $\Delta S_2$, and possibly constrains the values for
$\lambda_{r+1},...,\lambda_m$ to regions that are far from their
physical values. Secondly, non-coinciding optimal regulators also
could hint at a badly working truncation scheme, or badly chosen
$\lambda_m$. We emphasise again that searching for a solution of
\eq{eq:Rindep||} for some variable $\tilde I^{(i)}$ equals an
optimisation \eq{eq:mintI} only as long as the physical value $\tilde
I^{(\infty)}$ is not included in the range of possible values of
$\tilde I^{(i)}$. It is mainly for this reason that an
observable-independent optimisation is wished for.  \smallstep

\subsection{Stability criterion}\label{sec:stability}

The above mentioned problems are also directly related to the fact
that the preceeding use of \eq{eq:Rindep},\eq{eq:Rindep||} is not a
constructive one; it does not allow us to devise an optimal regulator
that limits the contribution of higher orders of the truncation by
construction. Moreover, an optimisation as in section~\ref{sec:PMS}
always involves considerable numerical effort. A constructive
optimisation criterion, directly based on the fundamental optimisation
condition \eq{eq:mintI} and on the structure of the functional RG, has
first been suggested in \cite{Litim:2000ci}. The construction there
also emphasises the link between optimisation, optimal convergence and
global stability of the flows.  We shall show later in
section~\ref{sec:functopt} that the criterion developed in
\cite{Litim:2000ci,Litim:2001up,Litim:2001fd,Litim:2001dt,Litim:2002cf}
relates to the local use of \eq{eq:Rindep}. \smallstep

The key point in \cite{Litim:2000ci} is the observation that
optimisation of any systematic expansion implies quickest convergence
of the expansion towards physics. Consequently we can turn the
question of optimisation into that of quickest convergence. The latter
allows to devise constructive optimisation conditions. In
\cite{Litim:2000ci} it was pointed out that for the standard flow
\eq{eq:standflowdGk} any such expansion includes an expansion in
powers of the propagator $G=1/(\Gamma_k^{(2)}[\phi]+R)$.  Hence
minimising the norm of the propagator $G$ relates to stability and
fastest convergence. Consider regulators introducing an IR cut-off
with $R=R(p^2)$ as discussed at the end of section~\ref{sec:setting}.
The norm implicitly used in \cite{Litim:2000ci} is the operator norm
on $L_2$: $\|G[\phi_0,R]\|= \sup_{\|\psi\|^{\ }_{L_2}=1} \{ \|
G[\phi_0,R]\psi\|^{\ }_{L_2}\}$, where $\|\psi\|^{\ }_{L_2}=(\int
|\psi|^2)^{1/2}$ is the $L_2$-norm. The norm $\|G[\phi_0,R]\|^{\ }_{L_2}$ 
is directly related to the biggest spectral value of $G$ at
$\phi_0$, and hence is sensitive on the growth of the maximum of $G^n$
for $n\to \infty$. A canonical choice for $\phi_0$ is a field
maximising $\|G[\phi,R]\|$ on the space of fields $\phi$. Within a
truncation scheme that uses an expansion in powers of the field
a natural choice for $\phi_0$ is the expansion point. Reformulating
the optimisation criterion of \cite{Litim:2000ci} in the present
setting leads to
\begin{eqnarray}\nonumber 
  &&\hspace{-1.3cm}\{R_{\rm stab}\}= \Bigl\{R \quad {\rm with}\quad 
  \|G[\phi_0,R]\|^{\ }_{L_2}\leq 
  \|G[\phi_0,R']\|^{\ }_{L_2}\, \\
  &&\forall\  
  R'\ {\rm and}\ R'(k_{\rm eff}^2) = R(k_{\rm eff}^2)=c\,
  k_{\rm eff}^2\Bigr\}\,.
\label{eq:optbyprop}\end{eqnarray} 
The normalisation constant $c$ is at our disposal. The condition
$R'(k_{\rm eff}^2)=c\, k_{\rm eff}^2$ is required for identifying a
parameter $k'(k_{\rm eff})$ at which the norm of the propagator is
taken. \Eq{eq:optbyprop} allows to construct optimised regulators for
general truncations schemes, even though the key demand of stability 
might necessitate supplementary constraints, see e.g.\ 
section~\ref{sec:beyond}.  At a given order it singles out a set of
stability inducing regulators as \eq{eq:optbyprop} does not restrict
the shape of $ R_{\rm stab}$. An optimisation with \eq{eq:optbyprop}
entails in the limit of large truncation order the PMS condition
\eq{eq:Rindep}, if the latter admits a solution \cite{Litim:2001fd}.
If the PMS condition has several solutions, by construction
\eq{eq:optbyprop} is likely to pick out that closer to the physical
value. \smallstep

The criterion \eq{eq:optbyprop} has very successfully been applied to
the derivative expansion \cite{Litim:2001dt,Litim:2002cf}, where also
the above statements have been checked. In its leading order, the
local potential approximation (LPA), a particularly simple optimised
regulator is provided by
\begin{eqnarray}\label{eq:litopt}
  R_{\rm opt}(p^2)= (k^2-p^2)\theta(k^2-p^2)\,,
\end{eqnarray}
where $\theta$ is the Heaviside step function. By now \eq{eq:litopt}
is the standard choice in the field. It is a solution of
\eq{eq:optbyprop} with $k_{\rm eff}^2 =\s012 k^2$ and $c=1$. As a
solution of \eq{eq:optbyprop} in LPA it only is optimised for the LPA
but not beyond, as has been already remarked in \cite{Litim:2001up}.
Beyond LPA a solution to \eq{eq:optbyprop} has to meet the necessary
condition of differentiability to the given order. The related
supplementary constraint is provided in \eq{eq:nth0con}. Solutions to
\eq{eq:litopt} with \eq{eq:nth0con} exist, being simple enhancements of
\eq{eq:litopt} \cite{Litinprep}. We add that \eq{eq:litopt} works within
truncation schemes where the full momentum dependence of correlation
functions is included from the onset.  \smallstep

\subsection{Functional optimisation}\label{sec:functopt}
In summary much has been achieved for our understanding as well as the
applicability of optimisation procedures within the functional RG.
Still, the situation is not fully satisfactory, in particular given
its key importance for the reliability of functional RG methods. In
the present section we exploit the functional equation \eq{eq:Rindep}
to devise an optimisation criterion based on stability as well as
discussing in more detail the link between stability-related criteria
and the PMS condition. We also aim at the presentation of fundamental
relations and concepts that are possibly helpful for making further
progress in this area.\smallstep

\subsubsection{Local optimisation}\label{sec:localopt} So far we have
only discussed the implications of \eq{eq:Rindep} in its integrated
form as done within the PMS optimisation in section~\ref{sec:PMS}.
Such a procedure always requires the integration of the flow and hence
involves considerable numerical effort. On the practical side, the
classes of regulators usually used for the PMS are not sufficiently
dense for resolving the local structure: for the standard choice of a
momentum regulator we parameterise quite generally $R(p^2)=p^2 r(x)$
with $x=p^2/k^2$. Then, a variation of $R$ is a variation of $r$ and
as such an integral condition as it implies a variation at all scales
$k$.  Consequently a resolution of the local (in $k$ and ${\bf a_i}$)
information of \eq{eq:Rindep} is only obtained for regulator classes
$\{R\}$ which include as differences $R_1-R_2$ smeared out versions
$\delta_\epsilon$ of the delta function in $k$: $(R_1-R_2)^{{\bf
    a}_1\cdots {\bf a}_n k}\propto \delta_\epsilon (k-k_{\rm eff})
\Delta R^{{\bf a}_1\cdots{\bf a}_n}$.  It is convenient to include
these variations functionally: evaluating \eq{eq:Rindep} for
variations local in $k$ we turn \eq{eq:Rindep} into a local condition
on $\tilde I[\phi,R]$.  As such it is the local form of the
integrability condition \eq{eq:Rindep} and can be read off from
\eq{eq:intflow} and \eq{eq:Rindep1},
\begin{eqnarray}\label{eq:localint} 
  \oint \delta \tilde I[\phi,R]=0\,, 
\end{eqnarray} 
the integral in \eq{eq:localint} describing a small closed curve in
the space of regulators. Within truncations, \eq{eq:localint} is a
non-trivial, physically relevant constraint. For example, gradient 
flows cease to be gradient flows within truncations that violate
\eq{eq:localint}. In turn, this property is kept intact if satisfying
\eq{eq:localint}. A consequence of \eq{eq:Rindep} and its local form
\eq{eq:localint} is
\begin{eqnarray} 
  \delta R^{{\bf a}_1\cdots {\bf a}_n k'}
  \0{\delta \tilde I[\phi,R(k)]}{\delta R^{{\bf a}_1\cdots {\bf a}_n k'}}
  = D_R \tilde I[\phi,R(k)]\,, 
\label{eq:Rindep3}\end{eqnarray} 
for all $\tilde I[\phi,R(k)]$ and variations $\delta R$ that vanish at
$\Lambda$. The right hand side in \eq{eq:Rindep3} accounts for a total
scale variation of the end-point $R(k)$ with $D_s$ as defined in
\eq{eq:D_R}.  We emphasise again that \eq{eq:localint} and
\eq{eq:Rindep3} are non-trivial constraints within truncations.
Moreover, at finite $k\neq 0,\infty$ the rhs in general does not agree
with $\delta\left(\ln\mu\right)\,D_{\mu}\tilde I[\phi,R(k)]$ even for
full flows, as already mentioned in section~\ref{sec:setting}:
firstly, a general variation w.r.t.\ $R$ leads to the flow
\eq{eq:RGflowtIk} with \eq{eq:D_R}, a special case being the one
parameter flow \eq{eq:flowtIk} with $\delta R= dk \partial_k R$ and
$D_s=\partial_t$.  Secondly, in the presence of two different
regulator functions $R,R'$ at some fixed scales $k,k'$ the two
regularised theories cannot completely agree as they differ by their
coupling to different composite operators $\Delta S[\phi,R]$ and
$\Delta S[\phi,R']$. Still it might be possible to identify
hyper-surfaces of regularised theories at the same physical cut-off
scale $k_{\rm eff}$. So far $k$ was just a parameter labelling
one-parameter flows, only its end-point $k=0$ (and to some extend
$R=\infty$) defining a specific theory.  For $k\neq 0$ this is a
priori not clear, the trivial example being two momentum
regularisations $R(p^2)$ and $R'(p^2)= R(c^2 p^2)/c^2$.  Obviously $k$
cannot be the physical cut-off scale in both cases. In this trivial
case it is simple to identify the relative effective cut-off scale for
$R,R'$ with $k=k_{\rm eff}$ and $k'(k_{\rm eff})=c k_{\rm eff}$. In
general the natural relation $k'(k)$ is less obvious, apart from not
being unique anyway. Nevertheless let us assume for the moment that we
have overcome this subtlety. Then we can define a variation of $R$ on
hyper-surfaces $\{ R_\bot\}_{k_{\rm eff}}=\{R( k(k_{\rm eff}))\}$
regularising the theory under investigation at the same physical
cut-off scale $k_{\rm eff}$.  Stability of the flow is achieved by
minimising its action on the set $\{R_\bot\}$ and \eq{eq:Rindep3}
translates into
\begin{eqnarray} \nonumber \left.\delta R_\bot^{{\bf a}_1\cdots {\bf
        a}_n k'} \0{\delta \tilde I[\phi,R]}{\delta R^{{\bf a}_1\cdots
        {\bf a}_n k'}} \right|_{R=R_{\rm stab}}=
  \delta \ln \mu \, D_{\mu}\tilde I[\phi,R_{\rm stab}]\,, \\
  \label{eq:prefinstable} \end{eqnarray} 
lifting \eq{eq:Rindep} to non-vanishing regulators.
\Eq{eq:prefinstable} is a non-trivial constraint already for full
flows. Subject to a given foliation of the space of theories with
$\{R_\bot\}$ for all cut-off scales $k_{\rm eff}$,
\eq{eq:prefinstable} entails maximal (in-)stability of the flow at its 
solutions $R_{\rm stab}$. With \eq{eq:Rindep3} we rewrite \eq{eq:prefinstable}
as
\begin{eqnarray} \label{eq:finstable} \left.  D_{R_\bot} \tilde
    I[\phi,R]\right|_{R=R_{\rm stab}}= 0\,,
\end{eqnarray}
where we have absorbed the RG rescaling on the rhs of
\eq{eq:prefinstable} in $D_{R_\bot}=D_R(\delta R=\delta R_\bot)$.  A
solution $R_{\rm stab}(k)$ of \eq{eq:finstable} is achieved by varying
the flows of variables $\tilde I$ in regulator space. In its form
\eq{eq:finstable} it cannot be used to construct regulators $R_{\rm
  stab}$.  To that end we have to rewrite \eq{eq:finstable} as a
criterion on the flow operator $\Delta S_2$. This is done as follows:
if a one-parameter flow $\tilde I_k[\phi]= \tilde I[\phi, R(k)]$ obeys
the constraint \eq{eq:finstable} for all $k$, so must $\partial_t
\tilde I_k$. Varying $\partial_t \tilde I_k$ with $\delta R_\bot$ it
follows with \eq{eq:flowtIk} and \eq{eq:finstable} that
\begin{eqnarray}\label{eq:dtfinstable} 
  \left.\left(
      D_{R_\bot}\Delta S_2[\phi,\dot R]
    \right)\, 
    \tilde I[\phi,R]\right|_{R=R_{\rm stab}}=0\,, 
\end{eqnarray}
where we have used that $\partial_t$ and $D_R$ commute up to
RG scalings.  For most practical purposes the RG scaling will be
neglected and \eq{eq:dtfinstable} boils down to
\begin{eqnarray} \label{eq:dtfinstableR} \left.\delta R_\bot^{{\bf
        a}_1\cdots {\bf a}_n} \0{\delta \Delta S_2[\s0{\delta}{\delta
        \phi},\dot R] }{\delta R^{{\bf a}_1\cdots {\bf a}_n}}\,\,
    \tilde I[\phi,R]\right|_{R=R_{\rm stab}}=0 \,.
\end{eqnarray} 
Finding a globally stable one-parameter flow $R_{\rm stab}(k)$ amounts
to demanding the validity of \eq{eq:dtfinstable} for all $\tilde I$
and $k$. This implies that the variation of $\Delta S_2$ in the
directions $\delta R_\bot$ has to vanish at all scales $k$ and all
index values ${\bf a_1}\cdots {\bf a_n}$, that is pointwise zero.
Clearly there is the danger of overconstraining the regulator. In
practical applications we limit ourselves to a restricted set of
$\tilde I$ for which we solve \eq{eq:dtfinstable}.  As any truncation
scheme is based on the assumption of dominance of certain degrees of
freedom the related $\{\tilde I_{\rm rel}\}$ should be taken. Then the
choices $R_{\rm stab}(k)$ lead to extrema of the action of $\Delta
S_2[\s0{\delta}{\delta \phi},\dot R_{\rm stab}]$ on $\{\tilde I_{\rm
  rel}\}$ for all scales $k$.  Such a flow, if it exists, is either
most stable (minimal $\Delta S_2$) or most unstable (maximal $\Delta
S_2$).  \Eq{eq:dtfinstable} implements the PMS condition
\eq{eq:Rindep} on $\{\tilde I_{\rm rel}\}$, as the $k$-flow vanishes
identically at $k=0$ and integrating \eq{eq:dtfinstable} over all
scales still is zero. We also emphasise that \eq{eq:dtfinstable}
defines local (in-)stability. We could have global extrema at the
boundary of the hyper-surface $\{R_\bot\}$ defined with $k_{\rm
  eff}$.\smallstep

\subsubsection{Optimisation and effective cut-off
  scale}\label{sec:opt+eff}
So far we have not fixed the hyper-surfaces $\{ R_\bot\}$ which
amounts to the definition of a metric on the space of regularised
theories.  Before embarking on a discussion of natural definitions of
such metrics we would like to elucidate the subtleties within a
simple example: assume we restrict ourselves to the set of regulators
given by a specific flow $R_{\rm base}(k)$ and possibly momentum
dependent RG rescaling of $R_{\rm base}(k)$.  Then the definition of a
natural (relative) physical cut-off scale is uniquely possible; the
set of regulators $\{R_\bot\}_k$ is defined by those regulators with
correlation functions $\tilde I[\phi,R]$ that only differ by
RG rescalings (fixed physics) from $\tilde I[\phi,R_{\rm base} ]$.
Note in this context that the RG scalings also change the field
$\phi$.  The $\{R_\bot\}_k$ cover the restricted space of regulators
we started with, and by definition \eq{eq:dtfinstable} is satisfied
for all $R\in\{R_\bot\}_k$. This should be the case as their physical
content is indistinguishable. In turn, if we had chosen another foliation 
the result would have been different. Then, necessarily $R(k), R(c
k)\in \{R_\bot\}_k$ for at least one regulator $R$ and
\eq{eq:dtfinstable} differentiates between them even though the
one-parameter flows $R(c k)$ and $R(k)$ are the same. Suitable foliations 
are those where the hyper-surfaces $\{R_{\bot}\}$ do not contain such
pathologies.  \smallstep

So far $k$ is only a parameter that provides a scale ordering without
identifying physical scales (except for $k=0$).  Consequently we have
to answer the question of how to define the distance $d$ of two points
$R$ and $R'$ in theory space given by their set of correlation
functions $I[\phi,R]$, $I[\phi,R']$, or more generally $\CO[\phi,R]$,
$\CO[\phi,R']$. To that end we define
\begin{eqnarray}\label{eq:prenorm}
  d_{{\cal O}}[R,R']=\sup_{\phi\in \CS}\{
  \|\CO[\phi,R]-\CO[\phi,R']\|\}\,,   
\end{eqnarray} 
where the supremum is taken in an appropriate space of fields $\CS$,
and we have to specify an appropriate norm $\|.\|$. A natural choice
for $\CS$ is the configuration space of the theory under
investigation. However, the definition \eq{eq:prenorm} only is useful
if $d_\CO$ is finite for almost all $R,R'$.  This can be achieved by
turning $\CO\to f(\CO)$ in an operator or functional that has a
spectrum that is bounded from below and above, e.g.\ $\CO\to
1/(C+|\CO|^2)$ with positive constant $C$.  Alternatively, one can
restrict the space of fields $\phi$, e.g.\ with $\phi\in {\cal
  S}_C=\{\phi|\,|\,\|\CO[\phi,R]\|\,,\,\|\CO[\phi,R']\|< C\}$.  Here,
the constant $C<\infty$ is introduced to get rid of singular fields
with $\CO[\phi,R]=\infty$ that possibly would render the distance
$d=\infty$ for all $R$, $R'$. Obviously allowing for these fields
would spoil the construction. We could also evaluate the norm in
\eq{eq:prenorm} for a specific configuration $\phi=\phi_0$ with
$\CS=\{\phi_0\}$. This is an appropriate choice if $\phi_0$ could be
singled out by the truncation scheme, e.g.\ as the expansion point in
an expansion in powers of the field. \smallstep

As general flows \eq{eq:RGflowtIk} for $\tilde I_k$ depend on
$\Gamma_k^{(n)}$ via $\Delta S_2$ which is the crucial input for the
optimisation, a natural choice for $\CO$ is the effective action
$\CO[\phi,R]= \Gamma[\phi,R]-\Gamma[0,R]$ \footnote{In
  \eq{eq:standflowGk} we have put an integration constant to zero,
  here we choose it to be $-\Gamma[0,R]$. At finite temperature the
  effective action $\Gamma$ cannot be renormalised that way as 
  $\Gamma[0,R]$ is related to the thermal pressure.}, or its second
derivative $\Gamma^{(2)}$.  Of course, any correlation function
$\tilde I_k$ (or set of correlation functions) that entails the full
information about the theory and has no explicit regulator dependence
is as good as the above suggestion. From now on we drop the subscript
${}_{\CO}$, keeping it only if discussing a specific choice for $\CO$.
The distance $d$ between two regularisation paths $R(k),R'(k')$ of a
theory at the effective cut-off scale $k=k_{\rm eff}$ is given by
\begin{eqnarray}\label{eq:norm}
  d[R,R'](k)=\min_{k'}\, d[R(k),R'(k')]\,,  
\end{eqnarray} 
which implicitly defines the relative effective cut-off scale $k'(k)$
as that $k'$ for which the minimum \eq{eq:norm} is obtained
\begin{eqnarray}\label{eq:keffective}
  d[R(k),R'(k'(k))]=d[R,R'](k)\,.
\end{eqnarray}  
In general $d[R,R'](k)=d[R',R](k'(k))\neq d[R',R](k)$. A priori,
$k'(k)$ is not necessarily continuous. Indeed one can even construct
pathological regulators that lead to discontinuities in $k'(k)$.  In
most theories such subtleties are avoided by using regularity
restrictions on the regulators $R(k)$ such as monotony in $k$: $R(k)\leq
R(k')$ for $k< k'$. \smallstep

The basic building block of the flow operator $\Delta S_2$ is the
full propagator $G=1/(\Gamma^{(2)}+R)$, and it would seem natural to use
$d_G$. However, $d_G[R,R']$ does not qualify directly for measuring
the distance: for physically close regularisations $R,R'$ the distance 
$d_{\Gamma^{(2)}}[R,R']$ is necessarily small 
\footnote{More precisely this applies to the distance 
$d_{f(\Gamma^{(2)})}[R,R']$ where the function $|f(x)|$ is bounded
  from above.}. Then, $d_G[R,R']$ is determined by the difference
$(R-R')$ evaluated in the regularised regime which has no physical
implication. Still, $d_G$ can be turned into a simple relation for the
effective cut-off scale $k_{\rm eff}$ with
\begin{eqnarray}\label{eq:beispiel} 
  d_{G,\rm sup}[R,\infty]=\|G[R]\|^{\ }_{\rm sup}=
\0{1}{Z_\phi}k_{\rm eff}^{\dim_G}\,,  
\end{eqnarray}
with 
\begin{eqnarray}\label{eq:supremumsnorm}
  \|G[R]\|^{\ }_{\rm sup} =\sup_{\phi} \{\| 
  G[\phi,R]\|^{\ }_{L_2}\}\,, 
\end{eqnarray} 
where the supremum is taken in configuration space. The norm $\|.\|^{\
}_{L_2}$ is the operator norm on $L_2$ already used for the criterion
\eq{eq:optbyprop}. In \eq{eq:beispiel} $\dim_G$ is the momentum
dimension of $G$, e.g.\ $\dim_G=-2$ for bosons and $ \dim_G=-1$ for
fermions. $Z_\phi$ is the wave function renormalisation of the field
$\phi$, and makes the definition of $k_{\rm eff}$ invariant under RG
rescalings. In most cases the norm \eq{eq:supremumsnorm} will be
evaluated in momentum space where it reads explicitly
\begin{eqnarray}\label{eq:normmoment}
&&\hspace{-1.3cm}
\|G[R]\|^{\ }_{\rm sup} =\sup_{\phi, \|\psi\|^{\ }_2=1} \left\{ 
\left(\int_p \bigl| G[\phi,R]\psi \bigr|^2(p)\right)^{1/2}\right\}\,.  
\end{eqnarray} 
Note that the use of $Z_\phi$ is not necessary as long as one uniquely
fixes the endpoint of the flows, the theory at vanishing regulator.
If one allows for simultaneous RG rescalings of the flow trajectories
the prefactor in \eq{eq:beispiel} arranges for an RG invariant $k_{\rm
  eff}$. For including relative RG rescalings of trajectories the
supremum in \eq{eq:beispiel} also has to be taken over
RG transformations. For most practical purposes these more general
scenarios are not of interest. \smallstep

The expression $k_{\rm eff}^{\dim_G}$ relates to the biggest spectral
value the propagator $G[\phi,R]$ can achieve for all fields $\phi$.
Therefore $k_{\rm eff}$ is the smallest relevant scale and hence is
the effective cut-off. In the limit $k\to0$ the effective cut-off
scale $k_{\rm eff}$ tends towards the smallest mass scale in the
theory \footnote{In a regime with anomalous momentum scaling
  $G\propto p^{\dim_G-2\kappa_\phi}$ one should rather define
  $\|G[R]\|_{\rm sup}= k_{\rm eff}^{\dim_G-2\kappa_\phi}/Z_\phi$ with
  dimensionful $Z_\phi$.}  .\smallstep

As an example we study a scalar theory with $R$ in the leading order
derivative expansion: $\Gamma_k[\phi]=\int \left(
\s012 \phi p^2\phi +V_k[\phi]\right)$. 
For regulators providing a momentum cut-off we can adjust $k$ 
as a physical cut-off scale by taking as a reference regulator the
sharp cut-off 
\begin{eqnarray}\label{eq:sharp}
  R_{\rm sharp}(p^2)=p^2(1/\theta(p^2-k^2)-1)\,.
\end{eqnarray}
For $R_{\rm sharp}$ it is guaranteed that $k^2$ is the momentum scale
below which $\phi$ modes do not propagate. Inserting \eq{eq:sharp} in
\eq{eq:beispiel} with $Z_\phi=1$, the effective cut-off scale is 
\begin{eqnarray}\label{eq:keffex} 
  k_{\rm eff}(k)=\sqrt{k^2+{V^{(2)}_{k,\rm min}}}\,, 
\end{eqnarray}
where $V^{(2)}_{k,\rm min}$ is the minimal value for $V^{(2)}_{k}$.
Hence, in theories with a mass gap the effective cut-off scale $k_{\rm
  eff}$ does not tend to zero but settles at the physical mass scale
of the theory. In the present example $k^2_{\rm eff}(k=0)=
V^{(2)}_{0,\rm min}$, the minimum of the second derivative of the full
effective potential. Note that the full effective potential is convex and 
hence $V^{(2)}_{0,\rm min}\geq 0$. \smallstep

\subsubsection{Optimisation criterion}\label{sec:janscriterion}
The analysis of the previous two sections allows to put forward a
general optimisation criterion in a closed form:
\begin{subequations}\label{eq:optcriterion}
\begin{eqnarray}\label{eq:opt0}
  \left.D_{R_\bot} \tilde I[\phi,R]\right|_{R=R_{\rm stab}}= 0\,,  
\end{eqnarray}
with 
\begin{eqnarray} \label{eq:setbot}\{R_{\bot}\}=  \left\{R\ {\rm with}\ \|
  G[R]\|^{\ }_{\rm sup}=\0{1}{Z_\phi} k_{\rm eff}^{\dim_G}\right\}\,,
\end{eqnarray} 
\end{subequations} 
where $\tilde I[\phi,R]$ are correlation functions in the given order
of the truncation. The norm $\|.\|_{\rm sup}^{\ }$ and the effective
cut-off $k_{\rm eff}$ have been introduced in \eq{eq:beispiel}. For
the sake of completeness of the definition \eq{eq:optcriterion} we
recall its properties here: $\dim_G$ is the momentum dimension of $G$,
and the effective cut-off $k_{\rm eff}$ is related to the biggest
spectral value of the propagator $k_{\rm eff}^{\dim_G}/Z_\phi $
. The norm in \eq{eq:optcriterion} is
the supremum of the $L_2$ operator norm,
\begin{eqnarray}\label{eq:supnorm}
  \hspace{-.5cm}
  \|G[R]\|^{\ }_{\rm sup}= \sup_{\phi} \{\| G[\phi,R]\|_{L_2}^{\ }\}\,, 
\end{eqnarray} 
see also \eq{eq:normmoment}. If the theory or the truncation scheme
admits a natural expansion point $\phi_0$, the supremum in
\eq{eq:supnorm} might be substituted by evaluating the propagator at
$\phi_0$, e.g.\ a configuration $\phi_0$ for which the minimum of the
effective potential is achieved. \smallstep

As shown in section~\ref{sec:localopt}, the constraint in
\eq{eq:optcriterion} can be rewritten as the constraint of minimal
action of $\Delta S_2$, \eq{eq:dtfinstable}:
\begin{eqnarray}\label{eq:reoptcriterion}
  \left.\left(
      D_{R_\bot}\Delta S_2[\phi,\dot R]
    \right)\, 
    \tilde I[\phi,R]\right|_{R=R_{\rm stab}}=0\,.
\end{eqnarray}
The criterion \eq{eq:optcriterion} is not bound to specific truncation
schemes. The trivial starting point at $R\equiv \infty$ is evaluated
for $k_{\rm eff}(R\equiv\infty)=\infty$ (assuming $d_g<0$), the
end-point at $R\equiv 0$ represents the mass gap of the theory,
$k_{\rm eff}(R\equiv 0)= (\| 1/\Gamma^{(2)}\|^{\ }_{\sup})^{1/\dim_G}$.
The monotone parameter $k_{\rm eff}$ defines the effective cut-off scale and
interpolates between the classical theory at $k_{\rm
  eff}= \infty$ and the full theory at $k_{\rm eff}(0)$. If the theory 
undergoes a phase transition, in particular if it is first
order, the monotony of $k_{\rm eff}(k)$ within truncations is at
stake. If this happens it hints at a truncation scheme that is not
well-adapted.  Nonetheless it can be dealt with in
\eq{eq:optcriterion}, it simply demands a more careful comparison of
regulators at an effective cut-off scale defined by \eq{eq:supnorm}.
Indeed, such pathologies can be avoided if restricting the space of
regulators to those with monotony in $k$, $R(k)\leq R(k')$ for $k<k'$
which entails that regulators implement a true mode (scale) ordering.
There are further secondary regularity constraints, but we do not want
to overburden the criterion \eq{eq:optcriterion} with technicalities.
\smallstep

The general form of the optimisation criterion \eq{eq:optcriterion} is
achieved by substituting $\|G\|_{\rm sup}$ by a general norm $d_{\CO}$
as defined in \eq{eq:norm}. For example, an interesting option can be
found in \cite{Polonyi:2001uc}.  In most cases the norm
\eq{eq:supnorm} applied to $G$ supposedly is the natural choice: the
propagator $G$ is the key input in $\Delta S_2$, any iterative
truncation scheme involves powers of $G$ and hence the importance of
its supremum is enhanced within each iteration step \footnote{First
  investigations within LPA reveal the suggested equivalence of
  different choices for $d_{\CO}$, see also \cite{JDL}.}.  \smallstep

Even in its form \eq{eq:reoptcriterion} the optimisation criterion
\eq{eq:optcriterion} shows some dependence on the correlation function
under investigation. Bearing in mind the discussion about
observable-independent optimisation we apply this idea to
\eq{eq:reoptcriterion}. First of all, $\Delta S_2$ depends on
$\Gamma^{(2)}$ (and possibly higher derivatives of $\Gamma$).  The
functional optimisation implies \eq{eq:optcriterion} for these
correlation functions which maximises the physics content of $\Delta
S_2$. Consequently the derivative in \eq{eq:reoptcriterion} is taken at 
\begin{subequations}\label{eq:indepopt}
\begin{eqnarray}\label{eq:zerorder}
  \left.D_{R_\bot}\Gamma^{(2)}\right|_{R_{\rm stab}}=0\,. 
\end{eqnarray} 
\Eq{eq:zerorder} facilitates the evaluation of \eq{eq:reoptcriterion}
as it only requires the evaluation of derivatives w.r.t.\ the explicit
$R$-dependence.  An optimisation for almost all relevant correlation
functions $\tilde I$ within a given truncation order implies the
vanishing of the operator $D_{R_\bot}\Delta S_2[\phi,\dot R]$ on the
span of these $\tilde I$. Assuming that we can embed this span in a
normed vector space $\CV_{\tilde I}$ we arrive at a
correlator-independent optimisation
\begin{eqnarray}\label{eq:indepoptcrit}
  \left\|
    D_{R_\bot}\Delta S_2[\phi,\dot R]
  \right\|_{R=R_{\rm stab}}=0\,,  
\end{eqnarray}
\end{subequations}
with \eq{eq:zerorder} with the operator norm $\|.\|$ on $\CV_{\tilde
  I}$. The optimisation \eq{eq:indepopt} minimises the action of
$\Delta S_2$ on correlation functions $\tilde I$ within a given
truncation order. The representation \eq{eq:indepopt} allows for a
clear understanding of the result of the optimisation with the example
of the two-point function. \Eq{eq:zerorder} entails that for optimal
regulators $R_{\rm stab}$ the spectrum of $\Gamma^{(2)}$ at the
effective cut-off scale $k_{\rm eff}$ is as close as possible (for the
set of regulators $R_\bot(k_{\rm eff})$ ) to that of the full
two-point function at $k=0$: the physics content of $\Gamma^{(2)}$ is
optimised. It also implies a monotone evolution of the spectral
values of $\Gamma^{(2)}$ for optimal regulators. In case
$\Gamma^{(2)}$ has negative spectral values at the initial scale, e.g.
a non-convex potential, the above investigations lead to one
$k$-independent spectral value, up to RG rescalings. \smallstep

The criterion \eq{eq:optcriterion}, \eq{eq:indepopt} can be rewritten
as a simple criterion on the full propagator and the full vertices.
For its importance and for the sake of simplicity we concentrate on
the standard flow \eq{eq:standflowtIk} with
\begin{eqnarray}\nonumber 
  \Delta S_2 &=& 
  (G\,\dot R\, G)_{\bf bc}\s0{\delta}{\delta \phi_{\bf c}}
  \s0{\delta}{\delta \phi_{\bf b}}\\ 
  &=&\partial_t|_{\Gamma^{(2)}} (G-G_0)_{\bf bc} 
  \s0{\delta}{\delta \phi_{\bf c}}
  \s0{\delta}{\delta \phi_{\bf b}} \,, 
\label{eq:standdeltaS_2}\end{eqnarray}
where $G_0$ is an appropriate $R$-independent normalisation, that
leads to well-defined insertions for correlation functions $\tilde I$
if applying $(G-G_0) \s0{\delta^2}{\delta \phi^2}$. In the presence of
a mass gap a possible choice is e.g.\ $G_0=G[\phi,R=0]$. The partial
$t$-derivative at fixed $\Gamma^{(2)}$ commutes with $D_{R_\bot}$ at
$R_{\rm stab}$. There, $D_{R_\bot}=D_{R_\bot}|_{\Gamma^{(2)}}$. Now we
use that the second functional derivative ${\delta^2}/{\delta\phi^{\bf
    a}\delta\phi^{\bf b}}$ does not vanish on almost all $\tilde I$. 
Therefore a vanishing norm \eq{eq:indepoptcrit} implies 
\begin{eqnarray}\label{eq:indepoptcrit2}
  \left\|\partial_t|_{\Gamma^{(2)}}\, 
    D_{R_\bot}|_{\Gamma^{(2)}} (G-G_0)
  \right\|_{R=R_{\rm stab}}=0\,. 
\end{eqnarray}
The norm in \eq{eq:indepoptcrit2} derives from the operator norm on
$V_{\tilde I}$, and hence is related to the truncation scheme. A
solution of $\left\| D_{R_\bot}|_{\Gamma^{(2)}}
  (G-G_0)\right\|_{R=R_{\rm stab}}=0$ for all $k$ implies a solution
of \eq{eq:indepoptcrit2}. Consequently we search for extrema on the
spectrum of the positive operator $G$. Now we use that the positive
operator $G$ vanishes identically for $R=\infty$ and tends towards the
full propagator $G[\phi,R=0]$ with positive spectrum at vanishing
regulator. Then with \eq{eq:indepoptcrit} and \eq{eq:indepoptcrit2} we
conclude that optimal flows maximise $G$ at a given $k_{\rm eff}$ for
all spectral values, with the constraint that $-\partial_t G\geq 0$ is
a positive operator. The latter constraint guarantees that the
maximisation is globally valid for all $k$. We conclude that optimal
flows are those where $G[\phi,R]$ is already as close as possible to
the full propagator for a given cut-off scale $k_{\rm eff}$. This
criterion can be cast into the form
\begin{subequations}\label{eq:functopt}
\begin{eqnarray}\label{eq:indepoptcrit3}
  \hspace{-.3cm}d_{\theta_\lambda(G)}[R_{\rm stab},0]
  =\min_{R_\bot} d_{\theta_\lambda(G)}[R_\bot,0]\,, 
\end{eqnarray}
for all $\lambda\in \R^+$ with $\{R_\bot\}$ as defined in
\eq{eq:optcriterion}, and $\theta_\lambda$ is defined via its action
on eigenvectors $|\psi_{\lambda_G}\rangle$ of $G$
\begin{eqnarray}\label{eq:theta_l}
  \hspace{-.5cm} \theta_\lambda(G)|\psi_{\lambda_G}\rangle 
  =\Bigl[\lambda+ 
    (\lambda_G-\lambda)\, \theta(\lambda-\lambda_G)\Bigr] 
  |\psi_{\lambda_G}\rangle \,,  
\end{eqnarray}
with Heaviside step function $\theta(x)$\footnote{$\theta_\lambda$
  is required to be a bounded operator. Hence for general norms used
  in $d[R,R']$ \eq{eq:theta_l} has to be modified, see e.g.\
  section~\ref{sec:beyond}.}. The operator $\theta_\lambda$ used in
\eq{eq:indepoptcrit3} resolves the full spectral information of $G$. 
The criterion \eq{eq:indepoptcrit3} entails the constraint that
$G[\phi,R_{\rm stab}]$ takes the closest spectral values (according to
the norm) to the full propagator $G[\phi,0]$ for all $R\in
\{R_\bot\}$, starting from the boundary condition $G[\phi,\infty]=0$,
or alternatively at $G[\phi,0]$.  This implies a minimisation of the
flow, as well as monotony of the spectral values of $G$ in $k$:
$G[\phi,0]\geq G[\phi,R]$. These considerations enable us to
reformulate \eq{eq:indepoptcrit3} without relying on the full
propagator $G[\phi,0]$.  We are led to
\begin{eqnarray}\nonumber 
\|\theta_\lambda(\Gamma^{(2)}[R_{\rm stab}]+R_{\rm stab})\|=\min_{R_\bot}
\|\theta_\lambda(\Gamma^{(2)}[R_{\bot}]+R_{\bot})\|\\ 
\label{eq:indepoptcrit4}\end{eqnarray}
\end{subequations}
for all $\lambda\in\R^+$. We remark that in deducing
\eq{eq:indepoptcrit4} from \eq{eq:indepoptcrit3} we have again used
$-\partial_t G\geq 0$ and $\Gamma^{(2)}[R=0]\geq 0$. If the distance
$d$ is defined with the $L_2$-norm in the given order of the
truncation, \eq{eq:indepoptcrit4} is also conveniently written as 
\begin{eqnarray}\label{eq:optL2}
  \hspace{-.3cm}d_{\theta_\lambda(G)}
  [R_{\rm stab},\infty]=\max_{R_\bot} 
  d_{\theta_\lambda(G)}[R_\bot,\infty]\,.
\end{eqnarray}
Note that in general \eq{eq:indepoptcrit4} can be written as
\eq{eq:optL2} and some supplementary constraints depending on the norm
used in \eq{eq:indepoptcrit4}, see e.g.\ section~\ref{sec:beyond}.
For each norm these supplementary constraints are straightforwardly
derived from \eq{eq:indepoptcrit3}. 

\Eq{eq:functopt} is a simple optimisation procedure independent of the
correlation functions $\tilde I$ under investigation. It already works
without computations of full flow trajectories. In its form the
criterion \eq{eq:optcriterion} has already been successfully applied
to Landau gauge QCD \cite{Pawlowski:2003hq,Pawlowski:2004ip}, see also
section~\ref{sec:fixed}. We emphasise again that the appropriate norm
relates to the truncation used.  The above analysis extends to general
regulators. There, one also has to take into account the evolution of
higher vertices $\Gamma^{(n)}$.  Their properties under
$R_\bot$-variations at $R_{\rm stab}$ derive from \eq{eq:zerorder} by
taking field-derivatives.  Spectral considerations are more involved
but it can be shown that an optimisation for general regulators
implies \eq{eq:functopt}. \smallstep

We close the section with some comments concerning the generality of
\eq{eq:optcriterion}, the existence of solutions, and its connection
to the criterion \eq{eq:optbyprop} \footnote{For its connection to the
  PMS condition \eq{eq:Rindep||} we refer the reader to the discussion
  below \eq{eq:dtfinstableR}.}:\smallstep

the definition of the set ${R_\bot}$ in \eq{eq:setbot} guarantees the
existence of $R_{\rm stab}$ for a general expansion scheme: within any
given truncation scheme the set of $\{R_\bot\}$ is bounded by 
possibly smooth modifications of the sharp cut-off and the optimal
cut-off \eq{eq:litopt} as functions on the spectrum of $\Gamma^{(2)}$
and for spectral values $\lambda(\Gamma^{(2)})\leq Z_\phi k_{\rm
  eff}^{-d_G}$. Together with positivity and monotony of the
regulators $R$ this proves the existence of a stable solution of
\eq{eq:optcriterion}, if neglecting the $R_\bot$-variation of
$\Gamma^{(2)}_k$. Indeed such a procedure defines a further truncation
scheme on top of that at hand.  Note also that possibly one has to
introduce a $\lambda$-ordering: we search for a solution to
\eq{eq:indepoptcrit3}, \eq{eq:indepoptcrit4} for a given $\lambda$ on
the sub-space of solutions to \eq{eq:indepoptcrit3},
\eq{eq:indepoptcrit4} for $\lambda'<\lambda$.  \smallstep
    
The argument above fails for generalisations of regulator functions
where the demand of positivity and monotony of the regulator are
dropped.  Still, for reasonable choices the set ${R_\bot}$ sweeps out
basically the area bounded by, possibly smooth modification, of the
sharp cut-off and the optimal cut-off \eq{eq:litopt}. However, it is
not guaranteed anymore that the boundary curves are themselves in
${R_\bot}$.  Therefore, a strict extremisation for all momenta
(spectral values) as demanded in \eq{eq:optcriterion} might fail for
generalisations of \eq{eq:optcriterion}. More details will be provided
elsewhere.\smallstep
 
Both criteria, \eq{eq:optbyprop} and \eq{eq:optcriterion}, are based
on the same key idea of global stability. In \eq{eq:optbyprop} the set
of regulators $\{ R_\bot\}$ is defined by normalising the regulators
at some momentum.  Then the inverse gap $\|G[R,\phi_0]\|^{\
}_{L_2}$ of the full propagator is minimised. In \eq{eq:optcriterion}
the set of regulators $\{ R_\bot\}$ is defined as those with the same
maximal spectral value (inverse gap) $\|G[R]\|^{\ }_{\rm sup}$ and the
action of the flow operator $\Delta S_2$ is minimised.  With
\eq{eq:optbyprop} one is comparing regulators with different effective
cut-off scales but, roughly speaking, close physics content.  Then,
optimal regulators are those where this physics content is achieved
for the biggest effective cut-off scale. In turn, with
\eq{eq:optcriterion} we compare regulators leading to the same
effective cut-off scale and single out those that lead to correlation
functions as close as possible to those in the full theory. \step

\setcounter{equation}{0}
\section{Applications to functional methods}\label{sec:applications}

In this chapter we discuss immediate structural consequences of the
setting developed so far. First of all this concerns the interrelation
of functional methods like the general flows studied here,
Dyson-Schwinger equations
\cite{vonSmekal:1997is1,vonSmekal:1997is2,Alkofer:2000wg,Lerche:2002ep,Fischer:2002hn,Alkofer:2003jj,Fischer:2003rp,Roberts:2000aa,Maas:2005hs},
stochastic quantisation
\cite{Zwanziger:2003cf,Zwanziger:2002ia,Zwanziger:2001kw}, and the use
of $N$PI effective actions
\cite{Luttinger:1960ua,Baym:1962sx,Cornwall:1974vz,Verschelde:2000dz,vanHees:2001ik,vanHees:2002bv,Blaizot:2003br,Blaizot:2003an,Berges:2004hn,Berges:2005hc,Cooper:2004rs,Cooper:2005vw,Buchbinder:ha,Mottola:2003vx,Arrizabalaga:2002hn,Carrington:2003ut,Berges:2004pu}.
All these methods have met impressive success in the last decade, in
particular if it comes to physics where a perturbative treatment
inherently fails.  Here, we discuss structural similarities as well as
functional relations between these approaches that open a path towards
a combined use as well as non-trivial consistency checks of respective
results. We also highlight the important aspect of practical
renormalisation schemes that can be derived from general flows for
either DS equations or $N$PI methods. However, given the scope of the
present work we only outline the relevant points, leaving a more
detailed analysis to future work. \smallstep

\subsection{Functional RG and DS equations}\label{sec:DSEs}

\subsubsection{DSEs as integrated flows}\label{sec:intDSE} Formally
Dyson-Schwinger equations \eq{eq:DS} are integrated flows.  They
constitute finite functional relations between renormalised Green
functions as well as bare vertices. They have been successfully used
for the description of the infrared sector of QCD formulated in Landau
gauge, initiated in \cite{vonSmekal:1997is1,vonSmekal:1997is2}, for a
review see \cite{Alkofer:2000wg}. This approach is also tightly linked
to a similar analysis in stochastic quantisation
\cite{Zwanziger:2003cf,Zwanziger:2002ia,Zwanziger:2001kw}.

More recently, these investigations have been extended to finite
temperature QCD, e.g.\ \cite{Maas:2005hs} and the review
\cite{Roberts:2000aa}. The formal finiteness of the DS equations is
more intricate if solving them within truncations
\cite{vonSmekal:1997is1,vonSmekal:1997is2,Alkofer:2000wg,Lerche:2002ep,Fischer:2002hn,Alkofer:2003jj,Fischer:2003rp,Roberts:2000aa,Maas:2005hs}.
Here, we discuss Dyson-Schwinger equations and their flow in the
presence of a standard regulator coupled to the fundamental fields.
This allows us to construct a general consistent BPHZ-type
renormalisation of DS equations from integrated flows being valid
beyond perturbation theory. The extension of the results to the
general setting is straightforward. \smallstep

Recall the DS operator $\hat I$ given in \eq{eq:DS} with
$\hat\phi=\hat\varphi$, the source $J$ coupled to the fundamental
fields: $\hat I_{\rm DSE} = J-\0{\delta S}{\delta \hat\phi}$.
Inserting this into \eq{eq:Ik} leads to
\begin{subequations}\label{eq:DSk}
\begin{eqnarray}\label{eq:DSIk}
  \tilde I_{\rm DSE}^a[\phi,R]=\Gamma{}^{,a}[\phi,R]
  -\langle S^{, a}[\hat\phi]
  \rangle\equiv 0
\end{eqnarray}
with 
\begin{eqnarray}\label{eq:hatDSIk}
  \hat I_{\rm DSE}^a[J,\s0{\delta}{\delta J},R] 
  =J^a -\0{\delta S}{\delta \hat\varphi_a}- 
  2 R^{ab}\0{\delta}{\delta J^b}\,.
\end{eqnarray} 
\end{subequations}
Note that $\langle S^{, a}[\hat\varphi]\rangle$ in \eq{eq:DSIk} has to
be read as a function of $\phi^a$. The flow of $\tilde I_{\rm DSE}$ is
given by \eq{eq:flowtIk} and reads
\begin{eqnarray}\label{eq:flowDSk}
  \left(\partial_t+\Delta S_2[\s0{\delta}{\delta \phi},\dot R]\right) 
  \tilde I_{\rm DSE}=0\,.
\end{eqnarray} 
The first term in the DSE \eq{eq:DSk} $\Gamma_k{}^{,b}$ already
satisfies \eq{eq:flowDSk}, see \eq{eq:flowofdGk}.  This leaves us with
the separate flow
\begin{eqnarray}\label{eq:flowDSkpartial}
  \left(\partial_t+\Delta S_2[\s0{\delta}{\delta \phi},\dot R]\right) 
  \langle S^{, a}\rangle=0\,. 
\end{eqnarray} 
\Eq{eq:flowDSkpartial} also follows directly from considering $\hat I=
S^{, a}[\s0{\delta}{\delta J}]$. By construction the corresponding
correlation function $\tilde I$ satisfies the flow equation
\eq{eq:flowtIk} and is given by $\tilde I[\phi,R] =\langle S^{,
  a}\rangle$.  From the above identities we also relate
$t$-derivatives of $\Gamma_k{}^{,a}$ and $\langle S^{, a}\rangle$,
i.e.\
\begin{subequations}\label{eq:flowDSrelate}
\begin{eqnarray}\label{eq:flowDSrelate1}
  \partial_t \Gamma_k{}^{,a}+
  \Delta S_2[\s0{\delta}{\delta \phi},\dot R]\langle S^{,a}\rangle=0\,, 
\end{eqnarray}
as well as 
 \begin{eqnarray}\label{eq:flowDSrelate2}
   \partial_t\langle S^{,a}\rangle+
   \Delta S_2[\s0{\delta}{\delta \phi},\dot R]\Gamma_k{}^{,a}=0\,.
\end{eqnarray}
\end{subequations}
\Eq{eq:flowDSrelate} highlights the aspect of the functional RG as a
differential DSE. The use of the above identities
\eq{eq:DSk},\eq{eq:flowDSk} and \eq{eq:flowDSkpartial} is twofold.
Firstly they allow us to relate DSEs and flow equations in similar
truncations, hence providing non-trivial consistency checks for both
approaches. Secondly they open a path towards a combined use of
functional RGs and DSEs dwelling on the advantageous features of both.
For example, an infrared analysis within both functional approaches
usually provides a set of possible solutions whose intersection is
possibly unique. In QCD this can be directly achieved by a fixed point
analysis of \eq{eq:flowDSrelate1} along the lines in
\cite{Pawlowski:2003hq,Pawlowski:2004ip}.  \smallstep

\subsubsection{Renormalisation}\label{sec:DSErenorm}
Furthermore the flow equation in its integrated form can be used to
set up an explicit renormalisation procedure within general truncation
schemes. Such a renormalisation is not necessarily multiplicative but
generalises the BPHZ renormalisation of perturbation theory to general
expansions. As it relies on a functional equation for the effective
action its consistency is guaranteed by construction. Hence it is
possible to derive consistent subtraction schemes for Dyson-Schwinger
equations from the integrated flow in a given truncation. \smallstep

We illustrate the above statements within the standard flow
\eq{eq:standflowWG} for the effective action.  Assume that we have
solved the theory within the $i$th order of a given general truncation
scheme, leading to $\Gamma^{(i)}_k$. Generally the flow can be written
as
\begin{eqnarray}\label{eq:nextstep}
  \partial_t\Gamma=\dot R^{ab} G_{ab}= -\s012 \partial_t (\ln G)_{aa} 
  -\s012 \dot \Gamma_k^{,ab} G_{ab}\,.
\end{eqnarray} 
In its integrated form this leads to 
\begin{eqnarray}\label{eq:int}
  \Gamma_k=\Gamma_\Lambda -\s012 \left.(\ln G)_{aa}\right|^k_\Lambda 
  -\s012 \int_\Lambda^k dt\,\dot \Gamma_k^{,ab} G_{ab}\,.
\end{eqnarray} 
The integrated flow \eq{eq:int} represents an integral equation for
the effective action $\Gamma_{k}$ with the boundary condition
$\Gamma_\Lambda$. Note that its solution for a given $k$ requires its
solution for $k'\in [k,\Lambda]$. As such it constitutes a
Dyson-Schwinger equation.  It provides an explicit (re)normalisation
procedure involving two different aspects. Firstly the choice of a
finite boundary condition $\Gamma_\Lambda$ implicitly renormalises the
theory: it ensures finiteness. The renormalisation conditions for the
full effective action, i.e. fixing the relevant operators (of
$\Gamma_0$) at some renormalisation scale $\mu$ translate to similar
conditions for $\Gamma_{k}$ for all $k$. In particular its choice at
$k'=0$ relates to an appropriate normalisation at $k=\Lambda$.  As can
be seen from the representation of the integrated flow in \eq{eq:int}
the renormalisation is done in a BPHZ-type way with subtractions
$-\s012 \ln G(k'=\Lambda)+ (\Gamma_\Lambda-S)$, the $t$-integral also
comprises some sub-leading subtractions. \smallstep

With \eq{eq:int} we have resolved the notorious consistency problem
for explicit renormalisation procedures within Dyson-Schwinger
equations.  Practically it can be solved within an iteration of
$\Gamma_{k}$ about some zeroth iteration step $\Gamma_{k,0}$ for
$k\in\{\Lambda,0\}$, e.g.\ $\Gamma_{k,0}=S_{\rm cl}$, the classical
action. This works for paths $R(k)$, for which the initial condition
$\Gamma_\Lambda$ is sufficiently close to the classical action, an
example being regulators $R$ implementing a momentum regularisation
with $\Lambda$ setting a high momentum scale. \smallstep

An interesting option are non-trivial $\Gamma_{k,0}$ that already
incorporate some non-trivial physics content of the theory under
investigation. If the zeroth iteration step is already close to the
full solution the numerical effort is minimised. Accordingly such a
procedure benefits from any information already collected by other
means about the physics content. In comparison to the standard
(numerical) solution of DS-equations involving momentum integrations
one has to perform an additional $t$-integration. In general this is
bound to increase the numerical costs. However, this additional
integral comes with the benefit that now the integrand is localised in
momenta and $t$ which stabilises the numerics. Indeed, the above ideas
have been used for resolving the infrared sector of QCD within the
Landau gauge thus furthering the evidence for the
Kugo-Ojima/Gribov-Zwanziger confinement scenario in this gauge
\cite{Pawlowski:2003hq,Pawlowski:2004ip}, and providing a
general consistent renormalisation procedure for related DS-studies
\cite{Alkofer:2000wg,Lerche:2002ep}. This aspect will
be further discussed in section~\ref{sec:optimalatwork}. We also
remark that the present analysis can be extended to the stochastic 
quantisation
\cite{Zwanziger:2001kw,Zwanziger:2002ia,Zwanziger:2003cf}. There it
helps that we do not rely on an explicit path integral
representation. This shall be detailed elsewhere.\smallstep 

Still the question arises whether \eq{eq:int} can be used more
directly for setting up a renormalisation procedure for functional
equations in the full theory at $k=0$, solved iteratively within a
given general truncation scheme
\begin{eqnarray}\label{eq:reptrunc}
  \Gamma_k^{(i)}[\phi,R]=\Gamma_k^{(i-1)}[\phi,R]
  +\Delta^{(i)} \Gamma_k[\phi,R]
  \,,
\end{eqnarray} 
as introduced in \eq{eq:truncexpansion} for general $\tilde I_k$.
Assume we have managed to construct regulators $R$ that lead to a
suppression of modes in the path integral related to orders $i> i_k$
of our truncation scheme.  As an example we take the derivative
expansion. Here we can use regulators that suppress at $k=\Lambda$ all
momentum-dependent fields, $i_\Lambda=0$.  By decreasing $k$ we add
more and more derivatives, $i_k\to\infty$ with $k\to 0$, either
continuously switching on their effects or adding more and more
derivatives in discrete steps.  \smallstep

If $R$ implements the truncation in discrete steps the flow only is
non-zero at the discrete set of $k_i$. Integrating the flow from
$k_i<k_1<k_{i+1}$ and $k_{i+1}<k_2<k_{i+2}$ we arrive at
\begin{eqnarray}\nonumber 
  \Gamma^{(i+1)}&=& 
  \Gamma^{(i)} -\s012 \left((\ln G)^{(i+1)}_{aa}-
    (\ln G)^{(i)}_{aa}\right)\\ 
  && \qquad -\s012 \int_{k_1}^{k_{2}} dt\,  \dot \Gamma^{,ab} G_{ab}\,. 
\label{eq:inti}\end{eqnarray} 
\Eq{eq:inti} recursively implements the renormalisation at a given
order $i+1$ of the truncations by subtraction of appropriate terms of
the order $i$. Naively the integral in \eq{eq:inti} can be performed
as $\dot \Gamma^{,ab}$ only is non-zero at $k_{i+1}$.  However, this
has to be done carefully for similar reasons to those that do not allow for
a naive integration of sharp-cut-off flows: at $k_{i+1}$, the flow
$\dot \Gamma^{,ab}$ is singular and $G$ jumps. Nonetheless, as in the
case of the sharp cut-off \eq{eq:inti} can be easily integrated within
explicit iteration schemes. For example, perturbation theory within
BPHZ-renormalisation can be reproduced with \eq{eq:inti} but it
extends to general schemes as well as general functional relations and
correlation functions $\tilde I$ of the theory that require explicit
renormalisation if it comes to truncations. \smallstep

\subsection{Composite operators and $N$PI flows}
\label{sec:compositeflows}
The analysis of the last section extends naturally to flows in the
presence of composite operators, in particular to flows of $N$PI
effective actions \cite{Luttinger:1960ua,
  Baym:1962sx,Cornwall:1974vz}. Flows with the coupling to composite
operators have been considered in e.g.\
\cite{Polonyi:2000fv,Polonyi:2001se,Polonyi:2004pp,Wetterich:2002ky,Schutz:2004rn,Salmhofer1,Dupuis:2005ij}. Flows for the 2PI effective 
action have been studied in 
\cite{Polonyi:2004pp,Wetterich:2002ky,Dupuis:2005ij}. 
\smallstep

In the presence of sources for composite operators the renormalisation
of these operators has to be taken into account. In particular, the
construction of practical consistent renormalisation schemes within
truncations poses a challenge, see e.g.\
\cite{Verschelde:2000dz,vanHees:2001ik,vanHees:2002bv,Blaizot:2003br,Blaizot:2003an,Berges:2004hn,Berges:2005hc,Cooper:2004rs,Cooper:2005vw}.
Such a renormalisation has to respect the symmetry and symmetry
breaking pattern of the theory under investigation. We discuss the use
of general flows for the construction of consistent subtraction
schemes in general truncations by extending the renormalisation ideas of
the last section. We also discuss the direct relation between flows in
the presence of composite operators and $N$PI effective actions,
relying on the interpretation of the regulator $R$ as a source for a
composite operator. \smallstep

\subsubsection{Linear flows}\label{sec:linearflows} The structure of
the flows \eq{eq:flowIk},\eq{eq:flowtIk} always allows us to reduce
the order of derivatives in $\Delta S_k$ at the expense of introducing
further tensorial currents. In general we have
\begin{eqnarray}\nonumber 
  &&\hspace{-.4cm}\left(\s0{\delta}{\delta J^{{{\bf a}_1
          \cdots {\bf a}_n} }}
    \s0{\delta}{\delta J^{{{\bf a}'_1}\cdots {{\bf a}'_m}}}\right)^i
  \Bigl[e^{J^{{\bf 
        b}_1\cdots {\bf b}_n}\hat\phi_{{\bf b}_1\cdots {\bf b}_n}
    +J^{{\bf b}_1
      \cdots {\bf b}_{m}}\hat\phi_{{\bf b}_1\cdots {\bf b}_m}}\\\nonumber 
  &&  
  \hspace{1.5cm} \times e^{
    J^{{\bf b}_1\cdots {\bf b}_{n+m}}\hat\phi_{{\bf b}_1\cdots {\bf b}_n}
    \hat\phi_{{\bf b}_{n+1}\cdots  
      {\bf b}_{m+n}}}\Bigr] 
  \\\nonumber  
  & &\hspace{-.6cm}=  
  \left(\s0{\delta}{\delta J^{{\bf a}_1\cdots {\bf a}_{n+m}}}\right)^i 
  \Bigl[e^{J^{{\bf b}_1\cdots {\bf b}_n}\hat\phi_{{\bf b}_1\cdots {\bf b}_n}
    +J^{{\bf b}_1\cdots {\bf b}_{m}}\hat\phi_{{\bf b}_1\cdots {\bf b}_m}}\\ 
  &&  
  \hspace{1.5cm}\times e^{
    J^{{\bf b}_1\cdots {\bf b}_{n+m}}\hat
    \phi_{{\bf b}_1\cdots {\bf b}_n}\hat\phi_{{\bf b}_{n+1}\cdots  
      {\bf b}_{m+n}}}
  \Bigr] \,, 
  \label{eq:reduce} \end{eqnarray} 
with ${\bf a}_{n+j}={\bf a}'_j$. \Eq{eq:reduce} is valid 
for all $i\in \N$. We also could have substituted only a part 
of the derivatives, obviously the relation is not unique. 
In case the source term $ J^{{\bf b}_1\cdots
  {\bf b}_{n+m}}\hat\phi_{{\bf b}_1\cdots {\bf b}_n}\hat
\phi_{{\bf b}_{n+1}\cdots {\bf b}_{m+n}}$ was not present in
the Schwinger functional $W[J]$ it has to be added.  Note that the
derivatives w.r.t.\ $t$ are taken at fixed arguments $J$ and $\hat\phi$
respectively.  Hence the reduction to lower powers of derivatives is
accompanied by holding the corresponding Green functions fixed. With
\eq{eq:reduce} a part of the regulator term \eq{eq:reg} with $n$th
order derivatives, is reduced to order $n-m+1$ by adding a further 
source term to $W[J]$ 
\begin{eqnarray}\label{eq:addsource} 
  J^{\bf a}\hat\phi_{\bf a}\to J^{\bf a}\hat\phi_{\bf a}+
  J^{{\bf a}_1\cdots  {\bf a}_m} \hat\phi_{{\bf a}_1}
  \cdots \hat\phi_{{\bf a}_m}=
  J^{\bf a'}\hat\phi_{\bf a'}\,,
\end{eqnarray} 
where
\begin{eqnarray}
  \gamma^{\bf a' b'}=(\gamma\oplus (\otimes\gamma)^m)^{\bf a' b'} \,,
\end{eqnarray}
with enlarged multi-indices ${\bf a'}={\bf a},{\bf a_1}\cdots {\bf
  a}_m$ and $\gamma=(\gamma^{\bf ab})$. \Eq{eq:addsource} implies
$\hat\phi_{{\bf a}_1\cdots {\bf a}_m}=\hat\phi_{{\bf a}_1}\cdots
\hat\phi_{{\bf a}_m}$. With \eq{eq:addsource} we are led to
\begin{eqnarray}\nonumber 
  &&\hspace{-1cm}\left(R^{{\bf a}_{1}\cdots {\bf a}_{n}} 
    \0{\delta}{\delta J^{{
          \bf a}_{1}}}\cdots  \0{\delta}{\delta J^{{
          \bf a}_{n}}}\right)^i e^{J^{{\bf a}'}\hat\phi_{{\bf a}'}}\\ 
  &&\hspace{-.5cm}= 
  \left(  R^{{\bf a'}_{1}\cdots {\bf a'}_{n-m+1}}\0{\delta}{\delta J^{{
          \bf a'}_{1}}}\cdots  \0{\delta}{\delta J^{{
          \bf a'}_{n-m+1}}}\right)^i e^{J^{\bf a'}\hat\phi_{\bf a'}}\,, 
  \label{eq:regred} \end{eqnarray} 
with $R^{{\bf a}'_{1}\cdots {\bf
    a}'_{n}}\hat\phi_{{\bf a}'_1}\cdots \hat\phi_{{\bf a}'_{n-m}}=
R^{{\bf a}_{1}\cdots {\bf a}_{n}}\hat\phi_{{\bf a}_1}\cdots
\hat\phi_{{\bf a}_n}$.
The above relation is not unique, and we could have further reduced
the order of derivatives by identifying additional products
$\hat\phi_{{\bf a}_1\cdots {\bf a}_m}=\hat\phi_{{\bf a}_1}
\cdots\hat\phi_{{\bf a}_m}$ for $n-m\geq m$. By recursively using
\eq{eq:addsource},\eq{eq:regred} with general $m$ we can substitute
$\Delta S$ by an expression with only quadratic derivative terms, and the 
flow reduces to the standard form of the flow equation \eq{eq:standflowtIk}. 
Reducing $\Delta S$ one step further we arrive at first order
derivatives w.r.t.\ $J$ and \eq{eq:flowtIk} boils down to
\begin{eqnarray}\label{eq:0} 
  \partial_t\tilde I_k[\phi]=0\,.
\end{eqnarray}
It seems that \eq{eq:0} is rather trivial but it should be read as a
fixed point equation for the flow. When evaluating $\tilde I_k^{\bf
  a}=\gamma^{{\bf a }}{}_{\bf b} (J^{\bf b}-R^{{\bf
    b}})=\Gamma_k{}^{{,\bf a}}$ resulting from $\hat I_k^{\bf
  a}=\gamma^{{\bf a}}{}_{\bf b} J^{\bf b}$ the flow \eq{eq:0} reads
\begin{eqnarray}\label{eq:linflow} 
  \partial_t \Gamma_k{}^{,\bf a}[\phi]=\dot R^{\bf a}\,,
\end{eqnarray} 
where the partial $t$-derivatives is taken at fixed fields $\phi_{\bf
  a}$.  \Eq{eq:linflow} yields upon integration
\begin{eqnarray}\label{eq:standard} 
  \partial_t \Gamma_k[\phi] = 
  \dot R^{{\bf a}}  \phi_{{\bf a}} \,,
\end{eqnarray}
which also can be read off from \eq{eq:flowGkexplicit}. If
$R^{a_1\cdots a_n} =0$ for $n\neq 2$, \eq{eq:standard} resembles the
standard flow equation with $G\to \phi_{a_1 {a}_2}$, in particular for
$\hat\phi=\hat\varphi$.  However, even for general $n$ its integration
is trivial: we exploit that for $k=0$ the regulator vanishes, $R=0$
and get
\begin{eqnarray}\label{eq:trivint} 
  \Gamma_k[\phi]=\Gamma_0[\phi] +R^{{\bf a} } \phi_{{\bf a}}
  =\Gamma[\phi]+\Delta S_k[\phi]\,.
\end{eqnarray} 
\Eq{eq:trivint} can directly be obtained by evaluating the Legendre
transformation \eq{eq:Gammak} for the present scenario. For regulator
terms linear in $\phi$, $\Delta S_k[\phi]= R^{{\bf a}} \phi_{{\bf
    a}}$, there is a simple relation between the Schwinger functional
of the full theory and that of the regularised theory: $W_k[J]=W_0[J-
R]$. Moreover $\Delta S'_k[\phi]=0$. With these observations we can
rewrite \eq{eq:Gammak} for linear $\Delta S_k$ as
\begin{eqnarray}\nonumber 
  \Gamma_k&=& \sup_{\mbox{\tiny $ J$}} \bigl(J^{{\bf a}} \phi_{{\bf a}} 
  -W[J-R]\bigr)\\ \nonumber 
  &=&\sup_{\mbox{\tiny $ J$}} 
  \bigl((J-R)^{{\bf a}} \phi_{{\bf a}} -W[J-R]\bigr)+\Delta S_k\\ 
  &=&\Gamma+\Delta S_k\,. 
\label{eq:transform1}\end{eqnarray} 
In \eq{eq:transform1} we have used that the supremum over the space of
functions $J$ is the same as that over the space of functions $J-R$.
Strictly speaking, the last equality in \eq{eq:transform1} is only
valid for the subset of regulators $R$ that can be absorbed in
currents $J$. \smallstep

From the above definitions and the flow \eq{eq:standard} we can step
by step resolve the composite operators $\phi^{\bf a}$ by using the
related equations of motion. Here we show how such a procedure can be
used to finally recover the regularised effective action
$\Gamma_k[\phi_a]$ in \eq{eq:Gammak} and the general flows
\eq{eq:flowtIk}. The equations of motion for $\phi_{{a}_1\cdots
  a_{n_i}}$ for $n_i\geq 2$ read
\begin{eqnarray}\label{eq:eom}
  \0{\delta \Gamma_k[\phi]}{\delta \phi_{{a_1\cdots a_{n_i}}}}=0\,,
  \qquad \forall n_i\geq 2\, . 
\end{eqnarray} 
Using the solution $\bar\phi(\phi_a)=(\phi_a,\bar\phi_{a_1 a_2},...,
\bar\phi_{a_1\cdots a_{n_N}})$ of \eq{eq:eom} in \eq{eq:trivint}, we
end up with the effective action \eq{eq:Gammak}. As $\Delta S'_k=0$
for linear regulators we have
\begin{eqnarray}\label{eq:rel}
  \Gamma_k[\phi_a]=\Gamma_k[\bar\phi]-\Delta S_k[\hat\phi(\phi_a)],  
\end{eqnarray}
where $\Delta S_k[\hat\phi(\phi_a)]= \sum_{i} R^{a_1\cdots a_{n_i}}
\hat\phi_{a_1\cdots a_{n_i}}[\phi_a]$. Due to the linearity of the
$t$-derivative the flow \eq{eq:standard} holds true also for the
effective action $\Gamma_k[\phi_a]$. This statement reads more
explicitly 
\begin{eqnarray}\nonumber 
  \partial_t\Gamma_k[\phi_a]&=&\partial_t|_{\bar\phi}\Gamma_k[\bar\phi]
  +\Gamma_k^{,\bf a}[\bar\phi]\, \partial_t 
  \bar\phi_{\bf a}[\phi_a]\\ 
  &=& \partial_t|_{\bar\phi}
  \Gamma_k[\bar\phi]\,.
\label{eq:NPIto1PI}\end{eqnarray} 
The second term on the rhs of the first line in \eq{eq:NPIto1PI}
vanishes due to the equations of motion \eq{eq:eom} for $n_i\geq 2$
and due to $\partial_t \bar\phi_a[\phi_a]=0$ for the fundamental field
$\bar\phi_a:=\phi_a$, that is not a solution to the related equations
of motion but a general field.  Hence the flow equation for the 1PI
effective action reads
\begin{eqnarray}\label{eq:dt1PIfromNPI} 
  \partial_t\Gamma_k[\phi_a]=\dot R^{\bf a} \bar\phi_{\bf a}[\phi_a]\,.
\end{eqnarray} 
The equations of motion \eq{eq:eom} relate the fields $\bar\phi_{\bf
  a}[\phi_a]$ to a combination of Green functions
\begin{eqnarray}\label{eq:phihatphi}
  \bar\phi_{\bf a}[\phi_a]=\langle \hat\phi_{{\bf a}}[\hat\phi_a]
  \rangle_{J^{\bf a}=(J^a,0)}\,. 
\end{eqnarray} 
The relations \eq{eq:phihatphi} can be written in terms of functional
$\phi$-derivatives as
\begin{eqnarray}\label{eq:resolveNPI} 
  \bar\phi_{\bf a}=
\left({\hat\phi_{{\bf a}}}[G_{ab}\s0{\delta}{\delta\phi_b}+\phi_a
    ]\right) \,. 
\end{eqnarray}
As an example we use \eq{eq:resolveNPI} for the two-point function
$\hat\phi_{a_1 a_2}= \hat\phi_{a_1}\hat\phi_{a_2}$ and $(\phi_{\bf
  a})=(\phi_a,\phi_{a_1 a_2} )$. It follows
\begin{eqnarray}\nonumber 
  \bar\phi&=& \left(\phi_a\,,\, 
    \left((G\s0{\delta}{\delta\phi}+\phi)_{a_1}  
      (G\s0{\delta}{\delta\phi}+\phi)_{a_2}\right)\,\right) \\ 
  &=& \left(\phi_a\,,\, G_{a_1 a_2}+\phi_{a_1}\phi_{a_2}\right)\,. 
  \label{eq:2point} \end{eqnarray}
Inserting \eq{eq:resolveNPI} into the flow \eq{eq:dt1PIfromNPI} we
recover the flow \eq{eq:flowGk}. The relation \eq{eq:resolveNPI} also
leads to the general flows \eq{eq:flowtIk} starting at the
trivial flow in \eq{eq:0}, $\partial_t\tilde I_k=0$. The flow for
$\tilde I_k[\phi_a]=\tilde I_k[\bar\phi(\phi_a)]$ reads
\begin{eqnarray}\label{eq:tIkNPIto1PI}
  \partial_t \tilde I_k[\phi_a]-\tilde I_k^{,\bf a}[\bar\phi]
  \,\partial_t  \bar\phi_{\bf a}=0\,, 
\end{eqnarray}
similarly to \eq{eq:NPIto1PI}. In \eq{eq:tIkNPIto1PI} we have
used \eq{eq:0}, there is no explicit $t$-dependence.  In
contradistinction to \eq{eq:NPIto1PI} the remaining term on the rhs of
\eq{eq:tIkNPIto1PI} does not vanish as general correlation functions
do not satisfy the equations of motion \eq{eq:eom}. Note also that the
fields $\bar\phi$ trivially satisfy the flows \eq{eq:tIkNPIto1PI}.
The fields $\bar\phi(\phi_a)$ belong to the correlation functions
$\tilde I_k$ and hence they obey the flow equation
\begin{eqnarray}\label{eq:flowbarphi}
  \partial_t \bar\phi_{\bf a}[\phi_a] +
  \Delta S_2[\phi_a,\dot R]\, \bar\phi_{\bf a}[\phi_a]=0\,. 
\end{eqnarray}
Inserting \eq{eq:flowbarphi} into \eq{eq:tIkNPIto1PI} we arrive at the
flow
\begin{eqnarray}\label{eq:tIkNPI1PI} 
  \partial_t \tilde I_k[\phi_a] +(\Delta S_2 \bar\phi_{\bf a})
  \tilde I_k^{,\bf a}[\bar\phi]=0\,, 
\end{eqnarray} 
which implies \eq{eq:flowtIk}. The latter statement follows only after
some algebra from \eq{eq:tIkNPI1PI}. For its proof one has to consider
that $\Delta S_2$ acts linearly on $\tilde I_k$ which it does not on
general correlation functions $\CO_k$ \footnote{The proof can be
  worked out for $N$-point functions \eq{eq:resolveNPI} from where it
  extends straightforwardly.}. However, it is more convenient to work
with the flow \eq{eq:flowIk} for $I_k[J^{\bf a}]$ and with the
definition $I_k[J^a]=I_k[J^{\bf a}= (J^a,0)]$. By using the
equivalence of $J$-derivatives \eq{eq:reduce} valid for the $I_k$, the
flow for $I_k[J^a]$ derives from that of $I_k[J^{\bf a}]$ as
$(\partial_t +\Delta S_1[J^{a},\dot R]) I_k[J^{a}]=0$, implying the
flow \eq{eq:flowtIk} for $\tilde I[\phi_a]$. It is worth noting that
truncated flows derived from either the representation \eq{eq:flowtIk}
or \eq{eq:tIkNPI1PI} differ. This fact can be used for consistency
checks of truncations as well as an improvement in case one of the
representations is better suited within a given truncation. \smallstep

Accordingly there is a close link between $N$PI formulations of the
effective action and general flows. Moreover, it is possible to switch
back and forth between these formulations, thereby combining their
specific advantages.\smallstep

\subsubsection{2PI flows}\label{sec:2PI}
As an explicit example we study the standard flow related to the
quadratic regulator term
\begin{eqnarray}\label{eq:quad} 
  \Delta S_k[\s0{\delta}{\delta J}]=R^{a b}
  \0{\delta}{\delta J^{a}}\0{\delta}{\delta J^{b}}\,, 
\end{eqnarray} 
which can be linearised in terms of 2PI quantities
\begin{eqnarray}\label{eq:2PI} 
  \hat\phi_{a_1 a_2}=\hat\phi_{a_1} \hat\phi_{a_2}\,,
\end{eqnarray}
where $\hat\phi_a$ is not necessarily a fundamental field.  For
$\hat\phi_{a_1 a_2}$ as defined in \eq{eq:2PI} the relation
\eq{eq:reduce} reads
\begin{eqnarray}\nonumber  
  &&\hspace{-1.5cm}\0{\delta}{\delta J^{a_1}}\0{\delta}{\delta J^{a_2}} 
  e^{J^b \hat\phi_b +J^{b_1 b_2}\hat\phi_{b_1}\hat\phi_{b_2}}\\ 
  &&=
  \0{\delta}{\delta J^{a_1 a_2}}
  e^{J^b \hat\phi_b +J^{b_1 b_2}\hat\phi_{b_1} \hat\phi_{b_2}}\,, 
\label{eq:linids}\end{eqnarray} 
Using \eq{eq:linids} we reduce \eq{eq:quad} to a linear regulator at
the expense of also keeping the corresponding $2$-point functions
fixed,
\begin{eqnarray}\label{eq:2point0} 
  \partial_t \phi_{a_1 a_2}= \partial_t (G+\phi_{a_1}\phi_{a_2})=
  \partial_t G =0\,. 
\end{eqnarray} 
We substitute $\Delta S_k$ in \eq{eq:Wk},\eq{eq:Ik} with  
\begin{eqnarray}\label{eq:linear} 
  \Delta S_k[\s0{\delta}{\delta J}]=
  R^{a_1 a_2}\0{\delta}{\delta J^{a_1}} 
  \0{\delta}{\delta J^{a_2}}\to R^{a_1 a_2}
  \0{\delta}{\delta J^{a_1 a_2}} \,, 
\end{eqnarray}
and are lead to \eq{eq:0}, $\partial_t \tilde I_k[\phi]=0$. The
effective action and its flow are functions of the field $\phi_a$ and
the two-point function $\phi_{ab}$:
\begin{eqnarray}\label{eq:tGamma}
  \Gamma_k[\phi_{\bf a}]=\Gamma[\phi_{\bf a}]+
  R^{ab} \phi_{a b}\,, 
\end{eqnarray}
with $(\phi_{\bf a})=(\phi_{a_1},\phi_{a_1 a_2})$ and
\begin{eqnarray}\label{eq:dtGamma}
  \partial_t\Gamma_k[\phi_{\bf a}]=
  \dot R^{ab} \phi_{a b}\,. 
\end{eqnarray} 
The flow \eq{eq:dtGamma} resembles the standard flow equation
\eq{eq:flowtIk} and follows directly from the definition of $\Gamma_k$
in \eq{eq:tGamma}. It also follows by integration w.r.t.\ $\phi$ from
\eq{eq:linflow} with $\s0{\delta \dot \Gamma_k}{\delta
  \phi_{ab}}=\gamma^{ab\, a'b'} \dot R_{a'b'}$ and $\s0{\delta \dot
  \Gamma_k}{\delta \phi_{a}}=0$.  The equation of motion in $\phi_{a
  b}$ according to \eq{eq:eom} is given by
\begin{eqnarray}\label{eq:eom2PI} 
  \left.\0{\delta \Gamma_k[\phi_{\bf a}]}{\delta \phi_{ab}}\right|_{
    \phi=\bar\phi}
  =0\,. 
\end{eqnarray}
Its solution \eq{eq:resolveNPI} reads $\bar\phi_{\bf a}
=(\phi_a,\bar\phi_{a_1 a_2})$ with
\begin{eqnarray}\label{eq:eom2PI1}
  \bar\phi_{ab}= G_{ab}+\phi_a\phi_b\,.\end{eqnarray} 
The above relations lead to the standard flow equation for the 1PI
effective action $\Gamma_k[\phi_a]=\Gamma_k[\phi_a,\bar\phi_{ab}]
-R^{bc}\phi_b\phi_c$ defined in \eq{eq:rel}.  With \eq{eq:eom2PI} it
follows that \cite{Polonyi:2004pp,Wetterich:2002ky,Dupuis:2005ij}
\begin{eqnarray}\nonumber 
  \partial_t \Gamma_k[\bar\phi(\phi)]&=&\partial_t|_{\bar\phi}
  \Gamma_k[\bar\phi]+\Gamma_{k}^{,\bf a}[\bar\phi]\,
  \partial_t\bar\phi_{\bf a}[\phi] \\\di 
  &=& \partial_t|_{\bar\phi}
  \Gamma_k[\phi,\bar\phi]\,.
  \label{eq:eom2PI2} \end{eqnarray}
Using the flow \eq{eq:dtGamma} in \eq{eq:eom2PI2} we arrive at 
\begin{eqnarray}\label{eq:reder} 
  \partial_t \Gamma_k[\phi_a] =
  \dot R^{bc} G_{bc}\,,
\end{eqnarray} 
the standard flow \eq{eq:standflowGk}. Hence linear flows of 2PI
quantities and its fixed point equations reflect the standard flow
equation and offer the possibility of using 2PI expansions as well as
results in standard flows.  \smallstep

\subsubsection{Renormalisation}\label{sec:comprenorm}
The setting in the present work hinges on the bootstrap idea that the
path integral, more precisely the Schwinger functional $W[J,R]$, is
finite and uniquely defined. Resorting to Weinberg's idea of
non-perturbative renormalisability \cite{Gomis:1995jp} this simply
implies the existence of a finite number of relevant operators in the
theory. If not only the fundamental fields $\hat\phi=\hat\varphi$ are
coupled to the path integral but also general composite operators
$\hat\phi^{\bf a}$ some care is needed.  As an example let us consider
$\hat\phi^4$-theory in $d=4$ dimensions in the presence of a source
for $\hat\phi^6(x)$. More generally we deal with a Schwinger
functional $W[J,R]$ with $J^{\bf a} \hat\phi_{\bf
  a}=J^a\hat\varphi_a+J^{a_1\cdots a_6} \hat\varphi_{a_1}\cdots
\hat\varphi_{a_6}$. The composite $\hat\varphi^6(x)$ operator is
coupled with the choice $J^{a_1\cdots a_6} \hat\varphi_{a_1}\cdots
\hat\varphi_{a_6}=\lambda_6 \int_x \hat\varphi^6(x)$.  However, at
face value we have changed the theory to a $\hat\varphi^6(x)$-theory
with coupling $\lambda_6$ that is not perturbatively renormalisable in
$d=4$. Still, within functional RG methods one can address the
question whether such the theory is consistent. In particular if the
theory admits a non-trivial ultraviolet fixed point the problem of
perturbative non-renormalisability is cured. Leaving aside the problem
of its UV-completion the flow equation can be used to generate the
IR-effective action from some finite initial condition. Then, the flow
equation introduces a consistent BPHZ-type renormalisation.
\smallstep

In turn, as long as the composite operator $\phi^{\bf a}$ is
renormalisable we deal with the standard renormalisation of composite
operators \cite{Collins:1984xc}. Moreover, functional RG flows can be
used to actually define finite generating functionals in the presence
of composite operators as well as practical iterative renormalisation
procedures \cite{Polonyi:2000fv,Polonyi:2004pp}. The general case is
covered by the RG equations \eq{eq:RGflowIk},\eq{eq:RGflowtIk} and the
full flows \eq{eq:RGflowtIk}. In particular we deal with a matrix
$\gamma_{\phi}{}^{\bf a}{}_{\bf c}$ of anomalous dimensions, and the
corresponding renormalisation conditions, for the general perturbative
setting see e.g.\ \cite{Collins:1984xc}. We resort again to the above
example of $\hat\varphi^4$-theory in $d=4$ but coupled to the 2-point
function: $J^{\bf a} \phi_{\bf a}=J^a\hat\varphi_a+J^{a_1 a_2}
\hat\varphi_{a_1}\hat\varphi_{a_2}$. We have extended the number of
(independent) relevant operators $\langle \hat\varphi^2(x)\rangle$,
$\langle (\partial\hat\varphi)^2(x)\rangle$ and $\langle
\hat\varphi^4(x)\rangle$ with $\langle \hat\phi(x,x)\rangle$ and
$\langle \hat\phi(x,x) \hat\varphi^2(x)\rangle$ and $\langle
\hat\phi^2(x,x) \rangle$, where
$\hat\phi(x,y)=\hat\varphi(x)\hat\varphi(y)$. The anomalous dimensions
of these operators are related by the matrix $\gamma_\phi$ and
coincide naturally on the equations of motions.  \smallstep

Apart from these more formal questions there is the important issue of
practical renormalisation, i.e.\ consistently renormalising the theory
order by order within a given truncation scheme. The general flows
\eq{eq:flowtIk} together with the considerations of this section allow
to construct such a renormalisation. Again we outline the setting within 
the 2PI effective action with ${\bf a}=a, a_1 a_2$ and 
$\hat\phi_{\bf a}=(\hat\varphi_a,\hat\varphi_{a_1}\hat\varphi_{a_2})$.
 As distinguished from the last
section~\ref{sec:2PI} we couple a quadratic regulator to the fields,
\begin{eqnarray}\label{eq:22pi}
  \Delta S[\hat\phi,R]=R^{\bf a\bf b}\hat\phi_{\bf a}\hat\phi_{\bf b}\,,
\end{eqnarray}
where we also allow for insertions of the operators $\hat\phi_a
\hat\phi_{b_1 b_2}$ and $\hat\phi_{a_1 a_2} \hat\phi_{b_1 b_2}$. The
regulator \eq{eq:22pi} leads to the standard flow \eq{eq:standflowtIk}
for general correlation functions, for the effective action it is given by
\eq{eq:standflowGk}. In the present case it reads
\begin{eqnarray}\nonumber 
  \partial_t\Gamma_k[\phi]&=&\dot R^{ab} G_{ab}+\dot R^{a b_1 b_2} 
  G_{a b_1 b_2}
  +\dot R^{a_1 a_2 b} G_{a_1 a_2 b}\\ 
  & & +\dot R^{a_1 a_2 b_1 b_2} 
  G_{a_1 a_2 b_1 b_2} G_{a_1 a_2 b_1 b_2}\,.
\label{eq:2PIflowex}\end{eqnarray}
In the first term on the rhs of \eq{eq:2PIflowex} we could also
identify $G_{ab}=\phi_{ab}-\phi_a\phi_b$, see \eq{eq:2point}.
2PI expansions relate to loop (coupling) expansions in the field
$\phi_{\bf a}$ and hence, via the equations of motion, to resummations
of classes of diagrams. For general expansion schemes we refer to the
results of section~\ref{sec:DSErenorm} that straightforwardly
translate to the present multi-index situation.\smallstep

We proceed by discussing an iterative loop-wise resolution of the flow
\eq{eq:2PIflowex} that leads to a BPHZ-type renormalisation of
diagrams as in the standard case. This analysis is not bound to the
2PI example considered above as the index ${\bf a}$ could comprise
higher $N$-point functions. From now on we consider the general case.
Still we keep the simple quadratic regulator \eq{eq:22pi}. Assume that
we have resolved the theory at $i$th loop order leading to a finite
$i$-loop contribution $\Gamma_k^{(i)}$, the full effective action
being $\Gamma_k=\sum_i \Gamma_k^{(i)}$. Then, the $i+1$st order reads
in differential form
\begin{eqnarray}\label{eq:i+1}
  \partial_t\Gamma_k^{(i+1)} = \dot R^{\bf a b} G_{\bf ab}^{(i)}\,,
\end{eqnarray}
and is finite. At one loop, $i=1$, its integration results in 
\begin{eqnarray}\label{eq:2PI1loop}
  \Gamma^{(1)}_k[\phi]=
  -\left.\s012 (\ln G)^{\bf a}{}_{\bf a}\right|_{\Lambda}^k
  +\Gamma^{(1)}_{\Lambda} \,, 
\end{eqnarray}
where the $\Lambda$-dependent terms arrange for a BPHZ-type
renormalisation procedure and, in a slight abuse of notation, $G$
stands for the classical propagators of the fields $\phi_{\bf a}$.
The superscript ${}^{(1)}$ indicates the one loop order, not the one
point function.  The subtraction at $\Lambda$ makes the rhs finite.
$\Gamma_{\Lambda}$ ensures the $\Lambda$-independence as well as
introducing a finite (re)-normalisation. For $i=2$ we have to feed
$\Gamma^{(1)}[\phi]$ and its derivatives into the rhs of the flow
\eq{eq:i+1}. Again the $t$-integration can be performed as the rhs is
a total derivative w.r.t.\ $t$. It is the same recursive structure
which reproduces renormalised perturbation theory from a loop-wise
integration of the 1PI flow. At two loop the flow \eq{eq:i+1} reads
\begin{eqnarray}\label{eq:2PI2loop1}
  \dot R^{\bf a b} G_{\bf ab}^{(2)}=-\dot R^{\bf a b} G_{\bf a c} 
  \,\Gamma^{(1),{\bf cd}}G_{\bf db}\,,
\end{eqnarray}
assuming no coupling dependence of $R$. The two-point function at one
loop, $\Gamma^{(1),{\bf cd}}$, is the second derivative of
\eq{eq:2PI1loop} w.r.t.\ the field $\phi_{\bf a}$, and
\eq{eq:2PI2loop1} turns into a total $t$-derivative. Finally we arrive
at the two-loop contribution
\begin{eqnarray}
  \nonumber 
  &&\hspace{-.4cm}\Gamma_k^{(2)}=\018 
  \Gamma^{,{\bf a}_1 {\bf a}_2{\bf a}_3 {\bf a}_4} 
  (G-G|_{\Lambda})_{{\bf a}_1 {\bf a}_2}(G-G|_{\Lambda})
  _{ {\bf a}_3 {\bf a}_4}\\\nonumber 
  &&\hspace{-.2cm} -\01{12}  
  \Gamma^{,{\bf a}_1 {\bf a}_2{\bf a}_3} 
  \Gamma^{,{\bf a}_1 {\bf a}_5{\bf a}_2} 
  (G-G|_{\Lambda})_{{\bf a}_1 {\bf a}_2} 
  (G-G|_{\Lambda})_{{\bf a}_3 {\bf a}_4}  \\\nonumber  
  &&\hspace{2.5cm}\times 
  \left( (G-G|_{\Lambda}) +3 G|_{\Lambda} \right)_{{\bf a}_5 
    {\bf a}_6} \\ 
  &&\hspace{-.2cm}+\012 \Gamma_{\Lambda, {\bf a}_1 {\bf a}_2}^{(2)}
  (G-G|_{\Lambda})_{{\bf a}_1 {\bf a}_2} + \Gamma_{\Lambda}^{(2)}\,. 
\label{eq:2PI2loop}\end{eqnarray}
Higher orders follow similarly. Such a procedure allows for a
constructive renormalisation of the theory under investigation, and
also facilitates formal considerations concerning the renormalisation
of general truncations schemes. The first two terms in
\eq{eq:2PI2loop} are already finite due to the subtractions. The terms
proportional to $3 G$ in the third line of \eq{eq:2PI2loop} and in the
4th line constitute finite (re-) normalisations.  \Eq{eq:2PI2loop}
stays finite if the vertices and propagators are taken to be full
vertices and propagators in the sense of an RG improvement. Within the
2PI example considered in \eq{eq:2PIflowex} the integrated flow
\eq{eq:2PI2loop} is the consistently renormalised result for the 2PI
effective action at two loop. It translates into a resummed
renormalised 1PI effective action by using the equation of motion
\eq{eq:eom} for the composite field $\phi_{ab}$. However, the above
result also applies to $N$PI effective actions or more general
composite operators coupled to the theory: the integrated flow
\eq{eq:2PI2loop} constitutes a finite BPHZ-type renormalised
perturbative expansion.  Moreover, the above method straightforwardly
extends to general expansion schemes: in general the integrated flow
constitutes a finite BPHZ-type renormalised expansion. The consistency
of the renormalisation procedure is guaranteed by construction.  \smallstep

The renormalisation conditions for the full theory are set implicitly
with the choice of the effective action at the initial cut-off scale
$\Lambda$. We emphasise that any RG scheme that derives from a
functional truncation to the flow \eq{eq:flowtIk}, and in the
particular the loop expansion \eq{eq:i+1}, is consistent with the
truncation.  Moreover, the iterative structure displayed in
\eq{eq:i+1}, \eq{eq:2PI1loop} and \eq{eq:2PI2loop} allows us to
discuss general renormalisation conditions in the present setting. By
adding the operator $\hat\phi_{ab}$ we have extended the number of
relevant vertices in the effective action and hence the number of
renormalisation conditions.  In case $\phi_{\bf a}$ includes only
marginal and irrelevant operators the renormalisation proof can be
mapped to that of the 1PI case.  \smallstep

The basic example is provided by $(\phi_{\bf a})=(\phi_a,\phi_{a_1
  a_2})$, where the field $\phi_{a_1 a_2}$ with $\hat\phi_{a_1
  a_2}=\hat\phi_{a_1}\hat\phi_{a_2})$ counts like $\phi_{a_1}
\phi_{a_2}$. RG conditions for e.g.\ the 2-point function and the 
4-point function 
\begin{eqnarray}\label{eq:RG-conditions1} 
\0{\delta \Gamma}{\delta \phi_a \delta \phi_b}\,, \qquad 
\0{\delta \Gamma}{\delta \phi_{a_1}\cdots  \delta \phi_{a_4}}\,, 
\end{eqnarray}  
trigger additional RG conditions for  
\begin{eqnarray}\label{eq:RG-conditions2} 
\0{\delta\Gamma}{\delta \phi_{ab}}\,, \qquad 
\0{\delta\Gamma}{\delta \phi_{a_1 a_2} \delta \phi_{a_3 a_4}}\,, \qquad 
\0{\delta \Gamma}{\delta \phi_{a_1 a_2} \delta \phi_{a_3} \delta \phi_{a_4}}\,.
\end{eqnarray}
Using the relation \eq{eq:reduce} between derivatives w.r.t.\ $\phi_a$
and $\phi_{ab}$ we are left with the same number of independent RG
conditions as in the 1PI case. In other words, the matrix $\gamma^{\bf
  a}{}_{\bf b}$ is highly symmetric. This symmetry can be imposed on
the level of $\Gamma_{\Lambda}$ and evolves with the flow as its
rhs only depends on (derivatives of) $\Gamma_{k}$. We observe that
formally any choice of $\Gamma_{\Lambda}$ independently fixes these
RG conditions at all scales (via the flow) but violates the relation
\eq{eq:reduce}.  A priori there is nothing wrong with such a
procedure that simply relates to an additional additive 
renormalisation (at 1PI level) and can be absorbed in a possibly
$k$-dependent rescaling of the 2PI fields. The
above discussion extends to the general case with fields $\phi_{\bf
  a}$.  We shall detail these observations and structures elsewhere
and close with the remark that for general truncation schemes that do
not admit a direct resolution of the flow as in perturbation theory,
the costs relate to an additional $t$-integration as already discussed
in the 1PI case of section~\ref{sec:DSEs}. \step

\setcounter{equation}{0}
\section{Applications to gauge theories}\label{sec:gauge} 

The generality of the present approach fully
pays off in gauge theories, and the present work was mainly triggered
by related investigations. In flow studies for gauge theories 
\cite{Reuter:1992uk,Reuter:1993kw,Reuter:1997gx,Bonini:1993kt,Bonini:1993sj,Ellwanger:1994iz,D'Attanasio:1996jd,Reuter:1996be,Pawlowski:1996ch,Bonini:1997yv,Bonini:1998ec,Falkenberg:1998bg,Litim:1998qi,Litim:1998wk,Litim:2002ce,Freire:2000bq,Igarashi:1999rm,Igarashi:2001mf,D'Attanasio:1996zt,D'Attanasio:1996fy,Ellwanger:1997wv,Ellwanger:1998th,Ellwanger:1995qf,Ellwanger:1996wy,Bergerhoff:1997cv,Pawlowski:2003hq,Pawlowski:2004ip,pawlowski,Fischer:2004uk,Gies:2002af,Gies:2003ic,Braun:2005uj,Simionato:1998te,Simionato:1998iz,Simionato:2000ut,Panza:2000tg,Ellwanger:2002sj} and gravity
\cite{Reuter:1996cp,Lauscher:2001ya,Litim:2003vp,Bonanno:2004sy,Lauscher:2005xz,Fischer:2006fz} 
with the standard quadratic regulator one has to deal with modified
Slavnov-Taylor identities 
\cite{Reuter:1992uk,Reuter:1993kw,Reuter:1997gx,Bonini:1993kt,Bonini:1993sj,Ellwanger:1994iz,D'Attanasio:1996jd,Reuter:1996be,Pawlowski:1996ch,Bonini:1997yv,Bonini:1998ec,Falkenberg:1998bg,Litim:1998qi,Litim:1998wk,Litim:2002ce,Freire:2000bq,Igarashi:1999rm,Igarashi:2001mf}. 
These identities tend towards the Slavnov-Taylor identities of the
full theory in the limit of vanishing regulator. It is crucial to
guarantee this limit towards physical gauge invariance. \smallstep

The subtlety of modified Slavnov-Taylor identities can be avoided for
thermal flows. This is achieved by either modifying the thermal
distribution \cite{D'Attanasio:1996zt,D'Attanasio:1996fy}, or by
constructing the thermal flow as a difference of Callan-Symanzik flows
at zero and finite temperature in an axial-type gauge
\cite{Litim:1998nf}. The resulting thermal flows are gauge invariant.
We remark that Callan-Symanzik flows in axial gauges at zero
temperature
\cite{Simionato:1998te,Simionato:1998iz,Simionato:2000ut,Panza:2000tg}
are formally gauge invariant, but the approach towards the full theory
at vanishing regulator has severe consistency problems. This problem
is related to the missing locality in momentum space combined with the
incomplete gauge fixing \cite{Litim:2002ce}. One expects a better
convergence for Callan-Symanzik flows within covariant or Abelian
gauges \cite{Ellwanger:2002sj}\smallstep

Alternatively one can resort to gauge-invariant degrees of freedom
\cite{Branchina:2003ek,Pawlowski:2003sk}, gauge-covariant degrees of
freedom
\cite{Morris:1999px,Morris:2000fs,Arnone:2005fb,Morris:2005tv,Rosten:2005ep},
or higher order regulator terms with regulators $R^{\bf a_1\cdots
  a_n}$ with $n>2$. Then, $N$-point functions directly relate to
observables and allow for the construction of gauge-invariant flows.
In general such a parameterisation is payed for with non-localities,
in particular in theories with a non-Abelian gauge symmetry.
\smallstep

In this chapter we discuss the structural aspects of the above
formulations.  In particular we deal with the question of convenient 
representations of symmetry identities that facilitates their
implementation during the flow. Moreover we discuss the related
question of adjusted parameterisations of gauge theories, and evaluate 
the fate of symmetry constraints in gauge-invariant formulations. \smallstep

\subsection{Parameterisation}\label{sec:parameter}
In gauge fixed formulations of gauge theories, and in particular in
strongly interacting regimes, the propagators and general Green
functions are only indirectly related to physical observables.
Firstly, only combinations of them are gauge invariant and secondly,
the relevant degrees of freedom in the strongly interacting regime are
not the perturbative ones \footnote{Basically by definition; the relevant
degrees of freedom should only weakly interact.}. Good choices are
observables that serve as order parameters; e.g.\ the Polyakov
loop\footnote{The definition \eq{eq:pol} only applies in the case of
  periodic boundary conditions for the gauge field.}
\begin{eqnarray}\label{eq:pol} 
  P(\vec x)=\Tr\, \CP \exp \int_0^\beta
  A_0(x) d\tau\,,
\end{eqnarray}
and its two-point function $\langle P(\vec x) P^\dagger (\vec
y)\rangle$ in the case of the confinement-deconfinement phase
transition. These observables fall into the class of $I_k$ defined in
\eq{eq:Ik}. For the Polyakov loop variable \eq{eq:pol} the
corresponding operator is $\hat I=P(\vec x)[A_0=\s0{\delta}{\delta
  J^0}]$ which implies $\hat I_k=\hat I$, see \eq{eq:hatIk}.  Hence
their flow can still be described in terms of field propagators and
vertices of the fundamental fields via \eq{eq:Ik},\eq{eq:tildeIk}. It
amounts to the following procedure: compute the flow of propagators
and vertices, even though partially decoupling in the phase
transition. Then, the flow of relevant observables $\tilde I$ is
computed from this input with the flow \eq{eq:flowtIk}, i.e.\ the
heavy quark potential from the flow of the Wilson loop or Polyakov
loop. Such a procedure allows for a direct computation of physical
quantities from the propagators and vertices of the theory in a given
parameterisation, and it applies to gauge fixed as well as gauge
invariant formulations.  It also emphasises the key r$\hat{\rm o}$le
played by the propagators of the theory, and matches their key
importance within the functional optimisation developed in
section~\ref{sec:optimal}.  \smallstep

One also can use appropriate fields $\hat \phi$ coupled to the theory.
In the above example of the confinement-deconfinement phase transition
a natural choice is provided by the gauge invariant field
$\hat\phi(x)=P(\vec x)$ with \eq{eq:pol}. Such a choice has to be
completed by additional $\hat\phi^{\bf a}$ that cover the remaining
field degrees of freedom. Alternatively one can integrate out the
remaining degrees of freedom and only keep that of interest. Another
interesting option are gauge covariant degrees of freedom, e.g.\
$\hat\phi_{\mu\nu}(x)=F_{\mu\nu}$ or $\hat\phi_{\mu\nu}(x)=\tilde
F_{\mu\nu}$, that is the dual field strength
\cite{Ellwanger:1997wv,Ellwanger:1998th}.  Both choices can be used to
derive (partially) gauge invariant effective actions, and aim at a
description of gauge theories in terms of physical variables. 
\smallstep

We emphasise that the above suggestions usually generate non-local and
non-polynomial effective actions even at the initial scale. We have to
keep in mind that gauge theories are formulated as path integrals over
the gauge field supplemented with a polynomial and local classical
action. Gauge fixing is nothing but the necessity to deal with a
non-trivial Jacobian that arises from the decoupling of redundant
degrees of freedom, and Slavnov-Taylor identities (STIs) carry the
information of this reparameterisation. If coupling gauge invariant or
gauge covariant degrees of freedom to the theory the necessity of
decoupling the redundant degrees of freedom remains, and hence the
symmetry constraints are still present. In a gauge invariant setting
the corresponding STIs turn into a subset of DSEs. Their relevance
might be hidden by the fact of manifest gauge invariance, but still
they carry the information about locality. In other words,
approximations to gauge invariant effective actions or general
correlation functions still can be in conflict with the Slavnov-Taylor
identities and hence violate physical gauge invariance. Indeed it is
helpful to explicitly gauge fix the theory within a choice that
simplifies the relation $\phi=\phi(A)$ for gauge-fixed fields $A$ as
it makes locality more evident in the variables $\phi$. For example,
in case of the confinement-deconfinement phase transition we choose
$\hat\phi(\vec x)=P(\vec x)$ defined in \eq{eq:pol}, and use the
Polyakov gauge or variations thereof, e.g.\
\cite{Reinhardt:1997rm,Lenz:1998qk,Ford:1998bt}.  \smallstep 

In summary we conclude that it is vital to study the fate of symmetry
constraints such as the Slavnov-Taylor identities for general flows,
be they gauge invariant or gauge variant. This is done in the next
three sections~\ref{sec:mWIs},
\ref{sec:gaugeinv},\ref{sec:chiral}.\smallstep

\subsection{Modified Slavnov-Taylor identities}\label{sec:mWIs} 

The propagators and vertices of a gauge theory are constrained by
gauge invariance of the theory. A non-trivial symmetry $I_k\equiv 0$
is maintained during general flows \eq{eq:RGflowIk},
\eq{eq:RGflowtIk}: if $I_k\equiv 0$ is satisfied at the starting
scale, its flow vanishes as it is proportional to $I_k$. In particular
this is valid for $D_s=\partial_t$. The corresponding flows include
that of modified Ward-Takahashi or Slavnov-Taylor identities for the
effective action \cite{D'Attanasio:1996jd,Litim:1998qi,Litim:2002ce},
and that of Nielsen identities \cite{Pawlowski:2003sk} for gauge
invariant flows \cite{Branchina:2003ek,Pawlowski:2003sk}. \smallstep

The above statements imply that the generator of the flow, $D_s$, 
commutes with the generator of the modified symmetry $\hat I_k$.
Within truncations this property does not hold, and it is not
sufficient to guarantee the symmetry at the starting scale.
Consequently a symmetry relation $I_k\equiv 0$ should be read as a fine-tuning
condition which has to be solved at each scale. This is technically
rather involved, and any simplification is helpful. Here we aim at a
discussion of different representations of symmetry constraints and
their flows. \smallstep

\subsubsection{STI} 
First we concentrate on a pure non-Abelian gauge theory with general
gauge fixing $\CF[A]$. For its chief importance we shall explain the
structure with sources coupled to the fundamental fields $\varphi$,
and a standard quadratic regulator term $R^{ab} \varphi_a\varphi_b$.
We keep the condensed notation and refer the reader to
\cite{Zinn-Justin:1993wc} for some more details. The Schwinger
functional is given by
\begin{eqnarray}
  e^{W[J,Q]}=\int \d\hat\varphi\, d\lambda \,e^{-S[\hat\varphi]
    +J^{\bf a}\hat\phi_{\bf a} 
    +Q^{\bf a} \,\Gs \hat\phi_{\bf a}}\,. 
\label{eq:brstW}\end{eqnarray} 
In \eq{eq:brstW} we have also included source terms $Q^{\bf a} \,\Gs
\hat\phi_{\bf a}$ for the symmetry variations of the fields as
introduced in section~\ref{sec:general}. Here $\Gs$ generates
BRST transformations defined below in \eq{eq:brstvar}. The fields
$\hat\phi_{\bf a}[\hat\varphi]$ depend on the fundamental fields
$\hat\varphi_a$ given by 
\begin{eqnarray}\label{eq:fundfields}
  (\hat\varphi_a)=(A_i\,,\,C_\alpha\,,\, \bar C_\alpha )\,, 
\end{eqnarray} 
where we have dropped the hats on the component fields. The component
fields in \eq{eq:fundfields} read more explicitly $A_i=A_a^\mu(x)$,
the gauge field, and $C_\alpha=C_a(x)$, $\bar C_\alpha=C_a(x)$, the
ghost fields. A more explicit form of the source term in the case of
$\hat\phi= \hat\varphi$ reads
\begin{eqnarray}\label{eq:source}
  &&\hspace{-.4cm}
  J^a\hat\varphi_a = J_i A_i+\bar J_\alpha C_\alpha 
  -J_\alpha \bar C_\alpha \\
  \nonumber 
  &&= \int_x\left( J^a_\mu(x) A^a_\mu(x)+\bar J^a(x)C^a(x)+
    \bar C^a(x) J^a(x)\right)\,. 
\end{eqnarray} 
The action $S$ in the path integral \eq{eq:brstW} is given by
\begin{eqnarray}\nonumber 
  &&\hspace{-1cm}S[\hat\varphi,\lambda]=
  S_{\rm YM}[\hat\varphi]\\
  &&-\omega(\lambda)+\lambda_\alpha \CF_\alpha(A) 
  -\bar C_\alpha M^{\alpha\beta}C_\beta\, ,
\label{eq:brstaction1}\end{eqnarray}
with
\begin{eqnarray}\label{eq:brstaction2}
  M^{\alpha\beta}= \CF^{\alpha}{}_{,i}D_i^\beta(A)\,, \qquad 
  \omega(\lambda)=\0\xi2 \lambda_\alpha \lambda_\alpha\,,
\end{eqnarray}
the latter equation for $\omega$ leading to the standard gauge fixing
term $\s01{2\xi} \CF^\alpha \CF_\alpha$ upon integration over
$\lambda$. Then, in a less condensed notation, \eq{eq:brstaction1}
turns into
\begin{eqnarray}\nonumber 
  &&\hspace{-1.5cm}S[\hat\varphi]=\014 \int_x F_{\mu\nu}^a  
  F_{\mu\nu}^a\\
  && -\0{1}{2\xi}\int_x \CF^a\CF^a- 
  \int_x \,\bar C^a \0{\partial \CF^a}{\partial A^\mu_b}
  D_\mu^{bc} C^c \,.
\label{eq:brstaction3}\end{eqnarray}
Matter fields and a Higgs sector can be straightforwardly added. The
action \eq{eq:brstaction1} is invariant under the BRST transformations
\begin{eqnarray}\label{eq:brstvar}
  (\Gs \hat\varphi)=( D_i^\alpha C_\alpha\, ,\, \s012 
  f_{\alpha\beta\gamma} C_\beta C_\gamma\,,\, \lambda_\alpha)\,, 
\end{eqnarray}
and $\Gs$ acts trivially on $\lambda$: $\Gs\lambda_\alpha=0$. The
operator $\Gs$ can be represented as a functional differential
operator on the fields $\hat\varphi,\lambda$ with
\begin{eqnarray} 
  \Gs=(\Gs \hat\varphi_a) \,\0{\delta}{\delta \hat\varphi_a}\,,
\label{eq:brstder}\end{eqnarray}
making the anti-commuting (Grassmann) property of $\Gs$ explicit.  The
invariance of the action, $\Gs S[\hat\varphi]=0$ can be proven
straightforwardly by insertion. Moreover, $\Gs$ is a differential with
$\Gs^2 \varphi=0$ allowing for a simple form of the symmetry
constraint.  The only BRST-variant term is the source term
$J^a\hat\varphi_a$. The related Slavnov-Taylor identity (STI) is cast into
an algebraic form with help of the source terms for the
BRST variations \eq{eq:brstvar} included in \eq{eq:brstW}. For
$\phi=\varphi$ this source terms reads  
\begin{eqnarray}\label{eq:brstsource}
  Q^a\,\Gs\hat\varphi_a=Q^i D_i^\alpha C_\alpha +\s012 \bar Q^\alpha 
  f_{\alpha 
    \beta\gamma} C_\beta C_\gamma  
  +Q^\alpha\lambda_\alpha\,,
\end{eqnarray}
where $Q^\alpha\, \Gs \bar C_\alpha=Q^\alpha\, \lambda_\alpha$ could
also be considered as a standard source term for the auxiliary field
$\lambda$. The general BRST source term reads
\begin{eqnarray}\label{eq:brstsourcegen}
  Q^{\bf a}\,(\Gs\hat\phi)_{\bf a}=Q^{\bf a}\,(
  \Gs\hat\varphi)_a \,\hat
  \phi_{\bf a}^{,a}[\hat\varphi]
  \,,
\end{eqnarray}
following with \eq{eq:brstder}. The Slavnov-Taylor identity follows
from
\begin{eqnarray}\label{eq:sti0} 
  \hspace{-.5cm}\int {\Gs}\,\Bigl( d\hat\varphi\,d\lambda\,
  \exp\{-S[\hat\varphi]+J^{\bf a}\hat\phi_{\bf a}+
Q^{\bf a} (\Gs\hat\phi)_{\bf a}\}
  \Bigr)\equiv 
  0\,. 
\end{eqnarray}
\Eq{eq:sti0} is of the form \eq{eq:genDS}.  It follows with
\eq{eq:brstder} after a partial functional integration and $(\Gs
\hat\varphi_a)^{,a}=D_{i,i}^\alpha C_\alpha +
f_{\alpha\alpha\beta}C_\beta=0$ (for compact Lie groups). Except for 
the source term $J^{\bf a}\hat\phi_{\bf a}$ all terms in \eq{eq:sti0} are
BRST-invariant: $\Gs d\hat\varphi=0$, $\Gs S[\hat\varphi]=0$, $\Gs
(Q^{\bf a}\,\Gs\hat\phi_{\bf a})=0$. The operator $\Gs$ commutes due
to its Grassmannian nature with bosonic currents $J$ and anti-commutes
with fermionic ones. For example, for the fundamental fields and
currents this entails that $\Gs$ commutes with $J^i$ but anti-commutes
with $J^\alpha, \bar J^\alpha$ and $\Gs J^a\hat\varphi_a = J^b
\gamma^a{}_b (\Gs\hat\varphi_a )$. Using all these properties in
\eq{eq:sti0} leads us to the Slavnov-Taylor identity
\begin{eqnarray}\nonumber 
  &&\hspace{-.5cm}\int d\hat\varphi\, d\lambda\, J^{\bf b} 
  \gamma^{\bf a}{}_{\bf b} (\Gs\hat\varphi_{\bf a} ) 
  \,\exp\{-S[\hat\varphi]+J^{\bf a}\hat\varphi_{\bf a}+Q^{\bf a}\, 
  \Gs\hat\varphi_{\bf a}\}\\
  &&\hspace{-.5cm} = 
  J^{\bf b} \gamma^{\bf a}{}_{\bf b}\s0{\delta }{\delta Q^{\bf a}} 
  e^{W[J,Q]}\equiv 0\,. 
\label{eq:sti}\end{eqnarray}
\Eq{eq:sti} is of the form $e^{W} I[J,Q]\equiv 0$ leading to
\eq{eq:symrelR} with $I$ defined in \eq{eq:I0} for
\begin{eqnarray}\label{eq:hatIs}
  \hat I_{\Gs}
  = J^{\bf b} \gamma^{\bf a}{}_{\bf b} \0{\delta }{\delta Q^{\bf a}}\,,
\end{eqnarray} 
The operator $\hat I_\Gs$ generates BRST transformations on the
Schwinger functional $W$. Accordingly the STI \eq{eq:sti} can be
written as
\begin{eqnarray}\label{eq:stiW}
  \hat I_\Gs W[J,Q]\equiv 0\,, 
\end{eqnarray} 
that is the Schwinger functional is invariant under
BRST transformations.  The STI \eq{eq:stiW} can be generalised to that
for correlation functions $I$. To that end we use that \eq{eq:stiW}
can be multiplied by any operator $\hat I$ from the left. We are led
to
\begin{subequations}\label{eq:stigen}  
\begin{eqnarray}\label{eq:stigen1}
  \CW_{\Gs, I}\equiv 0, \quad {\rm with} \quad 
  \hat\CW_{\Gs, I}=\hat I\, \hat I_{\Gs}\,,
\end{eqnarray}
where $\CW_I$ is derived from $\hat\CW_I$ with \eq{eq:I0}. The
symmetry relation \eq{eq:stigen} is a direct consequence of
\eq{eq:sti}, which is reproduced for $\hat I=1$. We can write the
correlation function $\CW_I$ in terms of $I$ as
\begin{eqnarray}\label{eq:stigen2}
  \CW_{\Gs,I}[J,Q]=\hat I_{\rm \Gs}\, I[J,Q]+\delta I_{\Gs,I}[J,Q]\,,
\end{eqnarray} 
with 
\begin{eqnarray}\label{eq:stigen3}
  \widehat{ \delta I}=[\hat I\,,\,
  J^{\bf b}\gamma^{\bf a}{}_{\bf b}\,\s0{\delta}{\delta Q^{\bf a}} ]\,.
\end{eqnarray}
\end{subequations}
For the derivation of \eq{eq:stigen2} we have used that $\hat I\, \hat
I_{\Gs} = \hat I_{\Gs}\,\hat I + \widehat{\delta I}$ as well repeatedly using 
$[\hat I_{\Gs}\,,\,W]=0$, which is the STI \eq{eq:sti}. \smallstep 

For $Q$-independent $\hat I$ the commutator $\widehat{\delta I}$
substitutes one of the $J$-derivatives in $\hat I$ by one w.r.t.\ $Q$.
Applied on $e^{W}$ this generates a (quantum) BRST transformation on
$\hat I$. Consequently we write
\begin{eqnarray}\label{eq:hatdeltaI}
  \widehat{ \delta I}_{\Gs,I}[J,\hat\phi]\,e^W=-\left(\Gs[\hat\phi]\, 
    \hat I[J,\hat\phi]\right) e^W\,,
\end{eqnarray} 
which we evaluate at $\hat\phi=\s0{\delta}{\delta J}$. Accordingly,
for BRST-invariant $\hat I[J,\hat \phi]$ the second term on the rhs of
\eq{eq:stigen2} disappears. Hence, if $I$ is the expectation value
of a BRST-invariant $\hat I[J,\hat\phi]$, the second term on the rhs
of \eq{eq:stigen2} vanishes and $I$ is BRST-invariant, $\hat I_\Gs\,
I=0$. \smallstep

We remark that \eq{eq:stigen}, as the flow \eq{eq:flowIk}, does not
directly encode the STI for the Schwinger functional.  This comes
about since we have divided out the STI for $W$, \eq{eq:stiW} in its
form $[\hat I_\Gs\,,\, W]$ in the derivation of \eq{eq:stigen}. In
turn, it has to be trivially satisfied. Indeed, for either $\hat I=1$
or $\hat I=W[J,Q]$, leading to $I=1$ and $I=W$, the STI \eq{eq:stigen}
is trivially satisfied. The situation is similar to that of the flow
\eq{eq:flowIk} where the flow of the Schwinger functional has been
divided out. Without resorting to the STI for $W$, \eq{eq:stiW}, the
STIs $\CW_{\Gs,I}$ derived with $\hat \CW_{\Gs,I}$ in \eq{eq:stigen1}
read
\begin{eqnarray}\label{eq:stigenalt}
  \CW_{\Gs,I}[J,Q]=\left(\hat I_{\Gs}-(\hat I_{\Gs}W)\right)\, 
  I[J,Q]+\delta I\,, 
\end{eqnarray} 
and, for $\hat I=1$ or $\hat I=W[J,Q]$ the STI for the Schwinger
functional, \eq{eq:stiW} follows. Hence, we shall refer to the STI
\eq{eq:stiW} as $\CW_{\Gs,1}=0$. Note also that its trivial resolution
does not imply that it is not encoded in the representation
\eq{eq:stigen2}.  Similarly to the derivation of its flow from the
general flow \eq{eq:flowIk}, the STI for the Schwinger functional
derives from $\hat I=\s0{\delta}{\delta J}$, inserted in
\eq{eq:stigen}.  We are led to $\s0{\delta}{\delta J}\hat I_\Gs
W[J,Q]=0$ which entails \eq{eq:stiW}. \smallstep

\subsubsection{mSTI}\label{sec:msti}

So far we have adapted the analysis of the STI in its algebraic form
to the present setting. Now we consider regularisations of the
Schwinger functional $W[J,Q,R]$ defined in \eq{eq:WR}, as well as
general operators $I[J,Q,R]$ defined in \eq{eq:IR}.  The operator
$\hat I_\Gs[J,\s0{\delta}{\delta J},\s0{\delta}{\delta Q},R]$
corresponding to $I_{\Gs}[J,Q,R]$ is derived from \eq{eq:hatIR} as
\begin{eqnarray}\label{eq:brsthatIkgen} 
  \hat I_{\Gs}
  = (J^{\bf b}- [ \Delta S\,,\, J^{\bf b}]) \gamma^{\bf a}{}_{\bf b} 
  \s0{\delta }{\delta Q^{\bf a}}\,,   
\end{eqnarray}
where the second term generates BRST transformations of the regulator
term $\Delta S$, and we have used that $\Delta S$ is bosonic. As an
example we compute \eq{eq:brsthatIkgen} for the standard flow,
$\hat\phi=\hat\varphi$ and a quadratic regulator term
$R^{ab}\hat\varphi_a\hat\varphi_b$. This leads us to the symmetry
operator
\begin{eqnarray}\label{eq:brsthatIk} 
  \hat I_{\Gs}
  = (J^{b}- 2 R^{cb}\s0{\delta }{\delta J^{c}}) \gamma^a{}_b 
  \s0{\delta }{\delta Q^a}\,, 
\end{eqnarray} 
where we have used the symmetry properties of $R$ in
\eq{eq:standRsyms} for standard flows. The STI for the Schwinger
functional \eq{eq:sti} turns into
\cite{Reuter:1992uk,Reuter:1993kw,Reuter:1997gx,Bonini:1993kt,Bonini:1993sj,Ellwanger:1994iz,D'Attanasio:1996jd,Reuter:1996be,Pawlowski:1996ch,Bonini:1997yv,Bonini:1998ec,Falkenberg:1998bg,Litim:1998qi,Litim:1998wk,Litim:2002ce,Freire:2000bq,Igarashi:1999rm,Igarashi:2001mf}
\begin{eqnarray}\label{eq:stiWR}
  \hat I_\Gs\, W[J,Q,R]=0\,, 
\end{eqnarray}
with $\hat I_\Gs$ defined in \eq{eq:brsthatIk}. It entails that only
the source terms $J^{\bf a}\phi_{\bf a}$ and the regulator term are
BRST-variant. The relation \eq{eq:stiWR} was coined modified
Slavnov-Taylor identity (mSTI) as it encodes BRST invariance at $R=0$,
and shows its explicit breaking via the regulator term at $R\neq0$.
 \smallstep

The general case with $\CW_I$ leads to the same general STI
\eq{eq:stigen} with all operators and correlation functions
substituted by their $R$-dependent counterparts defined in \eq{eq:IR}, 
\begin{subequations}\label{eq:stigenR}
\begin{eqnarray}\label{eq:stigenR1}
  \CW_{\Gs,I}[J,Q,R]\equiv 0\,,
\end{eqnarray}
with
\begin{eqnarray}\label{eq:stigenR2}
  \hspace{-.3cm}\CW_{\Gs,I}[J,Q,R]=\hat I_{\rm \Gs}\, I[J,Q,R]
  +\delta I\,.  
\end{eqnarray} 
\end{subequations}
The correlation function $\delta I[J,Q,R]$ is the $R$-dependent
counterpart derived from \eq{eq:stigen3} with \eq{eq:IR};
\begin{eqnarray}\label{eq:relR}
  \widehat{\delta I}[R]=e^{-\Delta S} 
  \,[\hat I\,,\,\hat I_{\Gs}]_{R=0}\,e^{\Delta S}\,,
\end{eqnarray}
where $\Delta S=\Delta S[\s0{\delta}{\delta J},R]$. Hence the second
term on the rhs of \eq{eq:stigenR2} still vanishes for a BRST-invariant
$\hat I[J,\s0{\delta}{\delta J},0]$. The modification of BRST
invariance is solely encoded in the modification of the BRST operator
$\hat I_\Gs$ in \eq{eq:brsthatIk}. The flow of \eq{eq:stigenR} is
governed by \eq{eq:flowIk}. \smallstep

The mSTI \eq{eq:stiWR} for the Schwinger functional follows as
$\CW_{\Gs,1}\equiv 0$ with the alternative representation
\eq{eq:stigenalt}.  As in the case without regulator, it also can be
derived from \eq{eq:stigenR} from $\CW_{\Gs,W_{,\bf a}}$.  Inserting $\hat
I=\s0{\delta}{\delta J}$ into \eq{eq:stigenR2} leads to $\CW_{W_{,\bf
    a}}= \s0{\delta}{\delta J^{\bf a}}\hat I_\Gs W[J,Q,R]\equiv 0$ and
hence to \eq{eq:stiWR}. \smallstep

As in the case of the flows we can turn the general mSTIs
\eq{eq:stigenR} into mSTIs for correlation functions $\tilde I$ in
terms of the variable $\phi$. The definition of the effective action
\eq{eq:GammaR} extends to the case with external currents $Q$:
\begin{eqnarray}\nonumber  
  &&\hspace{-1cm}\Gamma[\phi,Q,R]=J^{\bf a}(\phi,Q)\phi_{\bf a}-
  W[J(\phi,Q),Q,R]\\ 
  &&\hspace{1cm}-\Delta S'[\phi,R]\,, 
\label{eq:GammaRQ}\end{eqnarray} 
the source $J$ now depends on the fields $\phi$ and the source $Q$.
\Eq{eq:GammaRQ} entails that
\begin{eqnarray}\label{eq:dQWG}
  \0{\delta W}{\delta Q^{\bf a}}
  =-\0{\delta (\Gamma+\Delta S')}{\delta Q^{\bf a}}
  = -\0{\delta \Gamma}{\delta Q^{\bf a}}\,,
\end{eqnarray}
as $\Delta S'$ does not depend on $Q$ and the $Q$-dependence of $J$
cancels out. In \eq{eq:dQWG} the $Q$-derivatives of $W$ and $\Gamma$
are taken at fixed arguments $J$ and $\phi$ respectively.  The
correlation functions $\tilde I$ derive from \eq{eq:tildeIk} as
\begin{eqnarray}\label{eq:tIRQ}
  \tilde I[\phi,Q,R]=I[J(\phi,Q,R),Q,R]\,. 
\end{eqnarray} 
For the mSTIs $\tilde \CW_I\equiv 0$ we have to rewrite
$Q$-derivatives at fixed $J$ in terms of $Q$-derivatives at fixed
$\phi$. This reads
\begin{eqnarray}\label{eq:fixedJ}
  \hspace{-.2cm}\0{\delta}{\delta Q_\Gs} :=
  \left.\0{\delta}{\delta Q}\right|_J =
  \left.\0{\delta}{\delta Q}\right|_\phi-
  \gamma^{\bf a}{}_{\bf b} 
  \0{\delta\Gamma^{,\bf b}}{\delta Q} G_{\bf ac} 
  \0{\delta }{\delta \phi_{\bf c}}
  \,, 
\end{eqnarray}
where we have used \eq{eq:Jphik} and \eq{eq:djphi}.  With the above
relations we arrive at the modified Slavnov-Taylor identity
\begin{subequations}\label{eq:STIR}  
  \begin{eqnarray} \label{eq:STIR1} \tilde \CW_{\Gs,I}[\phi,Q,R]\equiv
    0\,,
\end{eqnarray} 
with  
\begin{eqnarray}\label{eq:STIR2}
  \hspace{-.5cm}
  \tilde \CW_{\Gs,I}[\phi,Q,R]= \hat I_\Gs\, \tilde I[\phi,Q,R]+
  \widetilde{\delta I}_{\Gs,I}[\phi,Q,R]\,,
\end{eqnarray}
where the operator $\hat I_\Gs$ is defined in \eq{eq:brsthatIk}. In
\eq{eq:STIR2} it acts on functionals of the variable $\phi$.  With
\eq{eq:fixedJ} it can be written as
\begin{eqnarray} \nonumber &&\hspace{-.5cm}\hat I_{\Gs}= \left(
    \0{\delta \Gamma}{\delta \phi_{\bf a}}-\Delta S^{,\bf a}[
    G\s0{\delta}{\delta \phi}+\phi,R]
    +\Delta S^{,\bf a}[\phi,R]\right)\0{\delta}{\delta Q_\Gs}\,.   \\
  &&\label{eq:STIR3} \end{eqnarray}
\end{subequations}
The sum of the $\Delta S$-terms in \eq{eq:STIR3} give the part of $\Delta
S^{,\bf a}[ G\s0{\delta}{\delta \phi}+\phi,R]$ with at least one
$\phi$-derivative acting to the right. The operator $\hat I_{\Gs}$
defined in \eq{eq:STIR3} generates BRST transformations while keeping
the regulator term fixed.  Consequently the mSTI \eq{eq:STIR} entails
that such a BRST transformation of $\tilde I$ is given by the explicit
BRST variation due to $\widehat{\delta I}$.  The correlation function
$\widetilde{\delta I}$ is the expectation value of $ \widehat{\delta
  I}$ defined in \eq{eq:relR}. Similarly to \eq{eq:hatdeltaI} we write
\begin{eqnarray}\label{eq:sdeltahatI}
  \widehat{\delta I}_{\Gs,I}[J,\hat\phi,R]=e^{-\Delta S}
  \left(\Gs[\hat\phi]\, \hat I[J,\hat\phi]
  \right)e^{\Delta S} \,. 
\end{eqnarray}
\Eq{eq:sdeltahatI} entails that $ \widehat{\delta I}$ vanishes for
BRST-invariant correlation function $\tilde I$. In this case $\hat
I_\Gs \tilde I\equiv 0$. Finally we remark that the representation
\eq{eq:stigenalt} of $\tilde \CW$ translates into
\begin{eqnarray}\label{eq:STIRalt}
  \tilde \CW_{\Gs,I}=\left(\hat I_{\Gs}[\s0{\delta}{\delta Q_\Gs}]
    +(\hat I_{\Gs}[\s0{\delta}{\delta Q}]\,
    \Gamma)\right)\, 
  \tilde I+\widetilde{\delta I}\,, 
\end{eqnarray}
where we have used $\hat I[\s0{\delta}{\delta Q}|_J]\, W[J,R]=- \hat
I[\s0{\delta}{\delta Q}|_\phi]\,\Gamma[\phi,R] $, following from
\eq{eq:brsthatIkgen}, \eq{eq:dQWG} and \eq{eq:fixedJ}.  \smallstep

We proceed with elucidating the general identity \eq{eq:STIR} with two
examples. Firstly we discuss the standard regularisation with a
quadratic regulator $R^{\bf ab}\phi_{\bf a}\phi_{\bf b}$.  Inserting
this into \eq{eq:STIR} we are led to
\begin{eqnarray}\label{eq:STIRstand}  
  \tilde \CW_{\Gs,I}= \left(\0{\delta \Gamma}{\delta \phi_{\bf a}}-  
    2 R^{\bf ba}\, G_{\bf cb} \0{\delta}{\delta \phi_{\bf  a}}\right)
  \0{\delta \tilde I}{\delta Q_\Gs^{\bf a}}+\widetilde{\delta I}\,. 
\end{eqnarray}
The second important example is provided by the mSTI for $\Gamma$. It
can be read off from the alternative representation for $\tilde
\CW_{\Gs,I}$ in \eq{eq:STIRalt} for $I=1$ ($\hat I=1$) leading to
\begin{eqnarray}\label{eq:mSTIGR}
  \hspace{-.3cm}\left(  \0{\delta \Gamma}{\delta \phi_{\bf a}}
    -\Delta S^{,\bf ba}[
    G\s0{\delta}{\delta \phi}+\phi,R]\,G_{\bf cb}
    \,\0{\delta}{\delta \phi^{\bf c}}\right)
  \0{\delta\Gamma}{\delta Q^{\bf a}}\equiv 0\,. 
\end{eqnarray}
We emphasise that the $Q$-derivative in \eq{eq:mSTIGR} is that at
fixed $\phi$ and not at fixed $J$. It is also possible to derive it
directly from \eq{eq:STIR} with $I=\phi$. For the standard regulator
the mSTI \eq{eq:mSTIGR} reads \cite{Ellwanger:1994iz}
\begin{eqnarray}\label{eq:brstIk} 
  \0{\delta \Gamma}{\delta \phi_{\bf a}}\,
  \0{\delta \Gamma}{\delta Q^{\bf a}}-  
  2 R^{\bf ab}\,\0{\delta \Gamma^{,{\bf c}}}{\delta Q^{\bf b}}
  \, G_{\bf ca}=0\,.
\end{eqnarray} 
The terms proportional to derivatives of $\Delta S'$ cancel in
\eq{eq:brstIk}. 
\smallstep

\subsubsection{Flows and alternative representations}\label{sec:flow&alt}

The compatibility of \eq{eq:STIR} with the flow is
ensured by the flow \eq{eq:flowtIk} for $\tilde \CW_I$,
\begin{eqnarray}\label{eq:flowsti}
  \left(\partial_t+\Delta S_2\right)\tilde \CW_I=0\,,  
\end{eqnarray} 
for the effective action and quadratic regulator see \cite{D'Attanasio:1996jd,Litim:1998qi,Litim:2002ce,Freire:2000bq}

\Eq{eq:flowsti} implies that a truncated solution to $\tilde I_{\rm
  STI}\equiv 0$ stays a solution during the flow if the flow is
consistent with the truncation. Then it suffices to solve the mSTI for
the initial condition $\Gamma[\phi,Q,R_{\rm in}]$, $\tilde
I[\phi,Q,R_{\rm in}]$. However, the search for consistent truncations
is intricate as \eq{eq:STIR} involves loop terms. It is worth
searching for alternative representations of the mSTI \eq{eq:STIR}
that facilitate the construction of such truncations.  For the sake of
simplicity we discuss this for the mSTI \eq{eq:brstIk} for the
effective action in the presence of quadratic regulator terms.  The
generalisation to correlation functions $\tilde I$ and general $\Delta S$ is
straightforwardly done by substituting the correlation function $\Gamma$ with
$\tilde I$ (leaving the $\Gamma$-dependence of $\hat I_\Gs$ unchanged)
as well as the quadratic regulator $R^{\bf ab}$ with a general $R$. 
We can cast \eq{eq:brstIk} into an algebraic form using the fact that
$-R$ serves as a current for $G$:
\begin{eqnarray}\label{eq:brstIkalg}  
  \tilde \CW_{\Gs,1}= \0{\delta \Gamma}{\delta \phi_{\bf a}}\,
  \0{\delta \Gamma}{\delta Q^{\bf a}}+
  2 R^{\bf ab}\,\0{\delta \Gamma^{,\bf c}}{\delta Q^{\bf b}}\, 
  \0{\delta \Gamma}{\delta 
    R^{\bf ca}}\,. 
\end{eqnarray}
The algebraic form of the STI \eq{eq:brstIkalg} can be used to ensure
gauge invariance in a given non-trivial approximation to $\Gamma$ by
successively adding explicitly $R$-dependent terms. Such a procedure
accounts for gauge invariance of classes of resummed diagrams. We add
that in most cases it implicitly dwells on an ordering in the gauge
coupling. We also remark that \eq{eq:brstIkalg} seems to encode a
preserved symmetry. This point of view becomes even more suggestive if
introducing anti-fields \cite{Igarashi:1999rm,Igarashi:2001mf}. Note
that in general the related symmetry transformation is inherently 
non-local. \smallstep

\Eq{eq:brstIkalg} constitutes an ordering in $R$. This can be made
explicit by fully relying on the interpretation of $R$ as a current.
There is a simple relation between $Q$-derivatives and
$J$-derivatives: BRST variations of the fundamental fields $\varphi$
are at most quadratic in the fields, see \eq{eq:brstvar}.  Hence, the
$\varphi$-order of the BRST transformation of a composite field $\Gs
\hat\phi$ is at most increased by one. Therefore, the source term
$Q^{\bf a}\,s\hat\phi_{\bf a}$ can be absorbed into a redefinition of
$J^{\bf a}$,
\begin{eqnarray}\label{eq:Q->JR}
  &&\hspace{-.0cm}
  J^{\bf a}\hat\phi_{\bf a}-R^{\bf ab}\hat\phi_{\bf a}\hat\phi_{\bf b}
  +Q^{\bf c}\,{\Gs}\hat\phi_{\bf c}=\\\nonumber 
  &&\hspace{-.5cm}
  \left(J^{\bf a}+Q^{\bf c}\,(\Gs\hat\phi_{\bf c})_{\hat\phi=0}^{,\bf a}
  \right)\hat\phi_{\bf a}
  -\left(R^{\bf ab}-\s012 Q^{\bf c} (\Gs\hat\phi_{\bf c})^{,\bf ba}\right) 
  \hat\phi_{\bf a}\hat\phi_{\bf b}\,.
\end{eqnarray} 
The tensors $(\Gs\hat\phi_{\bf c})^{,\bf ab}$ are the structure constants
of the gauge group as can be seen within the example of the fundamental
fields \eq{eq:fundfields} and their BRST variation \eq{eq:brstvar}.
With \eq{eq:Q->JR} we can rewrite $Q$-derivatives of $W$ and $\Gamma$
in terms of $J$,$R$-derivatives of $W$ and $R$-derivatives and fields
$\varphi$ for $\Gamma$. The key relation is
\begin{eqnarray}\label{eq:QR}
  \0{\delta \Gamma}{\delta Q^{\bf a}}=  
  -\Gs \phi_{\bf a}+ \s012 (\Gs\phi_{\bf a})^{,\bf cb}
  \0{\delta \Gamma}{\delta R^{\bf bc}}\,,
\end{eqnarray}
where we also have to admit source terms with source $-R$ for $A_i
C_\alpha$ and $C_\alpha C_\beta$. With \eq{eq:QR} we can substitute
the $Q$-derivatives in \eq{eq:brstIkalg} and eliminate $Q$. Then the
correlation function $\tilde \CW_{\Gs,1}[\phi,R]=\tilde
\CW_{\Gs,1}[\phi,0,R]$ reads
\begin{eqnarray}\nonumber 
  &&\hspace{-1.1cm}\tilde \CW_{\Gs,1}[\phi,R]=
  -\0{\delta \Gamma}{\delta \phi_{\bf a}}\left( \Gs\phi_{\bf a} + 
    \s012 (\Gs\phi_{\bf b})^{,\bf ed}\0{\delta \Gamma}{\delta 
      R^{\bf de}}\right)\\
  &&\hspace{-.3cm}-2 R^{\bf ab}\left( (\Gs\phi_{\bf b})^{,\bf c}+ 
    \s012 (\Gs\phi_{\bf b})^{,\bf ed}
    \0{\delta \Gamma^{,\bf c}}{\delta R^{\bf de}}
  \right)\0{\delta \Gamma}{\delta R^{\bf ca}}\,.
\label{eq:purealg}\end{eqnarray}
At $R=0$ the second line vanishes and we deal with the standard STI.
The parameterisation \eq{eq:STIR} and \eq{eq:brstIkalg} of the STI
emphasise the gauge symmetry and are certainly convenient within a
coupling expansion. The parameterisation \eq{eq:brstIkalg} and
\eq{eq:purealg} naturally relate to the 'importance-sampling' relevant
in the flow equation.  The latter, \eq{eq:purealg}, requires no
BRST source terms and hence reduces the number of auxiliary
fields/terms.  \smallstep

The derivation of \eq{eq:purealg} highlights the fact that
\eq{eq:STIRstand} also constitutes the Slavnov-Taylor identity for the
2PI effective action, e.g.\
\cite{Arrizabalaga:2002hn,Carrington:2003ut}. To that end we restrict
ourselves to ${\bf a}=a$ and quadratic regulators $R^{ab}$. With the
substitution $R^{ab}\to -J^{ab}$ we are led to the Slavnov-Taylor
identity for $\Gamma[\phi_a,Q,-J^{ab}]$. More explicitly we have
\begin{eqnarray}\label{eq:2PIJ}
  J^{ab}=-R^{ab}\,, 
\end{eqnarray} 
and 
\begin{eqnarray}\label{eq:bJ}
  J^{\bf a}\phi_{\bf a}=J^a\phi_a+J^{ab}\phi_a\phi_b\,, 
\end{eqnarray} 
with the implicit definition $\phi_{\bf
  a}=(\phi_a,\phi_{bc}=\phi_b\phi_c)$. We perform a second Legendre
transformation with
\begin{eqnarray}\nonumber 
  &&\hspace{-.8cm}\Gamma_{\rm 2PI}[\phi_a,\phi_{ab},Q]\\
  &&\hspace{-.3cm}=\sup_J
  \left(J^{ab}\phi_{ab}+\Gamma[\phi_a,Q,R^{ab}=-J^{ab}]\right) \,, 
\label{eq:G2PI}\end{eqnarray}
leading to $\s0{\delta \Gamma_{\rm 2PI}}{\delta \phi_{ab}}=J^{ab}$ and
$\phi_{ab}=G$. Note that $\Gamma[\phi_a,Q,R^{ab}]$ already includes the
standard subtraction $-J^{ab}\phi_a\phi_b$. We arrive at
\begin{eqnarray}\label{eq:STI2PI}
  \0{\delta \Gamma_{\rm 2PI}}{\delta \phi_a}\,
  \0{\delta \Gamma_{\rm 2PI}}{\delta Q^a}+  
  2 \0{\delta \Gamma_{\rm 2PI}}{\delta \phi_{ab}}\,
  \0{\delta \Gamma_{\rm 2PI}{}^{,c}}{\delta Q^b}\, \phi_{ca}\equiv 0. 
\end{eqnarray} 
The last term on the lhs of \eq{eq:STI2PI} accounts for the BRST
variation of $\phi_{ab}$ that derives from the BRST variations of its
field content $\hat\phi_a\hat\phi_b$. The BRST variation of
$\hat\phi_a\hat\phi_b$ can be added with a source term $Q^{ab} \Gs
(\hat\phi_a\hat\phi_b)$ in the path integral leading to $\Gamma_{\rm
  2PI}=\Gamma_{\rm 2PI}[\phi_a,\phi_{ab},Q_a,Q_{ab}]$.  Then we have
\begin{eqnarray}\label{eq:phiabrst}
  \0{\delta \Gamma_{\rm 2PI}}{\delta Q^{ab}}= 
  \0{\delta \Gamma_{\rm 2PI}{}^{,c}}{\delta Q^b}\, \phi_{ca}+\phi_{bc}
  \0{\delta \Gamma_{\rm 2PI}{}^{,c}}{\delta Q^a}\,. 
\end{eqnarray} 
\Eq{eq:phiabrst} and the symmetry property $\phi_{ab}=\gamma^c{}_b
\phi_{ca}$ lead to \eq{eq:STI2PI}.  Collecting the fields into a
super-field $\phi_{\bf a} =(\phi_a,\phi_{bc})$, and $Q^{\bf
  a}=(Q^a,Q^{bc})$ with $\Gamma_{\rm 2PI}=\Gamma_{\rm 2PI}[\phi_{\bf
  a},Q^{\bf a}]$, we get an appealing form of the STI \eq{eq:STI2PI}
 \begin{eqnarray}\label{eq:brst2PI}
   \0{\delta \Gamma_{\rm 2PI}}{\delta \phi_{\bf a}}\,
   \0{\delta \Gamma_{\rm 2PI}}{\delta Q^{\bf a}}=0\,.
\end{eqnarray} 
In its spirit \eq{eq:brst2PI} is close to the mSTI written as a master
equation \cite{Igarashi:1999rm,Igarashi:2001mf}. As in \eq{eq:brst2PI}
the master equation emphasises the algebraic structure of the mSTI but
hides the symmetry-breaking nature of the identities. Nonetheless
algebraic identities are useful if constructing consistent truncations
as well as discussing minimal symmetry breaking due to quantisation in
the sense of Ginsparg-Wilson relations \cite{Ginsparg:1981bj}. \smallstep

As in \eq{eq:purealg} we can absorb $Q^a$-derivatives with help of
\eq{eq:Q->JR}, \eq{eq:QR}.  As the source $Q$ is a spectator of the
Legendre transformation \eq{eq:G2PI} we have $\s0{\delta
  \Gamma}{\delta Q}= \s0{\delta \Gamma_{\rm 2PI}}{\delta Q}$ and
\eq{eq:QR} reads for the 2PI effective action
\begin{eqnarray}
  \0{\delta \Gamma_{\rm 2PI}}{\delta Q^a}= -\left(\Gs \phi^a+\s012 
    (\Gs\phi^a)^{,bc}\phi_{bc}\right)\,, \label{eq:QR2PI}
\end{eqnarray}
where we have used that $R^{ab}=-J^{ab}$ and hence $ \s0{\delta
  \Gamma}{\delta R^{ab}}=-\phi_{ab}$.  Using \eq{eq:QR2PI} we arrive
at
\begin{eqnarray}\label{eq:brst2PI0}
\hspace{-.3cm}  -\0{\delta \Gamma_{\rm 2PI}}{\delta \phi_{a}}
  \left(\Gs\phi^a+(\Gs\phi^a)^{,bc}\phi_{bc}\right)+
  \0{\delta \Gamma_{\rm 2PI}}{\delta \phi_{ab}}
  \0{\delta \Gamma_{\rm 2PI}}{\delta Q^{ab}}=0\,.
\end{eqnarray} 
The BRST variation of $\hat\phi_{ab}$ involves
$\hat\phi_{cd}\hat\phi_e$ and $Q^{ab}$ is a source for a specific
tensor structure $\CT^{abcde} \phi_{cd}\phi_e$. Within regularisation
of the 2PI effective action that regularises three point functions the
source $Q^{ab}$ can be eliminated analogously to \eq{eq:purealg}. This
is an interesting option for $N$PI regularisations of gauge theories, in
particular in view of consistent approximations \cite{Mottola:2003vx,Arrizabalaga:2002hn,Carrington:2003ut,Berges:2004pu}. \smallstep

We close this section with a short summary of the derivation of STIs
without the use of BRST transformations.
To that end we integrate out the auxiliary field $\lambda$ and use the
classical gauge-fixed action \eq{eq:brstaction3}. In view of the
auxiliary nature of the ghost fields we derive identities that
describe the response of general correlation functions to
(infinitesimal) gauge transformations $\Gg_\omega$ of the physical
fields, the gauge field and possible matter fields. Gauge-invariant
correlation functions $I,\tilde I$ are invariant under these
transformations.
\begin{eqnarray}\label{eq:gfundfields}
  (\Gg_\omega \hat\varphi)_a=( (D \omega)_i\,,\, [\omega,  
  C]_\alpha\,,\, [\omega,\bar C]_\alpha )\,. 
\end{eqnarray}
The linear operator $\Gg$ is bosonic as distinguished to $\Gs$.  It
can be cast into the form \eq{eq:brstder} as a functional derivative
operator $\Gg=(\Gg \hat \varphi_{\bf a})\s0{ \delta}{\delta \hat
  \varphi_{\bf a}}$.  The related generator $\hat I_{\Gg}$ reads
\begin{eqnarray}\label{eq:hatIg}
  \hat I_\Gg=\left(
    J^{\bf a} (\Gg\hat\phi)_{\bf a}-
    (\Gg S[\hat\phi])\right)_{\hat\phi=\s0{\delta}{\delta J}}\,, 
\end{eqnarray} 
leading to the STI \eq{eq:stigen} for $\CW_{\Gs,I}$. Restricting
ourselves to $J$-independent $\hat I$'s \eq{eq:stigen3} reads
\begin{eqnarray}\label{eq:deltaI}
\widehat{\delta I}=(\Gg\hat I[\hat\phi])[\hat\phi=\s0{\delta}{\delta J}]\,.
\end{eqnarray}
In the presence of a regulator term the generator of symmetry
transformations turns into
\begin{eqnarray}\label{eq:hatIgR}
  \hat I_\Gg=\left(
    J^{\bf a} (\Gg\hat\phi)_{\bf a}-
    (\Gg (S+\Delta S))[\hat\phi]\right)_{\hat\phi=\s0{\delta}{\delta J}}\,, 
\end{eqnarray} 
leading to the mSTIs \eq{eq:stigenR} and \eq{eq:STIR} for
$\CW_{\Gg,I}$ and $\CW_{\Gs,I}$ respectively. We close this section
with exemplifying the mSTI $\tilde W_{\Gg,I}$ at $I=1$ and the
standard flow. Then, with the alternative representation
\eq{eq:STIRalt} we are led to \cite{Freire:2000bq}
\begin{eqnarray}\nonumber 
  &&\hspace{-1cm}
  \Gg \Gamma[\phi,R]= \left( \Gg \Bigl(
    \0{1}{2\xi} \CF^\alpha\CF^\alpha+\Delta S\Bigr)
    [G\s0{\delta}{\delta \phi}+\phi]\right)\\ 
  &&\hspace{-.8cm}-\Gg\Delta S'[\phi,R]-\left( 
    \Gg(\bar C_\alpha \0{\partial \CF^\alpha}{\partial A_i}
    D_i^\beta C_\beta)[G\s0{\delta}{\delta \phi}+\phi]\right) .
\label{eq:STIgRex}\end{eqnarray}
The right hand side of \eq{eq:STIgRex} reproduces the gauge variation
of the classical action $\Gg S[\phi]$ as well as loop terms.  The
highest loop order (in the full propagator) is given by the highest
order of the field $\phi$ in the gauge fixing term and the ghost term
in the classical action as well as the regulator term.  For linear
gauges and $\hat\phi=\hat\varphi$ the modified STI \eq{eq:STIgRex}
involves one loop (gauge fixing, ghost term) and two loop terms (ghost
term) apart from the regulator-dependent terms.  Thus a purely
algebraic form of the mSTI \eq{eq:STIgRex} can be achieved for
regulator terms with $R$ involving $R^{ab}$ and $R^{abc}$.\smallstep

\subsection{Gauge-invariant flows}\label{sec:gaugeinv} An interesting
option for flows in gauge theories is the construction of (partially)
gauge-invariant flows. The gain of such formulations is twofold.
Firstly they allow for a more direct computations of physical
observables. Observables are gauge-invariant as opposed to Greens
functions in gauge-fixed formulations. Secondly one can hope to avoid
the subtleties of solving the symmetry relations in the presence of a
regulator. However, gauge-invariant formulations come to a price that
also has to be evaluated: if the corresponding flows are themselves
far more complicated than the standard gauge-fixed flows the benefit
of no additional symmetry relations is, at least partially, lost.
Also, gauge invariance does not rule out the persistence of
non-trivial symmetry relations, mostly formulated in the form of
Nielsen identities or, alternatively, in the form of specific
projections of the general Dyson-Schwinger equations valid within such
a setting. \smallstep

In the present work we concentrate on gauge-invariant flows formulated
in mean fields and the effective action $\Gamma_k$. An alternative
construction of gauge-invariant flows is based on the Wilsonian
effective action $S_{{\rm eff}_k}$, \eq{eq:S_k}, formulated in Wilson
lines and using gauge-covariant regulators. For details we refer the
reader to
\cite{Morris:1999px,Morris:2000fs,Arnone:2005fb,Morris:2005tv,Rosten:2005ep}
and references therein.

\subsubsection{Background field flows}
The first and most-developed gauge-invariant flow originates in
the use of the background field formalism. We couple the fundamental
fields to the currents, $\phi=\varphi$ with
\begin{eqnarray}\label{eq:fields} 
  \phi=(a_i, C_\alpha,\bar C_\alpha)\,,
\end{eqnarray} 
where the full gauge field is defined as 
\begin{eqnarray}\label{eq:A=a+barA}
  A=\bar A+a\,.  
\end{eqnarray} 
The gauge field $A$ is split into a background field configuration
$\bar A$ and a fluctuation field $a$ coupled to the current. BRST
transformations and gauge transformations are defined by
\eq{eq:brstvar} and \eq{eq:gfundfields} respectively at fixed
background field $\bar A$, $\Gs \bar A=\Gg \bar A=0$. Note that the
covariant derivative reads $D=D(a+\bar A)$. Therefore, the mSTIs
\eq{eq:STIR} for $\CW_{\Gs,I}$ and $\CW_{\Gg,I}$ persist.  Within
appropriate gauges, e.g.\ the background field gauge $\CF=D(\bar A)
a$, there is an additional symmetry: the action \eq{eq:brstaction3} is
invariant under a combined gauge transformation of the background
field $\bar A \to \bar A+D(\bar A)\omega $ and the fluctuation field
$a\to [\omega,a]$. This invariance follows by using that the
fluctuation field $\phi$ in \eq{eq:fields} as well as the covariant
derivatives $D(A)$ and $D[\bar A$ transform as tensors under this
combined transformation. Defining background field transformations
\begin{eqnarray}\label{eq:bGg}
  \bar\Gg_\omega (\varphi\,,\,\bar A)=
  (-D(\bar A)\omega\,,\, 0\,,\, 0\,,\, D(\bar A)\omega)  \,,  
\end{eqnarray}
the transformation properties under the combined transformation are
summarised in
\begin{equation}\label{eq:g+bg}
  (\Gg+\bar\Gg)_\omega (\varphi\,,\,D(A)\,,\,D(\bar A)\,)
  =[\omega\,,\, (\varphi\,,\,D(A)\,,\,D(\bar A)\,)]\,.
\end{equation}
with $\Gg$ defined in \eq{eq:gfundfields}. As the action $S$ in
\eq{eq:brstaction3} with $\CF=D(\bar A)$ or similar choices can be
constructed from $(\varphi\,,\,D(A)\,,\,D(\bar A)\,)$ this leads
us to
\begin{eqnarray}\label{eq:g+bargS}
  (\Gg+\bar \Gg) S[\phi,\bar A]=0\,, 
\end{eqnarray} 
Then, the corresponding effective action $\Gamma[\phi,\bar A]$ is
invariant under the above transformation, in particular we define a
gauge-invariant effective action with
\begin{eqnarray}\label{eq:bGa}
  \Gamma[A]=\Gamma[\phi=0,A]\,.
\end{eqnarray} 
We have $ (\Gg+\bar \Gg) \Gamma[\phi,\bar A]=0$, where $\Gg,\bar\Gg$
act on $\phi=\varphi$ according to \eq{eq:gfundfields} and
\eq{eq:bGg}. This implies in particular $\Gg\Gamma[A]=0$.  The gauge
invariance of $\Gamma[\phi,A]$ persists in the presence of a regulator
if $\Delta S[a,R(\bar A)]$ is invariant under the combined
transformation of $a$ and $\bar A$. This is achieved for regulators
$R$ that transform as tensors under gauge transformations $\bar A \to
\bar A+D(\bar A)\omega $. This amounts to the definition of a
background field dependent $R(\bar A)$ with
\begin{eqnarray}\label{eq:barR}
  \bar \Gg R(\bar A)=[\omega\,,\, R(\bar A)]\,.
\end{eqnarray} 
For example, standard flows follow with the regularisation $\Delta
S=R^{ij}(\bar A) a_i a_j +R^{\alpha\beta}(\bar A) C_\alpha C_\beta$.
The invariance property $(\Gg+\bar\Gg)\Delta S=0$ follows immediately
from \eq{eq:g+bg} and \eq{eq:barR}. The relation \eq{eq:barR} is e.g.\
achieved for regulators in momentum space depending on covariant
momentum $D(\bar A)$.  Correlators $\tilde I$ still satisfy the
modified STI \eq{eq:STIR}, but additionally there is a modified STI
related to the background field gauge transformations \eq{eq:bGg}. The
related generator is
\begin{eqnarray}\label{eq:hatIbarg}
  \hat I_\Gg=\left(
    J^{\bf a} (\bar\Gg\hat\phi)_{\bf a}-
    (\bar\Gg (S+\Delta S))[\hat\phi]
  \right)_{\hat\phi=\s0{\delta}{\delta J}}\,, 
\end{eqnarray}
leading to mSTIs \eq{eq:stigenR} and \eq{eq:STIR} for $\CW_{\bar
  \Gg,I}$. For the effective action ($I=1$) the mSTI reads
\begin{eqnarray}\nonumber 
  \hspace{-.7cm}
  \bar\Gg \Gamma[\phi,R]&=& \left(\0{1}{2\xi} \bar\Gg (\CF^\alpha\CF^\alpha)
    [G\s0{\delta}{\delta \phi}+\phi]\right)\\ 
  &&-\left( 
    \bar\Gg(\bar C_\alpha \0{\partial \CF^\alpha}{\partial A_i}
    D_i^\beta C_\beta)[G\s0{\delta}{\delta \phi}+\phi]\right) \,.
\label{eq:STIbgRex}\end{eqnarray}
Adding \eq{eq:STIgRex} and \eq{eq:STIbgRex} we arrive at
\begin{eqnarray}\label{eq:g+bargG}
  (\Gg+\bar \Gg) \Gamma[\phi,\bar A,R]=0\,.
\end{eqnarray} 
The derivation makes clear that, despite background gauge
invariance \eq{eq:g+bargG}, the effective action $\Gamma[\phi,\bar A]$
still carries the BRST symmetry \eq{eq:STIR} displayed in $\CW_\Gs$ or
$\CW_\Gg$, where the background field is a spectator $s\bar A=0$. In
other words, the non-trivial relations between $N$-point functions of
the fluctuation field are still present. However, for $N$-point
functions in the background field they play no r$\hat{\mbox{o}}$le
which has been used for simplifications within loop computations.
\smallstep

Therefore it is tempting to use these features for the construction of
gauge-invariant flows. General flows within such a setting are still
provided by \eq{eq:flowtIk}. In particular with \eq{eq:flowGk} we
arrive at the flow of $\Gamma_k[A]$ as
\cite{Reuter:1992uk,Reuter:1993kw,Reuter:1997gx,Pawlowski:1996ch,Litim:1998wk,Pawlowski:2001df,Pawlowski:2002eb,Litim:2002ce,Freire:2000bq,Gies:2002af,Gies:2003ic,Braun:2005uj}
\begin{equation}
  \dot\Gamma_k[A]=(\Delta
  S[G\s0{\delta}{\delta \phi}
  +\phi\,,\,\dot R(A)])_{\phi=0}
  -\Delta S'[0,\dot R(A)]\,.
\label{eq:bflow}
\end{equation}
It has already been discussed in
\cite{Pawlowski:2002eb,Pawlowski:2001df,Litim:2002ce} that the flow
\eq{eq:bflow} is not closed as it depends on
\begin{eqnarray}\label{eq:flucprop}
  \0{\delta^2 \Gamma_k[0,\bar A,R]}{\delta \phi^2}\,,
\end{eqnarray} 
the propagator of the fluctuation field, and possibly higher
derivatives w.r.t.\ $\phi$ evaluated at vanishing fluctuation field
$\phi=0$. The lhs of \eq{eq:bflow} cannot be used to compute this
input as it only depends on $\bar A=A$. Moreover, as has been stated
above, these $N$-point functions still satisfy the modified
Slavnov-Taylor identities discussed in the last section. The
differences between $\Gamma^{(2)}[A]$ and the fluctuation propagator
\eq{eq:flucprop} become important already at two loop. The correct
input \eq{eq:flucprop} at one loop was used to compute the universal
two loop $\beta$-function which cannot be reproduced by using
$\Gamma^{(2)}[A]$ \cite{Pawlowski:2002eb,Pawlowski:2001df}. Still one
can hope that qualitative features of the theory are maintained in
such a truncation. Then, a measure for the quality of such a
truncation is given by the difference between a derivative w.r.t.\
$\bar A$ and one w.r.t.\ $a$ of the effective action. This relation
reads \cite{Pawlowski:2001df,Pawlowski:2002eb,Litim:2002ce}
\begin{eqnarray}\nonumber 
  &&\hspace{-.4cm} \left(\0{\delta}{\delta \bar A}
    -\0{\delta}{\delta a}\right)\Gamma[\phi,\bar A,R]=
  \left\langle\left(\0{\delta}{\delta \bar A}
      -\0{\delta}{\delta \hat a}\right) (S+\Delta S)\right\rangle\\ 
  &&\hspace{.1cm}=
  \left(\left(\bigl(\0{\delta}{\delta \bar A}
      -\0{\delta}{\delta \hat a}\bigr) (S+\Delta S)\right)
    [G\0{\delta}{\delta a}
    +\phi,\bar A]\right)\,,
\label{eq:Niback}\end{eqnarray} 
and can be understood as a Nielsen identity. \Eq{eq:Niback} relates
Green functions of the background field with that of the fluctuation
field.  The latter satisfy mSTIs whereas the former transform as
tensors under gauge transformations reflecting gauge invariance.
Hence, \eq{eq:Niback} encodes the mSTIs. Note also that the background
field dependence stemming from the regulator should be understood as a
parameter dependence and not as a field dependence \footnote{For
  infrared diverging regulator $R(\bar A)$ even the computation of the
  one loop $\beta$-function requires a subtraction of the field
  dependence of $R(\bar A)$
  \cite{Pawlowski:2001df,Pawlowski:2002eb,Litim:2002ce}.}. An
improvement of the current results in gauge theories 
\cite{Reuter:1992uk,Reuter:1993kw,Reuter:1997gx,Reuter:1996be,Pawlowski:1996ch,Bonini:1998ec,Falkenberg:1998bg,Litim:1998wk,Pawlowski:2001df,Pawlowski:2002eb,Litim:2002ce,Freire:2000bq,Gies:2002af,Gies:2003ic,Braun:2005uj} 
requires an implementation of the Nielsen identity \eq{eq:Niback}
beyond perturbation theory. \smallstep

It is possible to enhance background field flows to fully
gauge-invariant flows with standard STIs by
identifying the background field with a dynamical field. There are two
natural choices: $\bar A=\hat A$ \footnote{This choice can be only 
used in the regulator.} and $\bar A=\langle \hat A\rangle=A$.
The latter leads to the definition of the effective action as a higher
order Legendre transform. Then we have additional terms to those
\eq{eq:flowtIk} as
\begin{eqnarray}\label{eq:leghigh} 
  \Gamma^{,a}= J^a-\left\langle\0{\delta (S +
      \Delta S)}{\delta \bar A}
  \right\rangle\,.
\end{eqnarray}
With \eq{eq:leghigh} we get additional terms in the relations between
$\phi$-derivatives of $\Gamma$ and $J$-derivatives of $W$.
\Eq{eq:leghigh} is actually implementing the Nielsen identity
\eq{eq:Niback} on the level of the Legendre transformation. This
entails that in particular the basic relations \eq{eq:Jphik} and
\eq{eq:propk} receive modifications originating in \eq{eq:leghigh}.
As an example we study the standard flow for the effective action
which reads
\begin{eqnarray}\nonumber 
\hspace{-1cm}  \dot \Gamma[A,\phi]&=& \dot R^{ab} W_{,ab}\\
  &=&\dot R^{ab}G_{ab}+
  \left(\0{\delta (S +\Delta S)}{\delta \bar A} {\rm -terms}\right)\,,
\label{eq:gaugestand}\end{eqnarray}
where the propagator $G$ is defined with $G=1/(\Gamma^{(2)}+\Delta
S^{(2)})$.  The propagator $G$ of the dynamical field transforms as a
tensor under gauge transformations reflecting gauge invariance.
However, it can be shown in a perturbative loop expansion that
effectively the flow equation can be rewritten as that in the
background field formalism: the effective propagator
$W^{(2)}+(W^{(1)})^2$ behaves as that of the fluctuation field in the
background field formulation. This is already indicated in
\eq{eq:leghigh}. The correction terms involve the same correlation functions 
already relevant in the Nielsen identity \eq{eq:Niback}.  So still we
deal with non-trivial symmetry identities. Nonetheless the above
formulation furthers the knowledge about truncation schemes that
expand about $\Gamma^{,a}= J^a$, or alternatively about
$(\s0{\delta}{\delta \bar A} -\s0{\delta}{\delta a})\Gamma[\phi,\bar
A,R]=0$. Details shall be provided elsewhere. \smallstep

The other suggestion $\bar A=A$ relates to the use of a regulator term $\Delta
S[A,R(A)]$. Such a regulator term can be written as $\widehat{\Delta
  S}[A,\hat R]=\Delta S[A,R(A)]$, where $\hat R^{a_1\cdots a_n}=\Delta
S^{a_1\cdots a_n}[0,R(0)]/(n!)$ is the $n$th expansion coefficient in
a Taylor expansion of $\Delta S[A,R(A)]$ in the gauge field $A$. This
flow is covered by the general flow \eq{eq:flowtIk} and involves all
loop orders in the full propagator. Again this effectively reduces to
the background field flow and comes at the expense of an infinite
series of loop terms in the flow. In this context we remark that the
latter set-back is avoided within the Polchinski equation. This
follows in the present setting with \eq{eq:S_k} and the flow
\eq{eq:flowIk} for the Schwinger functional.  \smallstep

\subsubsection{Geometrical effective action} \step 

We have seen in the last section that the flow of the gauge-invariant
effective action within the background field formulation 
is not closed. In the process of curing this problem
we encounter the persistence of non-trivial symmetry relations,
conveniently summarised in \eq{eq:Niback}. Both aspects originate in
the fact that the sources are coupled to fields that do not transform
trivially under gauge or BRST transformations.  Hence the question
arises whether one can do better. Within the framework of the
geometrical or Vilkovisky-DeWitt effective action the fields $\phi$
coupled to the sources are scalars under gauge transformations.

Then, gauge-invariant flows can be formulated
\cite{Branchina:2003ek,Pawlowski:2003sk}. We do not want to go in the
details of the general construction that can be found in
\cite{Pawlowski:2003sk}. The configuration space is provided with a
connection $\Gamma_V$ (Vilkovisky connection) which is constructed
such that the disentanglement between gauge fibre and base space is
maximal. The gauge fields $A_i$ are substituted by geodesic normal
fields $\phi_i$ that are tangent vectors at a base point (background
field) $\bar A$. As a consequence the geodesic fields $\phi_\alpha$
tangential to the fibre drop out of the path integral, only the fields
$\phi_A$ tangential to the base space remain and are gauge-invariant.
This construction is lifting up the relation between fluctuation field
and background field \eq{eq:A=a+barA}. The linear background relation
can be read as the limit in which the connection $\Gamma_V$ is
neglected.  The full relation reads schematically
\begin{eqnarray}\label{eq:geomabarA}
  \phi_i=A_i-\bar A_i+{\Gamma_{\!\mbox{\tiny $\rm V$}}\,}_i{}^{jk}\, 
  \phi_j \phi_k 
  +O(\phi_i^3)\,,
\end{eqnarray} 
with $\Gg \phi^A=0=\bar \Gg \phi^A$. This is used to construct a gauge-invariant effective action $\Gamma[\phi^A,\bar A,R]$ which is gauge-invariant under both sets of gauge transformations $\Gg$ and $\bar
\Gg$ \cite{Pawlowski:2003sk}. Again a gauge-invariant effective action
in one field can be defined as $\Gamma[A,R]=\Gamma[\phi=0,A,R]$. The
flows of $\Gamma[\phi,A,R]$ and $\Gamma[A,R]$ are given by
\eq{eq:flowGk} and \eq{eq:bflow} respectively, both being gauge-invariant flows. We still have a Nielsen identity equivalently to
\eq{eq:Niback}. In the underlying theory without regulator term it
reads
\begin{eqnarray}
\Gamma_{,i}+
\Gamma_{,a}\langle \phi^a{}_{;i}\rangle=0\,, 
\label{eq:Ni0}
\end{eqnarray}
where $\phi^a{}_{;i}$ stands for the covariant derivative with 
the Vilkovisky connection $\Gamma_V$. The related symmetry operator 
is provided by 
\begin{eqnarray}\label{eq:hatIN}
\hat I_\Gn=\0{\delta}{\delta \bar A}+J^{a}\hat\phi_a{}_{;i}
[\s0{\delta}{\delta J}]\,.
\end{eqnarray}
With \eq{eq:hatIR} this turns into 
\begin{eqnarray}\nonumber 
&&\hspace{-1.3cm}\hat I_\Gn=\0{\delta}{\delta \bar A}-
\0{\delta\Delta S}{\delta \bar A}[G\s0{\delta}{\delta \phi}+\phi]+
\0{\delta\Delta S}{\delta \bar A}[\phi] \\ 
&&\hspace{-.7cm} +\left(J^{a}-\Delta S^{,ab}[G\s0{\delta}{\delta \phi}] 
G_{bc}\s0{\delta}{\delta \phi}\right)\hat\phi_a{}_{;i}
[G\s0{\delta}{\delta \phi}+\phi]\,,
\label{eq:hatINR}\end{eqnarray}
in the presence of the regulator term. For standard flows the choice 
$\CW_{\Gn,1}$ in \eq{eq:STIRalt} reproduces  
the Nielsen identity derived in \cite{Pawlowski:2003sk}, 
\begin{eqnarray} 
\hspace{-.5cm}\Gamma_{k,i}=  
\s012 
G^{ab}\, R_{ba,i}+\bigl(
\Gamma_{k,a}-R_{ab}\, G^{bc}
\s0{\delta}{\delta \bar\phi^c}\bigr) \langle \phi^a{}_{;i}\rangle\,. 
\label{eq:geomNi}\end{eqnarray} 
For more details and its use within truncation schemes we refer to 
\cite{Pawlowski:2003sk}. The formalism discussed above provides gauge-invariant flows that are closely linked to the background field formalism 
(in the Landau-DeWitt gauge) as well as to standard Landau gauge. This comes 
with the benefit that results obtained in the latter can be partially used 
within the present formalism. Indeed the present setting can be used to 
improve the gauge consistency of these results. We hope to report on 
results for infrared QCD as well as gravity in near future.\smallstep  

To conclude, we have discussed the various possibility of defining
gauge-invariant flows and their relations to gauge-fixed formulations.
These relations come with the benefit that it allows to
start an analysis in the gauge-invariant formulations on the basis of
non-trivial results already achieved in gauge-fixed settings, one does
not have to start from scratch. \smallstep

\subsection{Chiral symmetry and anomalies}\label{sec:chiral}

We want to close this chapter with a brief discussion of FRG flows in
theories with symmetries that are flawed by anomalies on the quantum
level, e.g.\
\cite{Pawlowski:1996ch,Bonini:1997yv,Bonini:1998ec,Igarashi:1999rm}.
A more detailed account shall be given elsewhere.  In particular a
discussion of the chiral symmetry breaking requires a careful
investigation of chiral anomalies. The deformation of the chiral
symmetry from a general RG transformation has already been considered
in \cite{Ginsparg:1981bj}, and leads to the Ginsparg-Wilson relation
\footnote{The derivation in \cite{Ginsparg:1981bj} makes no use of the
  lattice.}. This has been emphasised in \cite{Igarashi:1999rm}.  A
discussion of chiral symmetry breaking requires a careful
investigation of chiral anomalies. Integrated anomalies are tightly
linked to topological degrees of freedom like instantons via the index
theorem. FRG methods have been shown to be sensitive to topological
degrees of freedom \cite{Reuter:1996be,Pawlowski:1996ch}, an
interesting quantum mechanical example can be found in
\cite{Zappala:2001nv}. In the present section we consider the gauge
field action \eq{eq:brstaction3} together with a Dirac action
\begin{eqnarray}\label{eq:fermact}
  S_D[\phi]=\bar\psi_a (\dr+m)^{ab}\psi_b\,,
\end{eqnarray}
with a possible mass term and $\phi=(A,C,\bar C,\psi,\bar\psi)$.  The
Dirac operator $\dr$ reads
\begin{eqnarray}\label{eq:diracop}
  \dr^{ab}=\left(\p+\CP \A\right)^{ab}
\end{eqnarray}
with the free Dirac operator $\p$ and a coupling to the gauge field
with a possible projection $\CP$ either proportional to the identity
$\CP=\id$, or projecting on right- or left-handed Weyl fermions
$\CP_{\pm}= \s0{1\pm\gamma_5}{2}$. Here we consider
\begin{eqnarray}\label{eq:left}
  \CP=\CP_+=\0{1+\gamma_5}{2}\,,\qquad m=0\,,
\end{eqnarray}
a theory with left-handed Weyl fermions coupled to a gauge field, and
free right-handed Weyl fermions. The symmetry transformation that
leaves the action \eq{eq:fermact} invariant is given by
\begin{eqnarray}\label{eq:gaugeleft}
  \hspace{-.5cm}
  \Gg_+ \phi=(\Gg_+ A\,,\,\Gg_+ C\,,\,\Gg_+ \bar C\,,\, \omega \CP_-
  \psi,-\bar\psi  \CP_+\omega)\,.
\end{eqnarray} 
The transformations \eq{eq:gaugeleft} cover both, BRST transformations
with $\Gg_+=\Gs$ with $\omega=C$, and $\Gg_+=\Gg$ with gauge
transformation parameter $ \omega$. Here we stick to $\Gg_+=\Gg$. The
chiral anomaly comes into play since the fermionic path integral
measure $d\psi \,d\bar\psi$ is not left invariant under the
transformation \eq{eq:gaugeleft}. In other words, \eq{eq:gaugeleft} is
not unitary. We quote the result
\begin{eqnarray}\label{eq:noninv}
  \Gg(d\psi\, d\bar\psi)=\omega^\alpha \CA_\alpha\, d\psi\, d\bar\psi\,, 
\end{eqnarray}
with infinitesimal variation $\omega^\alpha \CA_\alpha$. The
non-Abelian anomaly $\CA$ reads
\begin{eqnarray}\label{eq:nAanomaly}
  &&\hspace{-1.1cm}\CA^\alpha(x)=\0{1}{24 \pi^2} \epsilon_{\mu\nu\rho\sigma} 
\tr\,
    t^\alpha \left(\partial_\mu A_\nu
    F_{\rho\sigma}
    -\s012 A_\nu A_\rho A_\sigma\right)\,.
\end{eqnarray}
Then, the generator of gauge transformations $\hat I_{\Gg}$ in
\eq{eq:hatIg} receives a further contribution and reads
\begin{eqnarray}\label{eq:hatIgan}
  \hat I_\Gg=\left(
    J^{\bf a} (\Gg\hat\phi)_{\bf a}-
    (\Gg S[\hat\phi])-\CA[\hat\phi]\right)_{\hat\phi
    =\s0{\delta}{\delta J}}\,.  
\end{eqnarray} 
and with \eq{eq:hatIR} in the presence of the regulator term we arrive
at
\begin{eqnarray}\label{eq:hatIganR}
  \hat I_\Gg=\left(
    J^{\bf a} (\Gg\hat\phi)_{\bf a}-
    (\Gg (S[\hat\phi]+\Delta S))-
    \CA[\hat\phi]\right)_{\hat\phi=\s0{\delta}{\delta J}}\,.  
\end{eqnarray} 
\Eq{eq:hatIganR} can also be read off from \eq{eq:hatIgR} since the
anomaly term $\CA[\hat\phi]$ commutes with $\Delta S$. \smallstep

We conclude with briefly discussing the $U_A(1)$-anomaly relevant for
anomalous chiral symmetry breaking.  We restrict ourselves to standard
flows with quadratic regulator.  The Dirac action \eq{eq:fermact} with
$\CP=\id$ is invariant under global axial $U_A(1)$-transformations.
The related Noether current is derived from the $U_A(1)$
transformations of the fermions
\begin{eqnarray}\label{eq:u1a}
  \Gg_A\psi=\omega \gamma_5 \psi\,,\qquad 
  \Gg_A\bar\psi=\bar\psi \gamma_5 \omega\,.
\end{eqnarray} 
The rest of the fields transforms trivially with $\Gg_A A=\Gg_A
C=\Gg_A \bar C=0$.  The related anomaly reads 
\begin{eqnarray}\label{eq:u1}
  \CA=\0{1}{32\pi^2}\epsilon_{\mu\nu\rho\sigma}\tr\, F_{\mu\nu} 
F_{\rho\sigma}\,.
\end{eqnarray}
The anomalous Ward identity for the effective action, $\tilde\CW_{\Gg_A,1}$,
in the presence of the regulator reads
\begin{eqnarray}\nonumber 
  &&\hspace{-1.5cm}(\Gg_A\phi)_a \Gamma^{,a} +(\Gg_A (S_D+\Delta S))
  [G\s0{\delta}{\delta \phi}+\phi]\\
  &&\hspace{1.9cm}=(\CA[G\s0{\delta}{\delta \phi}+\phi])\,.
\label{eq:anWI}\end{eqnarray}
The space time integral of \eq{eq:anWI} produces the (expectation
value of the) topological charge on the rhs, as well as the analytical
index of the modified Dirac operator on the lhs. In
\cite{Pawlowski:1996ch} it has been shown that the number of zero
modes stays the same for regulators with chiral symmetry. The 
chiral anomaly has been investigated in \cite{Bonini:1997yv}. In
general the lhs of \eq{eq:anWI} is computed directly from the
effective action.  Accordingly we can use \eq{eq:anWI} for testing the
potential of given truncations to the effective action for
incorporating the important topological effects. Additionally its
provides non-trivial relations between the couplings. For example, the
leading order effective action derived in \cite{Pawlowski:1996ch}
satisfies \eq{eq:anWI} up to sub-leading terms (in $1/k$).
\Eq{eq:anWI} can be used to determine coefficients and form of these
sub-leading terms, in particular in view of CP-violating effects.
\step

\setcounter{equation}{0}
\section{Truncation schemes and optimisation}\label{sec:optimalatwork}

The reliability of results obtained within the functional RG rely on
the appropriate choice of a truncation scheme for the physics under
investigation, as well as an optimisation of the truncation with the
methods introduced in section~\ref{sec:optimal}. The truncation has to
take into account all relevant operators or vertices. In theories with
a complicated phase structure this might necessitate introducing a
large number of vertices to the effective action in terms of the
fundamental fields. A way to avoid such a drawback is to
reparameterise the theory in terms of the relevant degrees of freedom 
\cite{Polonyi:2000fv,Gies:2001nw,Gies:2002hq,Harada:2005tw,Polonyi:2001uc,Schwenk:2004hm,Jaeckel:2002rm,Wetterich:2002ky,Schutz:2004rn,Salmhofer1,Dupuis:2005ij}.\smallstep

Fixed point quantities like critical exponents and general anomalous
dimensions have very successfully been derived within the flow
equation approach, mostly in the derivative expansion, see reviews 
\cite{Fisher:1998kv,Litim:1998nf,Morris:1998da,Bagnuls:2000ae,Berges:2000ew,
  Polonyi:2001se,Salmhofer:2001tr}. For the
evaluation of these results in view of quantitative reliability one
has to assess the problem of optimisation. To that end we evaluate the
consequences of the relation between RG scaling and flow for an
appropriate choice of classes of regulators. As an example for the
optimisation criterion developed in section~\ref{sec:optimal}, we
discuss functional optimisation within the zeroth order derivative
expansion. The unique optimised regulator is derived and its extension
to higher order of the truncation scheme is discussed. For explicit
results we refer the reader to the literature, in particular
\cite{Litim:2002cf}.\smallstep 

\subsection{Field reparameterisations}\label{sec:fieldreps}

The derivation of the flow in section~\ref{sec:flows} was based on a
bootstrap approach in which the existence of a renormalised Schwinger
functional in terms of the possibly composite fields $\phi$ was
assumed. This already took into account that the fundamental fields
$\varphi$ may not be suitable degrees of freedom for all regimes of
the theory under investigation. For example, we could consider fields
$\phi(\varphi)$ that tend towards the fundamental fields in the
perturbative regime for large momenta,
\begin{eqnarray}\label{eq:tendpert}
  \phi(\varphi)\stackrel{p^2\to\infty}{\longrightarrow} \varphi\,, 
\end{eqnarray}
while being a non-trivial function of $\varphi$ in the infrared. This
includes the bosonisation of fermionic degrees of freedom 
\cite{Gies:2001nw,Gies:2002hq,Jaeckel:2002rm,Schutz:2004rn}, e.g.\ in
low-energy QCD, where the relevant degrees of freedom are mesons and
baryons instead of quarks. More generally such a situation applies to
all condensation effects. \smallstep

In such a case the Green functions of $\varphi$ will show a highly
non-trivial momentum dependence or even run into singularities.
Moreover, physically sensible truncations to the effective action in
terms of $\varphi$ could be rather complicated. These problems can be
at least softened with an appropriate choice of $\phi$ that mimics the
relevant degrees of freedom in all regimes. Such a choice may be
adjusted to the flow by implementing the transition from $\varphi$ to
$\phi(\varphi)$ in a $k$-dependent way \cite{Gies:2001nw,Gies:2002hq}.
This can be either done by coupling the current and the regulator
to a $k$-dependent field $\hat\phi_k$, or by choosing a $k$-dependent
argument $\phi_k$ of the effective action $\Gamma_k$: \smallstep

The former option leads to additional loop-terms in the flow. The
relation \eq{eq:F=DeltaI} is modified as the full Schwinger functional
$W[J]$ couples to a $k$-dependent field $\hat \phi_k$ with $\partial_t
W[J]=J^{\bf a}\langle \partial_t\hat {\phi_k}_{\bf a}\rangle$, and the
flow operator $\Delta S_2$ changes as the regulator term has an
additional $k$-dependence via the field, $\Delta S_2[\phi,\dot R]\to
\Delta S_2[\phi,\dot R']$ where $R'$ is defined with
\begin{eqnarray}\label{eq:R'}
  \Delta S[\hat\phi_k,\dot R']=\Delta S[\hat\phi_k,\dot R]+\partial_t 
  \hat\phi_k{}_{\bf a}\,\Delta S^{,\bf a}[\hat\phi_k,\dot R]\,, 
\end{eqnarray}
where $\partial_t \hat\phi_k{}_{\bf a}=\partial_t \hat\phi_k{}_{\bf
  a}(\hat\phi_k)$.  With these modifications the derivation of the
flow can straightforwardly be redone.  \smallstep

The latter option keeps the flow \eq{eq:flowtIk} as the partial
derivative is taken at fixed argument $\phi$: $\partial_t \tilde
I_k=\partial_t|_\phi \tilde I_k$. For integrating the flow the total
derivative is required,
\begin{eqnarray}\label{eq:totalflow}
  \0{d\tilde I_k[\phi_k]}{d t}=-\Delta S_2[\phi_k,\dot R]\,
  \tilde I_k[\phi_k]+\partial_t {\phi_k}_{\bf a}\,\tilde I_k^{,\bf a}[\phi_k] 
  \,.
\end{eqnarray} 
We can also combine the above options. For the sake of simplicity we
restrict ourselves to the flow of the effective action which reads in
this general case
\begin{eqnarray}\nonumber 
  &&\hspace{-1cm} 
  \0{d \Gamma_k[\phi]}{d t}=(\Delta S[G\s0{\delta}{\delta \phi}
  +\phi\,,\,\dot R'])-\Delta S'[\phi,\dot R']\\ 
  &&\hspace{1cm} +\left(\partial_t \phi_{\bf a}-
    \langle \partial_t \hat\phi_{\bf a}\rangle\right)\Gamma_k^{,\bf a}\,.
\label{eq:compositeGk}\end{eqnarray}
In \eq{eq:compositeGk} we dropped the subscript ${}_k$ with
$\phi=\phi_k$.  The first term on the rhs is the expectation value of
$\Delta S[\hat\phi, \dot R']$ defined in \eq{eq:R'}. The second term
originates in the definition of $\Gamma_k$ in \eq{eq:GammaR}.  The
expectation value in the second line in \eq{eq:compositeGk} can we
written as $\langle \partial_t \hat\phi\rangle=
((\partial_t{\hat\phi}) [G\s0{\delta}{\delta \phi}+\phi])$, and $\dot
R'$ is defined in \eq{eq:R'}. We remark that \eq{eq:compositeGk} is
finite for $k$-dependences of $\hat\phi$ that are local in momentum
space. General $k$-dependences may require additional
renormalisation.  The flow \eq{eq:compositeGk} can be used in several
ways to improve truncations. \smallstep

A given truncation scheme can be further simplify in a controlled way
by expanding the effective action about a stable solution $\bar\phi$
of the truncated equations of motion, $\Gamma_k^{,\bf a}[\bar\phi]=0$.
Then the second line in \eq{eq:compositeGk} is sub-leading for
$\phi-\bar\phi$ small and can be dropped if restricting the flow to
the vicinity of $\bar\phi$. As this is an expansion about a minimum of
the effective action, such a truncation has particular stability. \smallstep 

The second line also vanishes for $\partial_t \phi- \langle \partial_t
\hat\phi\rangle=0$. Subject to a given $\phi$ we demand $\hat \phi$ to
satisfy
\begin{eqnarray}\label{eq:zero} 
  \langle \partial_t \hat\phi\rangle=\partial_t \phi\,.
\end{eqnarray} 
With \eq{eq:zero} the second line in \eq{eq:compositeGk} vanishes
identically and the flow reduces to the first line. The construction
of $\dot R'$ requires the knowledge $\partial_t \hat \phi(\hat\phi)$.
Within given truncations \eq{eq:zero} turns into a set of loop
constraints that accompany the flow. These constraints resolve the
dependences of the flowing composite fields $\phi_k$ on the
microscopic degrees of freedom. This is more information than required
for solving the flow.  Indeed, we also can use \eq{eq:zero} to
circumvent the necessity of finding $\partial_t\hat\phi(\hat\phi)$. We
write for the expectation value of the second term in \eq{eq:R'}
\begin{eqnarray}\nonumber 
  &&\hspace{-1.5cm}\Delta S^{,\bf b}[\s0{\delta}{\delta J}
  +\phi,R] 
  \gamma^{\bf a}{}_{\bf b}\langle 
  \partial_t \hat\phi_{\bf a}\rangle \\
  &&\hspace{.3cm}  
  =(\Delta S^{,\bf b}[G\s0{\delta}{\delta \phi}+\phi,R] 
  \gamma^{\bf a}{}_{\bf b}\, \partial_t\phi)\,, 
\label{eq:resolvephi_k}
\end{eqnarray}
where we have used \eq{eq:djphi} and \eq{eq:zero}. With \eq{eq:zero}
and \eq{eq:resolvephi_k} we can substitute all dependences on
$\hat\phi,\partial_t \hat\phi$ in the flow \eq{eq:compositeGk} by that
on $\phi_k,\partial_t \phi_k$. We are led to a closed flow for the
effective action
\begin{eqnarray}\nonumber 
  &&\hspace{-1cm} 
  \partial_t \Gamma_k[\phi]
  =(\Delta S[G\s0{\delta}{\delta \phi}
  +\phi\,,\,\dot R])-\Delta S'[\phi,\dot R'] \\ 
  &&\hspace{.4cm}
  +(\Delta S^{,\bf b}[G\s0{\delta}{\delta \phi}+\phi,R] 
  \gamma^{\bf a}{}_{\bf b}\, \partial_t\phi)-
  \partial_t \phi_{\bf a}\Gamma_k^{,\bf a} \,.
\label{eq:compositeGk0}
\end{eqnarray}
The first term in the second line keeps track of the $k$-dependence in
$\hat\phi_k$ necessary to satisfy \eq{eq:zero}. The last term carries
the $k$-dependence of $\phi_k$. For the standard quadratic regulator
\eq{eq:compositeGk0} reads 
\begin{equation} 
  \partial_t \Gamma_k[\phi]
  = G_{\bf bc}\dot R^{\bf bc}+2 R^{\bf ab} G_{\bf ac} 
\dot\phi_{\bf b}{}^{,\bf c}-
  \dot\phi_{\bf a}\Gamma_k^{,\bf a} \,.
\label{eq:compositeGk0s}
\end{equation}
We illustrate the above considerations within simple examples for
quadratic regulator terms \eq{eq:compositeGk0s}. Furthermore the
examples are based on linear relations between $\partial_t\phi$ and
$\phi$. Then \eq{eq:zero} can be resolved explicitly and up to
rescalings \eq{eq:compositeGk0s} simplifies to the standard case: 
we absorb a $t$-dependent wave
function renormalisation $Z_\phi^{1/2}$ into the field:
$\phi_k=Z_\phi^{1/2} \phi_0$ with $\partial_t \phi_k=\gamma_\phi
\phi_k$ with $\gamma_\phi=\partial_t \ln Z_\phi$. \Eq{eq:zero} is
satisfied with $\hat\phi_k= Z_\phi^{1/2}\hat \phi_0$. Then
\eq{eq:compositeGk0s} reduces to
\begin{eqnarray}\label{eq:reduce1}
  \left(\partial_t + \gamma_\phi \phi_{\bf a}
    \s0{\delta}{\delta \phi_{\bf a}}\right) 
    \Gamma_k[\phi]
    = G_{\bf bc}(\partial_t +2 \gamma_\phi) R^{\bf bc}\,, 
\end{eqnarray} 
which also can be obtained by explicitly using $\hat\phi_k=
Z_\phi^{1/2} \hat\phi$. The flow \eq{eq:reduce1} also makes explicit
that the transformation $\phi\to Z_\phi^{1/2} \phi$ is a RG rescaling.
This procedure can be used to fix the flow of vertices.\smallstep

Another simple example is the expansion of the effective action
$\Gamma_k[\phi]$ about its minimum at $\phi_{\rm min}(k)$, implying
$\phi\to \phi_k=\phi-\phi_{\rm min}(k)$.  Such a reparameterisation
guarantees that the minimum is always achieved for $\phi_k=0$. The
flow \eq{eq:flowtIk} only constitutes a partial $t$-derivative, as it
is defined at fixed fields $\phi$. With $\hat
\phi_k=\hat\phi-\hat\phi_{\rm min}(k)$ we satisfy \eq{eq:zero} and we
are led to \eq{eq:compositeGk0} with $\partial_t \phi_k=-\partial_t
{\phi_{\rm min}}$ with $\partial_t \phi_{ \bf b}{}^{,\bf c}=0$. 
The flow \eq{eq:compositeGk0} reduces to the standard flow, 
\begin{eqnarray}\nonumber 
  &&\hspace{-1cm}\partial_t \Gamma_k[\phi]=
  (\Delta S[G\s0{\delta}{\delta \phi}
  +\phi\,,\,\dot R])-\Delta S'[\phi,\dot R]\\ 
  &&\hspace{1cm}+
  \Gamma_k^{,\bf a}[\phi]\, (\partial_t {\phi_{\rm min}})_{\bf a}\,, 
  \label{eq:excomposite} \end{eqnarray} 
now describing a total $t$-derivative of the effective action $\Gamma_k$. 
For quadratic regulators $R^{\bf ab}$ it reads 
\begin{eqnarray}
  \partial_t \Gamma_k[\phi]= \dot R^{\bf ab} G_{\bf ab} +
  \Gamma_k^{,\bf a}[\phi]\, (\partial_t {\phi_{\rm min}})_{\bf a}\,.
\label{eq:standexcomposite} \end{eqnarray} 
The flow of the minimum $\phi_{\rm min}$ can be resolved 
with help of $\s0{d }{d t}(\Gamma^{,\bf a}_k[\phi_{\rm min}])=0$, and reads 
\begin{eqnarray}
  (\partial_t {\phi_{\rm min}})_{\bf a}=
  \left(\0{1}{\Gamma^{(2)}_k[\phi_{\rm min}]
    }\right)_{\bf ab}\partial_t \Gamma_k^{,\bf b
  }[\phi_{\rm min}]\,. 
  \label{eq:resolvemin} 
\end{eqnarray} 
As mentioned before, the examples used linear dependences of
$\partial_t \phi$ on $\phi$. Then \eq{eq:compositeGk0s} can also be
derived explicitly as $\hat\phi$ is known. In the general case this is
not possible, and \eq{eq:compositeGk0} or \eq{eq:compositeGk0s} are 
the fundamental flows. \smallstep

\subsection{RG scaling and optimisation}\label{sec:secRG&opt} 
The reliability of results obtained within functional RG flows hinges
on an appropriately chosen truncation scheme and a regulator choice
that optimises the given truncation scheme.  Without specifying the
truncation scheme the following observation can be made: the
renormalisation group analysis in section~\ref{sec:RG-flows} relates
the RG equation for the full theory with that in the presence of a
regulator. In particular we deduce from \eq{eq:RGflow1PI} and by 
identifying $s$ with the RG scale $\mu$, that the RG equation for the
regularised effective action reads
\begin{eqnarray}\label{eq:1PIRG}
  D_\mu\Gamma_k=\s012 G_{\bf bc} [(D_{\mu} + 
  \gamma_\phi) R]^{\bf bc}\,.
\end{eqnarray}
The right hand side of \eq{eq:1PIRG} entails the modification of the
RG properties in the presence of the regulator. In \eq{eq:1PIRG} we
have restricted ourselves to quadratic regulators. As explained in
detail in the context of optimisation in
chapter~\ref{sec:optimal}, for full flows without truncations
different choices of regulators, in particular those with different RG
properties, lead to a RG rescaling of fields and coupling in the full
effective action $\Gamma$. However, within truncations this
modification usually leads to a physical change of the end-point of
the flow. In turn, this problem is softened if restricting the class
of regulators to those with \cite{Pawlowski:2002eb,Pawlowski:2001df}
\begin{eqnarray}\label{eq:physicalR}
  (D_{\mu} + \gamma_\phi) R =0\,, 
\end{eqnarray} 
where $(\gamma_\phi R)^{\bf ab}=2 \gamma_{\phi}{}^{\bf a}{}_{\bf c}\,
R^{\bf cb}$. The constraint \eq{eq:physicalR} leads to
\begin{eqnarray}\label{eq:nomodRG}
  D_{\mu}\Gamma_k=0\,. 
\end{eqnarray}
For the class of regulators with \eq{eq:physicalR} the regularised
correlation functions satisfy the same RG equation as in the underlying full
theory, in particular this holds for the effective action,
\eq{eq:nomodRG}.  Apart from the general optimisation arguments made
above this facilitates the identification of anomalous dimensions and
critical exponents. Indeed, the choice \eq{eq:physicalR} with the
additional identification $t=\ln\mu$ allows for the straightforward
identification of $t$-running and RG running within fixed point
solutions at all orders of the truncation scheme. \smallstep

An explicit example for a class of regulators in standard flows that
satisfy \eq{eq:physicalR} is provided by
\cite{Pawlowski:2002eb,Pawlowski:2001df}
\begin{subequations}\label{eq:exR}
\begin{eqnarray}\label{eq:exR1}
R^{\bf ab}= \hat\Gamma^{,\bf ac}[\bar\phi] r^{\bf cb}\,, 
\end{eqnarray}
with 
\begin{eqnarray}\label{eq:exR2}
D_{\mu} r=0\,, 
\end{eqnarray}
\end{subequations}
where $ \hat\Gamma^{,\bf ab}$ is $\Gamma^{,\bf ab}$ evaluated at some
background field $\bar\phi$, with a possible subtraction.  The
subtraction can be used to normalise $ \hat\Gamma^{,\bf ab}$. It could
be proportional to $\Gamma^{,\bf ab}$ evaluated at some momentum,
e.g.\ at vanishing momentum. By construction \eq{eq:exR} satisfies
\eq{eq:physicalR} as the two-point function does, $(D_{s} +
\gamma_\phi)_{\bf ac}\hat\Gamma^{,\bf cb}=0$.  If evaluating the
standard flow \eq {eq:standflowGk} for the effective action at the
background field $\bar\phi$, it takes the simple form
\begin{eqnarray}\nonumber 
&&\hspace{-2cm} \dot \Gamma_k[\bar\phi]= 
\s012 \left(\0{1}{1+r}\right)_{\bf bc}\dot r^{\bf bc}\\
&&\hspace{-0.3cm}+
\s012 \left(\0{r}{1+r}\right)_{\bf bc}\left(\0{1}{\Gamma^{(2)}}
\partial_t\Gamma^{(2)}\right)^{\bf bc}\,, 
\label{eq:specflow}  \end{eqnarray} 
where for the sake of simplicity we have taken $\hat\Gamma^{\bf
  ,ab}[\bar\phi]=\Gamma^{\bf ,ab}[\bar\phi]$, that is no subtraction.
The first term on the rhs of \eq{eq:specflow} can be integrated
explicitly and contributes to the effective action only at
perturbative one loop order. The second term gives non-trivial
contributions if the spectral density changes. \Eq{eq:specflow} is 
a spectrally adjusted flow. \smallstep 

In most truncation schemes used in the literature \eq{eq:exR} simply
amounts to the multiplication of the wave function renormalisation
$Z_\phi$.  Then the propagator factorises $G[Z_\phi]=Z_\phi^{-1} G[1]$
which facilitates the computations. It is for the latter reason that
\eq{eq:physicalR} is a standard choice for regulators and it is a
fortunate fact that the simple structure of flows for the choice
\eq{eq:physicalR} goes hand in hand with better convergence towards
physics. \smallstep

\subsection{Integrated flows and fixed points} \label{sec:fixed} 

An optimisation with \eq{eq:indepoptcrit} requires the minimisation of
the norm of the difference between the regularised propagator and the
full propagator with the constraint of keeping a fixed gap, see
\eq{eq:indepoptcrit3}. This implies a fine-tuning of the regulator in
dependence of the two-point function, $\Gamma_{,\bf ab}$.  Here we
outline a way of solving the flow equation which naturally
incorporates such a task and hence minimises the additional numerical
effort. First we turn the flow \eq{eq:flowtIk} into an integral
equation
\begin{subequations}\label{eq:intflowre}
\begin{eqnarray}\label{eq:intflowtI} 
  \tilde I_0=\tilde I_{\Lambda}+\int_{\Lambda}^0\d t\, 
  \Delta S_2 \tilde I_{k}\,, 
\end{eqnarray}
where $\Lambda$ is the initial cut-off scale and the integrated flow 
for the effective action derives from \eq{eq:flowGk} as  
\begin{eqnarray}\nonumber
  &&\hspace{-1.9cm} 
  \Gamma_0=\Gamma_{\Lambda}+\int_{\Lambda}^0\d t\,\Bigl( 
  (\Delta S[G\s0{\delta}{\delta \phi}
  +\phi\,,\,\dot R])\\
  &&\hspace{2.6cm}  +\Delta S'[\phi,\dot R]\Bigr)\,.
\label{eq:intflowGa} \end{eqnarray}
\end{subequations}
\Eq{eq:intflowre} constitutes DSEs as already explained in
section~\ref{sec:DSEs}. As distinguished to standard DSEs they only
involve full vertices and propagators. Such a set of equations can be
solved within an iteration about an ansatz for the full flow
trajectory $\tilde I^{(0)}[\phi,R(k)]$.  The better such an ansatz 
fits the result, the less iterations are needed for convergence
towards the full result $\tilde I^{(\infty)}[\phi,R(k)]$. A benefit of
such an approach is that it facilitates an implementation of the
optimisation criterion \eq{eq:optcriterion} in its form
\eq{eq:indepoptcrit3}. After each iteration step we can prepare our
regulator according to \eq{eq:indepoptcrit3} for the next step. Such a
preparation is in particular interesting for truncations with a
non-trivial momentum dependence for propagators and vertices.
Furthermore the integral equations \eq{eq:intflowre} are likely to be
more stable in the vicinity of poles of the propagator. \smallstep

The integral form \eq{eq:intflowre} also is of use for an analysis of
asymptotic regimes and in particular fixed point solutions.  In
general functional RG methods have been very successfully used within
computations of physics at a phase transition. In particular critical
exponents can be accessed easily.  \smallstep

At $k=0$ the flows \eq{eq:flowtIk} have a trivial fixed point,
$\partial_t \tilde I|_{k=0}\equiv 0$. In case the theory admits a
mass-gap $\Lambda_{\rm gap}$, this can be used to resolve the theory
below this scale hence getting access to the deep infrared behaviour.
For the sake of simplicity we further assume dimensionless couplings.
The dimensionful case will be discussed elsewhere. Then, in the regime
\begin{eqnarray}\label{eq:k<<L}
  k^2\ll \Lambda^2_{\rm gap}\,, 
\end{eqnarray} 
the flow of correlation functions $\tilde I_k$ is parametrically
suppressed by powers of $k/\Lambda_{\rm gap}$,
\begin{eqnarray}\label{eq:0flow}
  \partial_t \tilde I_k = O(k/\Lambda_{\rm gap})\,.  
\end{eqnarray} 
\Eq{eq:0flow} applies in particular to the effective action and its
derivatives. It is convenient to parameterise the correlation
functions $\tilde I_k$ as
\begin{eqnarray}\label{eq:parameter}
  \tilde I_k=\tilde I_0(1+\delta\tilde I_k)\,.
\end{eqnarray} 
Inserting this parameterisation into the integrated flow
\eq{eq:intflowre} we arrive at an integral equation for $\delta\tilde
I_k$,
\begin{eqnarray}\label{eq:intflowdeltI} 
  \tilde \delta I_k=-\int_{k}^0\d t'\, 
  \Delta S_2\,\left( \tilde I_{0}(1+\delta  \tilde I_{k'})\right)\,,  
\end{eqnarray}
where $\Delta S_2$ depends on $\Gamma_k^{,\bf ab}$ (and its
derivatives) that admit the same parameterisation \eq{eq:parameter}.
Assume for the moment that $\delta \tilde I_{k'}$ and $\delta
\Gamma_k^{,\bf ab}$ on the rhs of \eq{eq:intflowdeltI} only depend on 
dimensionless ratios
\begin{eqnarray}\label{eq:ratios}
  \hat p_i= \0{p_i}{k}\,, 
\end{eqnarray} 
where the $p_i$ are momenta of the correlation functions $\tilde I_k$, e.g.\
external momenta of $n$-point vertices. This assumption reads
\begin{eqnarray}\label{eq:dimles}
  \delta \tilde I_k=\delta\tilde I[\hat p_1,...,\hat p_n]
  +O(k/\Lambda_{\rm gap})\,. 
\end{eqnarray}
Inserting \eq{eq:dimles} into the rhs of the integrated flow
\eq{eq:intflowdeltI} we deduce from a scale analysis that the
resulting $\delta I_k$ on the lhs can only depend on dimensionless
ratios $\hat p_i$. A good starting point for the iteration is $\tilde
I_k=\tilde I_0$ with $\delta \tilde I\equiv 0$.  Such a choice
trivially only depends on the ratios \eq{eq:ratios}. Hence this holds
true for each iteration step, and we have proven
\eq{eq:dimles}.\smallstep

Now we invoke the optimisation \eq{eq:optcriterion} with
$D_{R_\bot}\tilde I=\tilde I_0 D_{R_\bot}\delta \tilde I $, and we are
led to the constraint
\begin{eqnarray}\label{eq:zerodelI}
  \int_{k}^0\d t'\, 
  \Delta S_2\,\left( \tilde I_{0}(D_{R'_\bot} \delta \tilde 
    I_{k'})\right)_{R_{\rm stab}}=0\,.
\end{eqnarray}
For positive definite $\delta \tilde I$ a solution to \eq{eq:zerodelI} is
given by $\Delta S_2 \delta \tilde I=0$. In this context we remark
that $\delta\tilde I$ is not a correlation function $\tilde I$, and
the above resolution does not imply a vanishing flow of $\delta \tilde
I$.  An optimisation along these lines was put forward in the infrared
regime of QCD, for details see
\cite{Pawlowski:2003hq,Pawlowski:2004ip}.\smallstep

\subsection{Optimisation in LPA}\label{sec:optLPA}

We continue with a detailed analysis of the optimisation
\eq{eq:optcriterion}, \eq{eq:indepopt} in the LPA a scalar theory with
a single scalar field $a=x$. We shall show that within the LPA the
regulator \eq{eq:litopt} follows as the unique solution to
\eq{eq:reoptcriterion}, see also the more explicit form without
RG scaling, \eq{eq:dtfinstableR}.  For the sake of simplicity we use
the standard flow \eq{eq:standflowtIk} with $\Delta S_2 \tilde I_k =
(G\,\dot R\, G)_{bc} \tilde I_k^{,cb}$. In the LPA we have to evaluate
\eq{eq:reoptcriterion} for constant fields.  Moreover we consider
correlation functions $\tilde I_k$ that are functionals of $\phi$ and not
operators. For example, in the present truncation scheme relevant
correlation functions are provided by
\begin{eqnarray}\label{eq:obrel}
  \int d^d x\, \tilde I^{(n)}_{k,\rm diag}[\phi]=\langle 
  \int d^d x \,\phi^n(x) \rangle_{J^a=\Gamma_k^{,a}+R^{ab}\phi_b}\,,  
\end{eqnarray}
and combinations thereof. In LPA all quantities are evaluated for
constant fields $\bar\phi$. On the rhs of the standard flow
\eq{eq:standflowtIk} the second derivatives $\tilde I_k^{,ab}$ are
required. In LPA they are parameterised as 
\begin{eqnarray}\label{eq:LPAI}
  \tilde
  I^{(2)}_k[\bar\phi](p,q)=\CI_k(\bar\phi,p^2)\delta(p-q)\,.
\end{eqnarray}
We also
need the full propagator $G(p,q)=(\tilde I_k^{(2)}-(\tilde
I_k^{(1)})^2)[\bar\phi](p,q)$, which reads
$1/(\Gamma^{(2)}_k[\bar\phi]+R)(p,q)=1/(p^2+V''[\bar\phi]+R(p^2))\,
\delta(p-q)$. Inserting these objects into \eq{eq:dtfinstableR} we
arrive at
\begin{eqnarray} \label{eq:LPAdtfinstable} \left.\delta
    R_\bot^{{a}_1{a}_2} \0{\delta (G\,\dot R\, G)_{bc} }{\delta
      R^{{a}_1{a}_2}}\, \left(\0{\partial^2\tilde I_k[\bar
        \phi]}{(\partial\bar\phi)^2}\right )^{cb}\right|_{R=R_{\rm
      stab}}=0\,,
\end{eqnarray} 
which we recast in a more explicit form 
\begin{eqnarray}\label{eq:stabLPAexplicit}
  &&\hspace{-.1cm}
  \int \0{d^d p}{(2\pi)^d}\, \left.\delta R_\bot (p^2)\0{\delta}{
      \delta R(p^2)}\right|_{\CI_k} 
  \\ 
  &&\hspace{-.3cm}\times 
  \int  \0{d^d q}{(2\pi)^d}\left.  \frac{\CI_k(\bar\phi,q^2)}
    {(q^2+R(q^2)+V''[\bar\phi])^2} \partial_t R(q^2)
  \right|_{R=R_{\rm stab}}=0
  \,.
  \nonumber \end{eqnarray}
Now we use that a general regulator $R$ can be written as 
$R(q^2)=q^2 r(x)$ with $x=q^2/k^2$, 
if no further scale is present in $R$. This entails that 
$\partial_t R=q^2 \partial_t r(x)=q^2 (-2 x)\partial_x r(x)$. 
Furthermore we 
can rewrite the integration over $q$ as one over 
$x$: $d^d q/(2 \pi^2)^d
=d\Omega_d\, d x\, x^{(d/2-1)}/2$. With these identifications 
we get for the 
$q$-integral in \eq{eq:stabLPAexplicit} after partial 
integration 
\begin{eqnarray}\nonumber 
  &&\hspace{-1.2cm}\Omega_d \CI_k(\bar\phi,0)\delta_{2d}
  + \Omega_d\int_0^\infty dx\, x^{d/2-2}\CI_k\Bigl
  \{\Bigl( (d/2-1)\\ 
  & &\hspace{-1.2cm} +x\partial_x 
  \ln \CI_k \Bigr) \0{r+V''/x}{1+r+V''/x}-\0{V''/x}{(1+r
    +V''/x)^2}\Bigr\} \,.
  \label{eq:LPA2} \end{eqnarray} 
Now we are in a position to discuss the extrema
\eq{eq:LPAdtfinstable}. Searching for minimal flows
is equivalent to searching for 
$r$ that minimise the absolute value of the integrand in \eq{eq:LPA2} 
\begin{eqnarray}\nonumber 
  &&\hspace{-1.5cm}\min_{r}\Bigl|\bigl( (d/2-1)+x\partial_x 
    \ln \CI_k \bigr) \0{r+V''/x}{1+r+V''/x}\\
  &&\hspace{1.5cm}-\0{V''/x}{(1+r+V''/x)^2}\Bigr|\,,
\label{eq:minr}\end{eqnarray} 
where we have left out the overall factor $x^{d/2-2} \CI_k$. A simpler
condition is achieved by neglecting the model-dependent second term
proportional to $V''/x$ leading to
\begin{eqnarray}\label{eq:extcoef}
  \min_{r}
  \0{r+V''/x}{1+r+V''/x}\,.
\end{eqnarray} 
We proceed with the extremisation of the full integrand by taking the
$r$-derivative at fixed $\CI_k$ of the function in \eq{eq:minr}. We
arrive at
\begin{eqnarray}\label{eq:derminr} 
\hspace{-.5cm}  \0{\bigl( (d/2-1)+x\partial_x 
    \ln \CI_k\bigr) (1+r+V''/x)+V''/x}{(1+r+V''/x)^3}\,.
\end{eqnarray}
We remark that subject to $\left( (d/2-1)+x\partial_x \ln \CI_k
\right)>0$ and $V''/x>0$ the $r$-derivative \eq{eq:extcoef} is
positive. Note also that $r+V''/x>0$ cannot be obtained for all $x$
and $\bar\phi$ if the potential $V$ is not convex yet. This statement
holds for all regulators\footnote{All regulator functions have to
  decay with more than $1/x$, the exception being the mass regulator
  with $r=1/x$.}. However, for optimised $r$ the region $V''/x<0$ for
$x$ should have small impact on \eq{eq:LPA2}. If $d\geq 4$ we regain
positivity for vanishing or positive $\partial_x \ln \CI_k$.  Leaving
aside this subtlety we solve \eq{eq:minr} for positive regulators
$r$. As its derivative is positive, \eq{eq:derminr}, this amounts to
minimising $r$
\begin{eqnarray}\label{eq:stab1}
  r_{\rm stab}\leq r\,, \qquad \forall r,x\,.
\end{eqnarray} 
So far we have not used the definition of $\{R_\bot\}$ in
\eq{eq:optcriterion}.  With its use we are straightforwardly led to
\eq{eq:stab2}. Still we would like to evaluate how unique or natural
the choice ${R_\bot}$ is.  If $r$ was an arbitrary positive function
of $x$, \eq{eq:stab1} leads to $r(x)\equiv 0$. However, as $r$ has
been introduced as an IR-regularisation it is inevitably constrained:
it entails an IR-regularisation in momentum space only with
\begin{eqnarray}\label{eq:gap}
  x+x \,r(x)\geq  c
\end{eqnarray} 
for some positive constant $c$. For a proper IR-re\-gulari\-sation the
full propagator $G$ has to display a maximum $G\leq 1/c_{\rm min}$
with $c_{\rm min}=c+V_{\min}''>0$, where $V_{\min}''$ is the minimal
value of $V''$, possibly negative.  For momenta $x> c$ the solution of
\eq{eq:stab1} with \eq{eq:gap} is $r(x>c)\equiv 0$.  For $x<c$ we
saturate the inequality \eq{eq:gap} with $r(x)=c/x-1$. This leads to a
unique solution $r_{\rm stab}$ of \eq{eq:stab1} for $r \in \{r_\bot\}$
defined by \eq{eq:gap}.
\begin{eqnarray}\label{eq:stab2}
  r_{\rm stab}(x)= (c-x) \theta(c-x)\,, 
\end{eqnarray}
which is equivalent to \eq{eq:litopt}. Note that in between \eq{eq:gap}
and \eq{eq:stab2} we have implicitly introduced the set $\{R_\bot\}$
of \eq{eq:optcriterion} by keeping $c$ fixed while minimising $r$.
Still, such a procedure was naturally suggested by the computation.
\smallstep

Above we have restricted ourselves to correlation functions $\tilde I_k$ with
$\left( (d/2-1)+x\partial_x \ln \CI_k \right)>0$. If we discuss
optimisation on the set of $\int d^d x\,\tilde I^{(n)}_{k,\rm diag}$ ,
\eq{eq:obrel} they lead to $\CI^{(n)}_k\propto 1/(q^2+R+V'')^n$. For
large $n$ the contributions of $x\partial_x \ln \CI_k$ will dominate
the $x$-integral in \eq{eq:LPA2}. Minimising the absolute value of the
integral then amounts to solving \eq{eq:extcoef}, so we still have to
minimise $r$. Note also that this does not extremise the flow of all
correlation functions $\int d^d x\,\tilde I^{(n)}_{k,\rm diag}$. \smallstep

It is also interesting to speculate about the most instable regulator.
It is found by maximising the integrand in \eq{eq:LPA2} in the
regularised momentum regime. This is achieved for $r_{\rm
  instab}=\infty$. If we also demand that $r$ is monotone and that
the gap \eq{eq:gap} is saturated at some momentum, this leads to
\begin{eqnarray}\label{eq:instab}
r_{\rm instab}(x)=1/\theta(x-c)-1\,,
\end{eqnarray}
the sharp cut-off. Note that this argument concentrates on instability
of the low momentum region of the flow. For high momenta maximal
instability is obtained for the regulator $R_{\rm CS}=k_{\rm eff}^2$,
the mass cut-off. The related flow equation is an un-renormalised
Callan-Symanzik equation. Indeed, the results for critical exponents
for scalar models in LPA are worse for the mass regulator
\cite{Litim:2002cf} than that for the sharp cut-off. \smallstep

The stable and instable regulators \eq{eq:stab2} and \eq{eq:instab}
have been derived from \eq{eq:optcriterion} by dropping
correlator-dependent terms. The regulators \eq{eq:stab2} and
\eq{eq:instab} can also be derived from \eq{eq:indepoptcrit4} in a
very simple manner. In the present truncation \eq{eq:indepoptcrit4} 
has to be evaluated on $L_2$ and boils down to
\begin{eqnarray}\label{eq:stabsim}
  \0{1}{x+x r_{\rm stab}(x)+V''}\geq \0{1}{x+x r_{\bot }(x)+V''}\,,
\end{eqnarray}
which can be converted into \eq{eq:stab1}. This nicely shows the
advantage of a simple functional criterion. \smallstep

Beyond LPA we are led to integrals as in \eq{eq:stabLPAexplicit} that
also contain derivatives w.r.t.\ $q$. Then $r$ also has to be
differentiable to the given order. Such regulators exist, they are
simply differentiable enhancements of \eq{eq:stab2}. \smallstep

\subsection{Optimisation in general truncation
  schemes}\label{sec:beyond}

In a general truncation and higher
truncation order the correlation functions $\tilde I_k$ resolve more
structure of the flow operator $\Delta S_2$. Roughly speaking, a
solution to the functional optimisation criterion \eq{eq:optcriterion}
minimises the expansion coefficients of $\Delta S_2$ for a given
truncation scheme. For example, in higher order derivative expansion
the flow $\Delta S_2 \tilde I_k$ is projected on the part that
contains higher order space-time derivatives.  In momentum space and
resorting to the representation \eq{eq:functopt} of the functional
optimisation criterion \eq{eq:optcriterion}, this amounts to
differentiability of $G\psi(p)$ w.r.t.\ momentum at the given order.
Consequently the norm has to be taken in the space of differentiable
functions with
\begin{eqnarray}\nonumber 
  && \hspace{-.6cm}
  \|\psi\|_{n}^{2}
  =\sum_{|\alpha|\leq n}\0{n!}{(n-|\alpha|)!\, \alpha_1!\cdots 
    \alpha_d !} \left\|\0{\partial^{|\alpha|}\psi(p)}{\partial 
      p^{\alpha_1}
      \cdots \partial p^{\alpha_d}}\right\|_{L_2}^2\\ 
  &&\label{eq:H^n}\end{eqnarray}
where $\alpha\in \N^d$ and $|\alpha|=\sum \alpha_i$.  \Eq{eq:H^n}
defines the norm on Sobolev-spaces $H^n$ with $n\in \N$. Applied to
the functional optimisation criterion, and leaving aside the
intricacies discussed in section~\ref{sec:janscriterion} we arrive at
the following optimisation in $n$th order derivative expansion:
\begin{eqnarray}\nonumber 
  &&\hspace{-1cm} 
  \| \theta_\lambda (G[\phi_0,R_{\rm stab}])
  -\theta_\lambda (G[\phi_0,0])\|_n^{\ }\\
  &&\hspace{0cm} =\min_{R_\bot}\| 
  \theta_\lambda (G[\phi_0,R_{\bot}])
  -\theta_\lambda (G[\phi_0,0])\|_n^{\ }\,, 
  \label{eq:nth} \end{eqnarray}
for all $\lambda\in \R^+$. Here $\phi_0$ is either defined by the
minimum of the potential or it maximises the propagator.
$\theta_\lambda$ has to meet the requirement of boundedness w.r.t\ the
norm $\|.\|_n$, as already discussed below \eq{eq:theta_l}. This is
achieved by using a $n$th-order differentiable version of
\eq{eq:theta_l}.  We emphasise that the form of $\theta_\lambda$ is of
no importance for the present purpose. The optimisation with \eq{eq:nth}
seems to depend on the full two-point function
$\Gamma^{(2)}[\phi_0,R=0]$. Now we proceed with the specific norm
$\|.\|_{n}^{\ }$ as indicated in section~\ref{sec:janscriterion} below
\eq{eq:indepoptcrit3}. The constraint \eq{eq:nth} entails that the
spectral values of $G[\phi_0,R_{\rm stab}]$ are as close as possible
to that of the full propagator $G[\phi_0,0]$. Moreover it entails
maximal smoothness.  Hence \eq{eq:indepoptcrit3} can be reformulated
as
\begin{eqnarray}\nonumber 
  &&\hspace{-1cm}\| \theta_\lambda (\Gamma^{(2)}[\phi_0,R_{\rm stab}]
  +R_{\rm stab}])\|^{\ }_n\\
  &&\hspace{0cm}= 
  \min_{R_\bot} \| \theta_\lambda (\Gamma^{(2)}[\phi_0,R_{\bot}]
  +R_{\bot}])\|_{n}^{\ }\,, 
  \label{eq:nth0} \end{eqnarray}
for all $\lambda\in \R^+$. A solution of \eq{eq:nth0} provides a
propagator $G[\phi_0,R_{\rm stab}]$ which is as close as possible to
the full propagator $G[\phi_0,0]$ as well as having minimal
derivatives of order $i\leq n$. \Eq{eq:nth0} also leads to the 
supplementary constraint for the stability criterion \eq{eq:optbyprop}. 
The maximisation of the gap has to be supplemented by the minimisation 
of  
\begin{eqnarray} \label{eq:nth0con}
  \| \Gamma^{(2)}[\phi_0,R_{\rm stab}](p_0^2)
  +R_{\rm stab}(p_0^2)]\|^{\ }_n\,, 
\end{eqnarray} 
within the class of $R_{\rm stab}$ singled out by \eq{eq:optbyprop}.
Here $p_0$ is the momentum at which the propagator takes its maximum. 
For an implementation of \eq{eq:nth0con} see \cite{Litinprep}.
\smallstep

In truncation schemes that carry a non-trivial momentum and field
dependence
\cite{Reuter:1993kw,Reuter:1997gx,Ellwanger:1995qf,Ellwanger:1996wy,Litim:2002ce,Pawlowski:2003hq,Pawlowski:2004ip,Gies:2002af,Gies:2003ic,Fischer:2004uk,Blaizot:2004qa,Blaizot:2005xy},
functional optimisation suggests the use of background field dependent
regulators, or even regulators with a non-trivial dependence on the
full field. Evidently in the latter case structural truncations of the
flows are inevitable, see also \cite{Litim:2002xm,Litim:2002hj}. In
case momentum and field dependence are intertwined, as happens in the
interesting truncation scheme put forward in
\cite{Blaizot:2004qa,Blaizot:2005xy}, functional optimisation directly
implies the use of (background) field dependent regulators. \smallstep

We continue with a brief discussion of a peculiar case relevant for the
optimisation of QCD-flows in Landau gauge QCD as initiated in 
\cite{Pawlowski:2003hq,Pawlowski:2004ip}. In case the spectral values
$\lambda(p^2)$ of the full propagator are not monotone in momentum,
an optimised regulator does not resolve the theory successively in
momentum. This happens for the gluon propagator in Landau gauge QCD
\cite{vonSmekal:1997is1,Alkofer:2000wg,Pawlowski:2003hq,Fischer:2004uk}.
A propagator that is monotone in momentum violates the condition
$-\partial_t G\geq 0$ for some interval in $t$ and some spectral
values. This implies that the flow trajectory is not minimised for
these spectral values. In turn, an optimised gluonic regulator can be
constructed from
\begin{eqnarray} \nonumber 
  &&\hspace{-1cm}R_{A,\rm stab}(p^2)\sim(Z_\phi k^2_{\rm eff}
  -\Gamma_k^{(2)}(p^2))
  \theta[Z_\phi k^2_{\rm eff}-\Gamma_0^{(2)}(p^2)]\\ 
  && +
  (\Gamma_0^{(2)}(p^2)-\Gamma_k^{(2)}(p^2))
  \theta[\Gamma_0^{(2)}(p^2)-Z_\phi k^2_{\rm eff}]\,,
\label{eq:optimalQCD}\end{eqnarray}
where $\Gamma_0^{(2)}(p^2)$ is the full two-point function at
vanishing regulator, and a possibly smoothened step-function $\theta$.
Note that \eq{eq:optimalQCD} boils down to the regulator \eq{eq:gap}
within LPA. The practical use of the suggestion \eq{eq:optimalQCD}
calls for an iterative solution of the flow about a suggestion
$\Gamma_0^{(2)}$ as described in section~\ref{sec:fixed}. Apart from
guaranteeing the mSTI, it also necessitates an appropriate choice of
the renormalisation conditions.  The latter ensures UV finiteness of
such a flow.  We also remark that within this approach further terms
are required on the rhs of \eq{eq:optimalQCD} in order to guarantee
that the regulator vanishes if the cut-off scale tends to zero. The
gluonic regulator \eq{eq:optimalQCD} has to be accompanied with
appropriate choices for ghost and quark regulators $R_C$ and $R_q$
respectively. A combined optimisation in $(R_A,R_C,R_q)$ may lead to a
successive integrating out of fields as found already in the 
IR-optimisation in \cite{Pawlowski:2003hq,Pawlowski:2004ip}
\footnote{The proof of an extremum being global is intricate}. More
details will be provided elsewhere \cite{pawlowski}. \smallstep

In the light of the above results we add a further brief comment on the
physical interpretation of optimisation as introduced in
chapter~\ref{sec:optimal} . The optimisation criterion is constructed
from stability considerations. Stability implies minimal integrated
flows and hence quickest convergence towards physics. At each order of
the given truncation scheme the optimised propagators and correlation
functions are as close as possible to the full propagator and
correlation functions respectively. This minimises regulator
artefacts, and triggers a most rapid approach towards the full theory.
Moreover, optimised flows preserve the RG properties of the full
theory within the regularisation as well as gradient flows, see
\eq{eq:localint}.  The above arguments emphasise the close structural
relation of the optimisation criterion to the construction of both
improved and perfect actions in lattice theory \footnote{An adaptation
  of the criterion \eq{eq:optcriterion} for lattice regularisations
  leads to improved actions and operators at lowest order of an
  expansion scheme based on the lattice spacing.}. We emphasise that
the optimisation can be implemented within an iterative procedure
which leads to small additional computational costs.

\section{Conclusions}\label{sec:conclusions}
The present work provides some structural results in the functional RG
which may prove useful in further applications, in particular in gauge
theories. We have derived flows \eq{eq:genflowtIk} and their
one-parameter reductions \eq{eq:flowIk},\eq{eq:flowtIk} and
\eq{eq:RGflowtIk} valid for a general class of correlation functions
$\tilde I_k$ defined in \eq{eq:Ik} with \eq{eq:tildeIk}. This class of
correlation functions $\tilde I_k$ includes $N$-point functions as
well as Dyson-Schwinger equations, symmetry relations such as
Slavnov-Taylor identities, and flows in the presence of composite
operators, e.g.\ $N$-particle irreducible flows. The present
formulation also allows us to directly compute the evolution of
observables in gauge theories. For example, the flows \eq{eq:flowtIk},
\eq{eq:RGflowtIk} hold for the Wilson loop and correlation functions
of the Polyakov loop, see section~\ref{sec:parameter}. This is a very
promising approach to the direct computations of observables in the
non-perturbative regime of QCD, e.g.\ the order parameter of the
confinement-deconfinement phase transition. In
section~\ref{sec:fieldreps} we derived closed flows in the presence of
general scale-dependent reparameterisations of the theory. This
extends the options for scale-adapted parameterisations of the theory,
and is particularly relevant in the context of
rebosonisation.\smallstep

The functional framework developed here was used to systematically
address the important issue of optimisation, and to derive a
functional optimisation criterion, see section~\ref{sec:janscriterion}
\eq{eq:optcriterion},\eq{eq:indepopt},\eq{eq:functopt}. Optimal
regulators are those, that lead to correlation functions as close as
possible to that in the full theory for a given effective cut-off
scale. The criterion allows for a constructive use, and it is
applicable to general truncation schemes. It can be also used for
devising new optimised schemes, for examples see
section~\ref{sec:optimalatwork}, in particular
section~\ref{sec:optLPA},\ref{sec:beyond}. The use of optimisation
methods becomes crucial in more intricate physical problems such as
the infrared sector of QCD, and can be used to resolve the pending
problem of full UV-IR flows in QCD. \smallstep

Another important structural application concerns renormalisation
schemes for general functional equations, e.g.\ DSEs and $N$PI
effective actions.  The functional flows \eq{eq:genflowtIk} can be
used for setting up of generalised BHPZ-type renormalisation schemes
that are, by construction, consistent within general truncation
schemes, see sections~\ref{sec:DSErenorm},\ref{sec:comprenorm}.
Moreover, such subtraction schemes are very well adapted for numerical
applications.  \smallstep

The present setting also allows for a concise and flexible
representation of symmetry constraints, which is particularly relevant
in gauge theories. So far, the practical implementation of modified
Slavnov-Taylor identities was restricted to their evaluation for
specific momentum values. The present setting allows for a functional
implementation that possibly adapts more of the symmetry, see
section~\ref{sec:msti},\ref{sec:flow&alt}. This opens a path towards
improved truncation schemes in gauge theories relevant for a more
quantitative computation in strongly interacting sectors of QCD.  The
above analysis also applies the Nielsen identities for gauge invariant
flows of the geometrical effective action. \smallstep 

In summary we have presented structural results that further our
understanding of the Functional Renormalisation Group. These results
can be used to qualitatively and quantitatively
improve FRG applications.\\[-3ex]

{\bf Acknowledgement} \\[-3.55ex]

\noindent It is a great pleasure to thank R.~Alkofer, J.~Berges,
C.~Ford, S.~Diehl, H.~Gies, D.~F.~Litim, J.~Polonyi, U.~Reinosa,
B.-J.~Schaefer, L.~von Smekal and I.-O.~Stamatescu for numerous
interesting discussions and helpful comments on the manuscript. 
I acknowledge DFG support under contract GI328/1-2.


\renewcommand{\thesection}{}
\renewcommand{\thesubsection}{\Alph{subsection}}
\renewcommand{\theequation}{\Alph{subsection}.\arabic{equation}}

\section*{APPENDIX}

\setcounter{equation}{0}

\subsection{Metric}\label{app:notation}
This appendix deals with the non-trivial metric $\gamma$ in field
space in the presence of fermions. The ultra-local metric $\gamma$ is
diagonal in field space for scalars and gauge fields and is given by
the $\epsilon$-tensor in fermionic space.  For
$\varphi_a=(\psi,\bar\psi)_a$ the fermionic metric reads
\begin{eqnarray}\label{eq:fmetric}
(\gamma^{ab})=\btensor{(}{cc} 0 &-1\\ 1& 0\etensor{)}\,. 
\end{eqnarray}
For raising and lowering indices we use the Northwest-Southeast
convention,
\begin{eqnarray}\nonumber 
  \phi^a&=& \gamma^{ab} \phi_b\,,\\ 
  \qquad \phi_a&= &\phi^b\gamma_{ba}\,.
\label{eq:nwse}\end{eqnarray}
The metric has the properties
\begin{eqnarray}\nonumber 
  \hspace{-.3cm}\gamma_b{}^a&=&\gamma^{ac}\gamma_{bc}=\delta^{a}_b\,,\\
  \gamma^{a}{}_{b}&=&\gamma^{ac}\gamma_{cb}=
  (-1)^{ab}\delta^a_b\,,
\label{eq:metprops} \end{eqnarray}
where 
\begin{eqnarray} \label{eq:ab} (-1)^{ab} =\left\{\begin{array}{rl} -1,
      & \quad
      a\ {\rm and}\ b\ {\rm fermionic}\\
      1 & \quad {\rm otherwise.}\end{array}\right.
\end{eqnarray} 
\Eq{eq:metprops} extends to indices ${\bf a} =a_1\cdots a_n$ and ${\bf
  b}=b_1\cdots b_m$ with 
\begin{eqnarray} \label{eq:genab}
(-1)^{\bf ab} =\left\{\begin{array}{rl} -1 & \quad 
{\bf a}\ {\rm and}\ {\bf b}\ {\rm contain\ 
odd}\ \#\ {\rm of}\\ 
 & \quad {\rm fermionic\ indices,}\\[1ex]
1 & \quad {\rm otherwise.}\end{array}\right.
\end{eqnarray} 
For arbitrary vectors $\phi,\tilde\phi$ the properties
\eq{eq:metprops} lead to
\begin{eqnarray}\label{eq:order}
  \tilde \phi^a \phi{}_a=\phi^a \tilde \phi_a=\tilde \phi_a \phi^b
  \gamma^a{}_b =\phi_b  \tilde \phi^a \gamma^b{}_a\,.
\end{eqnarray}
Due to the Grassmann nature of the fermionic variables
$\psi,\tilde\psi$ the order is important $ \psi^i \tilde\psi_i=-
\psi_i\tilde\psi^i$. \smallstep

We close this appendix with an example. In general a (composite) field
$\phi$ consists of scalar components, gauge fields and fermions, the
fundamental field reads in components
\begin{eqnarray}\nonumber 
(\phi_i)&=& (\varphi,A,\psi,\bar\psi)\,,\\ 
(\phi^i)& =&  (\varphi,A,\bar\psi,-\psi)\,. 
\label{eq:efields} \end{eqnarray} 
The contraction of the fundamental $\phi$ with itself leads to 
\begin{eqnarray}\nonumber 
  &&\hspace{-1.2cm}\phi^a\phi_a = \phi_b\gamma^{ab}\phi_a=\int d^d x\,\Bigl(
  \varphi_{\alpha_n}(x)\varphi_{\alpha_n}(x)\\
  &&\hspace{.5cm}+A^\mu_\alpha(x) 
  A_\mu^\alpha(x)+2\bar 
  \psi_\alpha^\xi(x)\psi^\alpha_\xi(x)
  \Bigr)\,  
\label{eq:defofphi_i}\end{eqnarray}
where $n$ labels the number of scalar fields, $\alpha$ the gauge
group, and $\xi$ sums over spinor indices and flavours. The current
$J$ related to $\phi$ is given by
\begin{eqnarray}\nonumber 
(J_a)&=&(J_\varphi,J_A, J_{\bar\psi},J_{\psi})\\
 (J^a)&=&(J_\varphi,J_A, J_{\psi},-J_{\bar\psi})\,, 
\end{eqnarray}
which implies schematically 
\begin{eqnarray}
J^a \phi_a=(J_\varphi\varphi+J_A A+J_{\psi}\psi+\bar\psi 
J_{\bar\psi}). 
\end{eqnarray}
Moreover 
\begin{eqnarray}\label{eq:erelations}
J^a \phi_a=\phi^a J_a=J_a \phi^b \gamma^a{}_b=\phi_b J^a 
\gamma^b{}_a\,.  
\end{eqnarray}

\setcounter{equation}{0}
\subsection{Derivatives}\label{app:derivatives} 
We deal with derivatives of functionals $F[f]$ w.r.t.\ $f(x)=\phi(x)$
or $f(x)=J(x)$. Derivatives are denoted as
\begin{eqnarray}\label{eq:derivef}
F_{,a}[f]:= \0{\delta F[f]}{\delta f^a}\,,\qquad \qquad 
F^{,a}[f]:=\0{\delta F[f]}{\delta f_a}\,, 
\end{eqnarray} 
that is, derivatives are always taken w.r.t.\ the argument of the
functional $F$. \Eq{eq:derivef} implies
\begin{eqnarray}\label{eq:not} 
 F^{,a}[f]=\gamma^{ba}F_{,b}[\phi], \quad \qquad  
 F_{,a}[f] =  \gamma_{ab}F^{,b}[f], 
\end{eqnarray}
which has to be compared with \eq{eq:order}.  We also take derivatives
w.r.t.\ some (logarithmic) scale $s$, e.g.\ $s=t=\ln k$. The total
derivative of some functional $F$ splits into
\begin{eqnarray}\nonumber 
\0{d\, F[J]}{d s}& =& \partial_s F[J]+\partial_s J^a\,F_{,a}[J]\,,\\
\0{d\, F[\phi]}{d s}&=& \partial_s F[\phi]+
\partial_s \phi_a\,F^{,a}[\phi]\,,
\label{eq:derives}\end{eqnarray} 
i.e., $\partial_s F[\phi]=\partial_s|_\phi F[\phi]$ and $\partial_s
F[J]=\partial_s|_J F[J]$. Partial derivatives w.r.t.\ the logarithmic
infrared scale $t=\ln k$ we abbreviate with
\begin{eqnarray}\label{eq:dot} 
\dot F=\partial_t F. 
\end{eqnarray} 
General differential operators are similarly defined as
\begin{eqnarray} \nonumber 
\hspace{-.4cm}
D_s F[J]&=& (\partial_s + \gamma_{g}{}^{i}{}_j g_i\partial_{g_j}+ 
 \gamma_J{}^{\bf a}{}_{\bf b} J^{\bf b}
\s0{\delta}{\delta J^{\bf a}} ) F[J],\\  
\hspace{-.4cm} 
D_s F[\phi]&=&(\partial_s +\gamma_{g}{}^{i}{}_j g_i  \partial_{g_j}+ 
 \gamma_\phi{}^{\bf b}{}_{\bf a} \phi_{\bf b}
\s0{\delta} {\delta \phi_{\bf a}} ) F[\phi], 
\end{eqnarray} 
with partial derivatives according to \eq{eq:derives}. The definitions
of this appendix directly carry over to the case of multi-indices
${\bf a},\bf b$.

\setcounter{equation}{0}
\subsection{Definition of $\Delta S_n$}\label{sec:Delta S_n}
The part of $\Delta S$ that contains at least $n\geq 1$ derivatives
w.r.t.\ the variable $x$, e.g.\ $x[J]=J,\phi$, acting to the right, is
given by
\begin{subequations}\label{eq:DelSi}
\begin{eqnarray}\label{eq:DelSi1}
  \Delta S_n[x,\dot R]=
  \Delta S_{{\bf a}_1\cdots {\bf a}_n}[x,\dot R] \0{\delta}{ 
    \delta x_{ {\bf a}_1}}\cdots  
  \0{\delta}{ \delta x_{ {\bf a}_n}}\,,
\end{eqnarray}
with coefficient 
\begin{eqnarray}\nonumber 
  \hspace{-.6cm}
  && \hspace{-1cm}\Delta S_{{\bf a}_1\cdots {\bf a}_n}[x\,,\, 
  \dot R]\\
  &&\hspace{-.1cm}=\sum_{i\geq n}  
  (\Delta S_{{\bf a}_1\cdots {\bf a}_{i}}[x\,,\, 
  \dot R])\0{\delta}{ \delta x_{ {\bf a}_1}}\cdots  
  \0{\delta}{ \delta x_{ {\bf a}_{i-n}}}\,.   
\label{eq:DelSi2}\end{eqnarray}
\end{subequations}
The coefficients $\Delta S_{{\bf a}_1\cdots {\bf a}_n}$ are operators.
The functionals $(\Delta S_{\bf a_n\cdots a_1})$ are the coefficients
in a Taylor expansion of the operator $\Delta S$ in powers of
$\s0{\delta}{\delta x}$, 
absorbing $n$ derivatives w.r.t.\ $x$ of $\Delta S[ \s0{\delta}{\delta
  J}+\phi\,,\,\dot R]$. We emphasise that $(\Delta S_{{\bf a}_1\cdots
  {\bf a}_n}[x\,,\, \dot R])$ is a functional, it contains no
derivative operators. If interested in $x=J$, the expansion
coefficients $(\Delta S^{{\bf a}_1\cdots {\bf a}_n}[x\,,\, \dot R])$
boil down to the Taylor coefficients in an expansion of $\Delta S$ in
$\phi_{\bf a}$. They are the $n$th right derivatives of 
$\Delta S[x,\dot R]$ w.r.t.\ $x_{\bf a}$, evaluated at 
$x=\s0{\delta}{\delta J}+\phi$. 

\setcounter{equation}{0}
\subsection{Standard 1PI flows}\label{app:quadratic1PI} 
For quadratic regulators \eq{eq:standreg} and ${\bf a}=a$ the flow
\eq{eq:preflowtIk} reads more explicitly
\begin{eqnarray}\label{eq:standpreflowtIk}
&&\hspace{-.9cm}\partial_t \tilde I_k+\s012  (G\,\dot R\, G)_{bc} \tilde 
I_{k}{}^{,cb}-\Bigl((\partial_t J^a)\\\nonumber 
&&\hspace{-.5cm} +\s012  (G\,\dot R\, G)_{bc}\,\Gamma_k{}^{,cbd}
\,\gamma^a{}_d 
-\phi_b \dot R^{ba}\Bigr)\, G_{ad}\tilde I_k{}^{,d}=0\,, 
\end{eqnarray}
where 
\begin{eqnarray*}\label{eq:GRG} 
(G\,\dot R\, G)_{bc}
=  G_{ba}\dot R^{ad} G_{dc}\, .
\end{eqnarray*}  
For the derivation of \eq{eq:standpreflowtIk} we have to express
$\Delta S[\s0{\delta}{\delta J},\dot R]$ in terms of derivatives
w.r.t.\ $\phi$ with the help of \eq{eq:djphi}. For bosonic variables
this is straightforwardly done. If fermionic variables are involved the
ordering of terms becomes important. We shall argue that
\begin{eqnarray}\nonumber
  && \hspace{-1.2cm}\dot R^{ab}\,\s0{\delta}{\delta J^a} 
  \s0{\delta}{\delta J^b}\\\nonumber 
  &&\hspace{-.8cm}= G_{ac}\s0{\delta}{\delta \phi_c} 
  \dot R^{ab} G_{bd}\s0{\delta}{\delta \phi_d}\\
  &&\hspace{-.8cm}= G_{ac}
  \dot R^{ab} (\s0{\delta}{\delta \phi_c}  G_{bd}) 
  \s0{\delta}{\delta \phi_d}+G_{ca}
  \dot R^{ab}  G_{bd} 
  \s0{\delta}{\delta \phi_d}\s0{\delta}{\delta \phi_c}\,.
 \label{eq:tek1}\end{eqnarray} 
The only non-trivial term is the last one on the right hand side.
\Eq{eq:Rsyms} entails that for $a$ being bosonic (fermionic), $b$ is
bosonic (fermionic). If either $a$ or $c$ or both are bosonic we
conclude $G_{ac}=G_{ca}$.  Moreover either $\s0{\delta}{\delta
  \phi_c}$ and $ G_{bd} \s0{\delta}{\delta \phi_d}$ or both are
bosonic and \eq{eq:tek1} follows. If $a,c$ both are fermionic,
$\s0{\delta}{\delta \phi_c}$ and $ G_{bd} \s0{\delta}{\delta \phi_d}$
are fermionic (as $b$ is fermionic) and we have $G_{ac}=-G_{ca}$. It
follows that
\begin{eqnarray}\label{eq:tek2}
  \s0{\delta}{\delta \phi_c} G_{bd} 
  \s0{\delta}{\delta \phi_d} = (\s0{\delta}{\delta \phi_c} 
  G_{bd}) 
  \s0{\delta}{\delta \phi_d}- G_{bd} 
  \s0{\delta}{\delta \phi_d}\s0{\delta}{\delta \phi_c}\,. 
\end{eqnarray} 
Inserting \eq{eq:tek2} into \eq{eq:tek1} the right hand side follows.
We also conclude that for $b,c$ fermionic
\begin{eqnarray}\label{eq:tek3}
  \s0{\delta}{\delta \phi_c} G_{bd}=G_{be} \Gamma_k{}^{,ecf}
  G_{gd} \,\gamma^g{}_f \,.
\end{eqnarray} 
The factor $\gamma^g{}_f$ originates in \eq{eq:propk}, $G_{ac}
(\Gamma_k{}^{,cb}+R^{bc})= \gamma^b{}_a$.  Inserting \eq{eq:tek3} into
\eq{eq:tek1} we arrive at
\begin{eqnarray}\nonumber 
  &&\hspace{-1.2cm}\dot R^{ab}\,\s0{\delta}{\delta J^a} 
  \s0{\delta}{\delta J^b}= G_{ab}\dot R^{bc} G_{cd}\,
  \s0{\delta}{\delta \phi_d}\s0{\delta}{\delta \phi_a}\\
  &&\hspace{.8cm}
  -(G\,\dot R\, G)_{ad} 
  \Gamma_k{}^{,daf} \,\gamma^g{}_f G_{ge}\, 
  \s0{\delta}{\delta \phi_e}\,. 
  \label{eq:tek4} \end{eqnarray} 
with $(G\,\dot R\,G)_{ad}=G_{ab}\dot R^{bc}\, G_{cd}$.

\vfill

\eject

\end{document}